\documentclass[11pt]{article}
\newcounter{ourcount}
\usepackage{amsmath,amsfonts,amssymb,graphicx,epsfig,pdflscape,multirow,theorem,gensymb}
\usepackage{lscape,arydshln}
\usepackage[matrix,arrow,curve]{xy} 
\numberwithin{equation}{section}
\graphicspath{{./eps/}}
\usepackage[nosort]{cite}
\usepackage[usenames]{color}
\usepackage{pstricks}
\usepackage{pst-plot}
\usepackage[backref=false]{hyperref}
\usepackage[capitalise,noabbrev]{cleveref} 
\hypersetup{
colorlinks=true,
citecolor=red,
linkcolor=darkblue,
urlcolor=darkblue
}
\definecolor{darkblue}{rgb}{0,0,.8}
\definecolor{red}{rgb}{1,0,0}

\theorembodyfont{\itshape} 
\theoremheaderfont{\scshape}
\theoremstyle{plain}

\newtheorem{Proposition}{Proposition}[section]

\numberwithin{equation}{section}

\newcommand{\nc}{\newcommand}

\def\arxiv#1#2{\href{http://arxiv.org/abs/#1}{\textsf{arXiv:#1 #2}}}
\nc{\ir}{\mathrm{i}}
\nc{\eE}{\mathsf{e}}
\nc{\bib}{\bibitem}
\nc{\be}{\begin{equation}}
\nc{\ee}{\end{equation}}
\nc{\chit}{\raisebox{0.25ex}{$\chi$}}
\nc{\pdTL}{\mathsf{pdTL}}
\nc{\Dbh}{\mbox{\boldmath $\hat D$}}
\nc{\Dh}{\mbox{$\hat D$}}
\nc{\Dbb}{\mbox{\boldmath $\bar D$}}
\nc{\Dbm}{\mbox{\boldmath $\mathcal D$}}
\nc{\Dbt}{\mbox{\boldmath $\tilde{D}$}}
\nc{\Tbt}{\mbox{\boldmath $\tilde{T}$}}

\nc{\db}{\mbox{\boldmath $d$}}
\nc{\Ab}{\mbox{\boldmath $A$}}
\nc{\Bb}{\mbox{\boldmath $B$}}
\nc{\Cb}{\mbox{\boldmath $C$}}
\nc{\Db}{\mbox{\boldmath $D$}}
\nc{\eb}{\mbox{\boldmath $e$}}
\nc{\Fb}{\mbox{\boldmath $F$}}
\nc{\Fbt}{\mbox{\boldmath $\tilde{F}$}}
\nc{\fb}{\mbox{\boldmath $f$}}
\nc{\fbt}{\mbox{\boldmath $\tilde{f}$}}
\nc{\Gb}{\mbox{\boldmath $G$}}
\nc{\Hb}{\mbox{\boldmath $H$}}
\nc{\Ib}{\mbox{\boldmath $I$}}
\nc{\Jb}{\mbox{\boldmath $J$}}
\nc{\Kb}{\mbox{\boldmath $K$}}
\nc{\Lb}{\mbox{\boldmath $L$}}
\nc{\Mb}{\mbox{\boldmath $M$}}
\nc{\Pb}{\mbox{\boldmath $P$}}
\nc{\Qb}{\mbox{\boldmath $Q$}}
\nc{\Rb}{\mbox{\boldmath $R$}}
\nc{\Tb}{\mbox{\boldmath $T$}}
\nc{\Tbb}{\mbox{\boldmath $\bar T$}}
\nc{\Tbh}{\mbox{\boldmath $\hat T$}}
\nc{\Tbm}{\mbox{\boldmath $\mathcal T$}}
\nc{\tb}{\mbox{\boldmath $t$}}
\nc{\Ub}{\mbox{\boldmath $U$}}
\nc{\Vb}{\mbox{\boldmath $V$}}
\nc{\Wb}{\mbox{\boldmath $W$}}
\nc{\xb}{\mbox{\boldmath $x$}}
\nc{\yb}{\mbox{\boldmath $y$}}
\nc{\Zb}{\mbox{\boldmath $Z$}}
\nc{\Lambdab}{\boldsymbol{\Lambda}}

\nc{\stan}{\mathsf{W}}
\newrgbcolor{darkgreen}{0., 0.733333, 0.0621042}
\definecolor{lightblue}{rgb}{.7,.7,1}
\definecolor{lightestblue}{rgb}{.95,.95,1}
\definecolor{lightlightblue}{rgb}{.85,.85,1}
\definecolor{midblue}{rgb}{.7,.7,1}

\nc{\elegant}{1.5pt}
\nc{\moyen}{1.0pt}
\nc{\mince}{0.5pt}

\def\loopa{
\psframe[linewidth=.25pt](0,0)(1,1)
}
\def\loopb{
\psframe[linewidth=.25pt](0,0)(1,1)
\psarc[linewidth=1.5pt,linecolor=blue](0,1){.5}{-90}{0}
}
\def\loopc{
\psframe[linewidth=.25pt](0,0)(1,1)
\psarc[linewidth=1.5pt,linecolor=blue](1,0){.5}{90}{180}
}
\def\loopd{
\psframe[linewidth=.25pt](0,0)(1,1)
\psarc[linewidth=1.5pt,linecolor=blue](0,0){.5}{0}{90}
}
\def\loope{
\psframe[linewidth=.25pt](0,0)(1,1)
\psarc[linewidth=1.5pt,linecolor=blue](1,1){.5}{180}{270}
}
\def\loopf{
\psframe[linewidth=.25pt](0,0)(1,1)
\psline[linewidth=1.5pt,linecolor=blue](0.5,0)(0.5,1)
}
\def\loopg{
\psframe[linewidth=.25pt](0,0)(1,1)
\psline[linewidth=1.5pt,linecolor=blue](0,0.5)(1,0.5)
}
\def\looph{
\psframe[linewidth=.25pt](0,0)(1,1)
\psarc[linewidth=1.5pt,linecolor=blue](1,0){.5}{90}{180}
\psarc[linewidth=1.5pt,linecolor=blue](0,1){.5}{-90}{0}
}
\def\loopi{
\psframe[linewidth=.25pt](0,0)(1,1)
\psarc[linewidth=1.5pt,linecolor=blue](0,0){.5}{0}{90}
\psarc[linewidth=1.5pt,linecolor=blue](1,1){.5}{180}{270}
}
\def\loopid{
\psframe[linewidth=.25pt](0,0)(1,1)
\psarc[linewidth=1.5pt,linecolor=blue,linestyle=dashed,dash=2pt 2pt](1,0){.5}{90}{180}
\psarc[linewidth=1.5pt,linecolor=blue,linestyle=dashed,dash=2pt 2pt](0,1){.5}{-90}{0}
}
\def\loopej{
\psframe[linewidth=.25pt](0,0)(1,1)
\psarc[linewidth=1.5pt,linecolor=blue,linestyle=dashed,dash=2pt 2pt](0,0){.5}{0}{90}
\psarc[linewidth=1.5pt,linecolor=blue,linestyle=dashed,dash=2pt 2pt](1,1){.5}{180}{-90}
}

\def\facegrid#1#2{
\psframe[fillstyle=solid,fillcolor=lightlightblue,linewidth=0pt]#1#2
\psgrid[gridlabels=0pt,subgriddiv=1]#1#2}

\def\specialcircle#1{
\pspolygon[fillstyle=solid,fillcolor=black,linewidth=0.5pt](#1,#1)(#1,-#1)(-#1,-#1)(-#1,#1)
}

\def\dloopone{
\psframe[linewidth=.25pt](0,0)(1,1)
}
\def\dlooptwo{
\psframe[linewidth=.25pt](0,0)(1,1)
\psarc[linewidth=1.5pt,linecolor=blue](0,1){.5}{-90}{0}
}
\def\dloopthr{
\psframe[linewidth=.25pt](0,0)(1,1)
\psarc[linewidth=1.5pt,linecolor=blue](1,0){.5}{90}{180}
}
\def\dloopfou{
\psframe[linewidth=.25pt](0,0)(1,1)
\psarc[linewidth=1.5pt,linecolor=blue](0,0){.5}{0}{90}
}
\def\dloopfiv{
\psframe[linewidth=.25pt](0,0)(1,1)
\psarc[linewidth=1.5pt,linecolor=blue](1,1){.5}{180}{270}
}

\def\dloopsev{
\psframe[linewidth=.25pt](0,0)(1,1)
\psline[linewidth=1.5pt,linecolor=blue](0.5,0)(0.5,1)
}
\def\dloopeig{
\psframe[linewidth=.25pt](0,0)(1,1)
\psarc[linewidth=1.5pt,linecolor=blue](1,0){.5}{90}{180}
\psarc[linewidth=1.5pt,linecolor=blue](0,1){.5}{-90}{0}
}

\def\hexab{
\pspolygon[fillstyle=solid,linewidth=0.5pt,linecolor=black,fillcolor=lightblue](0,0)(0,1)(0.866,1.5)(1.73205,1)(1.73205,0)(0.866,-0.5)
}

\def\hexaw{
\pspolygon[fillstyle=solid,linewidth=0.5pt,linecolor=black,fillcolor=white](0,0)(0,1)(0.866,1.5)(1.73205,1)(1.73205,0)(0.866,-0.5)
}

\def\trida{
\pspolygon[fillstyle=solid,linewidth=0.5pt,linecolor=black,fillcolor=lightlightblue](0.866,0.5)(-0.866,0.5)(0,-1)
}
\def\tridb{
\pspolygon[fillstyle=solid,linewidth=0.5pt,linecolor=black,fillcolor=lightlightblue](0.866,0.5)(-0.866,0.5)(0,-1)
\psarc[linewidth=1.5pt,linecolor=blue](0,-1){0.866}{60}{120}
}
\def\tridc{
\pspolygon[fillstyle=solid,linewidth=0.5pt,linecolor=black,fillcolor=lightlightblue](0.866,0.5)(-0.866,0.5)(0,-1)
\psarc[linewidth=1.5pt,linecolor=blue](-0.866,0.5){0.866}{-60}{0}
}

\def\tridd{
\pspolygon[fillstyle=solid,linewidth=0.5pt,linecolor=black,fillcolor=lightlightblue](0.866,0.5)(-0.866,0.5)(0,-1)
\psarc[linewidth=1.5pt,linecolor=blue](0.866,0.5){0.866}{180}{240}
}

\def\triua{
\pspolygon[fillstyle=solid,linewidth=0.5pt,linecolor=black,fillcolor=lightlightblue](0.866,-1)(-0.866,-1)(0,0.5)
}
\def\triub{
\pspolygon[fillstyle=solid,linewidth=0.5pt,linecolor=black,fillcolor=lightlightblue](0.866,-1)(-0.866,-1)(0,0.5)
\psarc[linewidth=1.5pt,linecolor=blue](0,0.5){0.866}{240}{300}
}

\def\triuc{
\pspolygon[fillstyle=solid,linewidth=0.5pt,linecolor=black,fillcolor=lightlightblue](0.866,-1)(-0.866,-1)(0,0.5)
\psarc[linewidth=1.5pt,linecolor=blue](-0.866,-1){0.866}{0}{60}
}

\def\triud{
\pspolygon[fillstyle=solid,linewidth=0.5pt,linecolor=black,fillcolor=lightlightblue](0.866,-1)(-0.866,-1)(0,0.5)
\psarc[linewidth=1.5pt,linecolor=blue](0.866,-1){0.866}{120}{180}
}

\def\wobblyarc#1#2#3{
\psparametricplot[linecolor=blue,linewidth=1.2pt,plotpoints=1435]{#2}{#3}{0.05 40 #1 t mul mul sin mul #1 add t cos mul 0.05 40 #1 t mul mul sin mul #1 add t sin mul}
}

\def\wobblyarcthree#1#2#3{
\psparametricplot[linecolor=blue,linewidth=1.0pt,plotpoints=1435]{#2}{#3}{0.04 70 #1 t mul mul sin mul #1 add t cos mul 0.04 70 #1 t mul mul sin mul #1 add t sin mul}
}

\def\young#1#2{
\multiput(0,0)(.1,0){#2}{\psline[linewidth=0.02cm]{-}(0,0)(0,.1)(.1,.1)(.1,0)(0,0)}
\multiput(0,.1)(.1,0){#1}{\psline[linewidth=0.02cm]{-}(0,0)(0,.1)(.1,.1)(.1,0)(0,0)}
}

\renewcommand{\ge}{\geqslant}
\renewcommand{\le}{\leqslant}

\nc{\proof}{{\scshape Proof.\ }} 				
\nc{\eproof}{{\hfill \rule{0.5em}{0.5em}\medskip}}

\begin{document}

\topmargin -5mm
\oddsidemargin 5mm

\vspace*{-2cm}

\setcounter{page}{1}

\vspace{22mm}
\begin{center}
{\huge {\bf Fusion hierarchies, $\boldsymbol T$-systems and $\boldsymbol Y$-systems\\[0.2cm] for the dilute $\boldsymbol{A_2^{(2)}}$ loop models}}

\vspace{10mm}
{\Large Alexi Morin-Duchesne$^\dagger$, Paul A. Pearce$^{\ddagger\S}$}
\\[.4cm]
{\em {}$^\dagger$Universit\'e catholique de Louvain, Institut de Recherche en Math\'ematique et Physique}\\
{\em Chemin du Cyclotron 2, 1348 Louvain-la-Neuve, Belgium}
\\[.4cm]
{\em {}$^\ddagger$School of Mathematics and Statistics, University of Melbourne}\\
{\em Parkville, Victoria 3010, Australia}
\\[.4cm]
{\em {}$^{\S}$School of Mathematics and Physics, University of Queensland}\\
{\em St Lucia, Brisbane, Queensland 4072, Australia}
\\[.4cm]
{\tt alexi.morin-duchesne\,@\,uclouvain.be\qquad \tt papearce\,@\,unimelb.edu.au}
\end{center}


\vspace{8mm}
\centerline{{\bf{Abstract}}}
\vskip.4cm
\noindent 

The fusion hierarchy, $T$-system and $Y$-system of functional equations are the key to integrability for 2d lattice models. We derive these equations for the generic dilute $A_2^{(2)}$ loop models. 
The fused transfer matrices are associated with nodes of the infinite dominant integral weight lattice of $s\ell(3)$. For generic values of the crossing parameter $\lambda$, the $T$- and $Y$-systems do not truncate. For the case $\frac{\lambda}{\pi}=\frac{(2p'-p)}{4p'}$ rational so that $x=\eE^{\ir \lambda}$ is a root of unity, we find explicit closure relations and derive closed finite $T$- and $Y$-systems. 
The TBA diagrams of the $Y$-systems and associated Thermodynamic Bethe Ansatz (TBA) integral equations are not of simple Dynkin type. They involve $p'+2$ nodes if $p$ is even and $2p'+2$ nodes if $p$ is odd and are related to the 
TBA diagrams of $A_2^{(1)}$ models at roots of unity by a ${\Bbb Z}_2$ folding which originates from the addition of crossing symmetry. 
In an appropriate regime, the known central charges are $c=1-\frac{6(p-p')^2}{pp'}$. 
Prototypical examples of the $A_2^{(2)}$ loop models, at roots of unity, include critical dense polymers ${\cal DLM}(1,2)$ with central charge $c=-2$, $\lambda=\frac{3\pi}{8}$ and loop fugacity $\beta=0$
and critical site percolation on the triangular lattice ${\cal DLM}(2,3)$ with $c=0$, $\lambda=\frac{\pi}{3}$ and $\beta=1$. Solving the TBA equations for the conformal data will determine whether these models lie in the same universality classes as their $A_1^{(1)}$ counterparts. More specifically, it will confirm the extent to which bond and site percolation lie in the same universality class as logarithmic conformal field theories.

\vspace{.5cm}
\noindent\textbf{Keywords:} Loop models, fusion hierarchies, $T$-systems, $Y$-systems\\

\newpage
\tableofcontents

\newpage
\hyphenpenalty=30000

\setcounter{footnote}{0}

\section{Introduction}

In statistical mechanics, a classical 2d lattice model on the square lattice (whose face weights contain a distinguished parameter $u$ called the spectral parameter) is exactly solvable~\cite{BaxterBook1982} if the Boltzmann face weights satisfy the Yang-Baxter equation. This is a strong form of integrability that ensures the existence of one-parameter families of commuting transfer matrices on the cylinder and strip with a countably infinite number of conserved quantities. The solutions of the Yang-Baxter equation, and thereby the solvable 2d lattice models, are classified~\cite{Bazh85,Jimbo86} by Lie algebras. This classification extends to the associated 1d quantum Hamiltonians through the logarithmic derivative with respect to the spectral parameter evaluated at $u=0$. The solvable models classified by higher-rank Lie algebras are perhaps esoteric for physical applications to simple statistical systems. So, arguably, the solvable families of foremost importance are classified by the affine Lie algebras $A_1^{(1)}$, $A_2^{(1)}$ and the twisted affine Lie algebra $A_2^{(2)}$. The rank $r=1,2,2$ respectively indicates that, in spin language, the affine Lie algebras are built on the special linear Lie algebras $s\ell(r)$. 

\begin{table}[ht]
\begin{eqnarray}
{\begin{array}{clc}
\hline\hline\\[-10pt]
\mbox{\bf Algebra}&\mbox{\bf Type}&\mbox{\bf Critical Solvable Models}\\[4pt]
\hline\hline\\[-8pt]
&\mbox{Vertex}&\mbox{6-vertex~\cite{Lieb1967,BaxterBook1982,Baxter1972,Baxter1973,ZP95,BazhMang2007,FrahmMDP2019}}\\
A_1^{(1)}&\mbox{RSOS}&\mbox{ABF/FB~\cite{ABF84,FB85,KP92}}\\[2pt]
&\mbox{Loop}&\mbox{Dense TL Loop Models~\cite{Nienhuis82,BloteNienhuis89,YB95,PRZ2006,SAPR2009,MDPR2014,MDKP2017,FrahmMDP2019}}\\[4pt]
\hline\\[-10pt]
&\mbox{Vertex}&\mbox{15-vertex~\cite{KR82,BabelonEtAl1982,dV89,AlcarazMartins1990,dVGR94,KNS94,ZinnJustin1998}}\\
A_2^{(1)}&\mbox{RSOS}&\mbox{Vector Reps~\cite{JMO88,DFZ90}}\\[2pt]
&\mbox{Loop}&\mbox{Fully packed TL loop models~\cite{Resh91,DEI2016,MDPR2018}}\\[4pt]
\hline\\[-10pt]
&\mbox{Vertex} &\mbox{Izergin-Korepin 19-vertex~\cite{IK1981,WBN1992,AMN1995,VJS2014}}\\
A_2^{(2)}&\mbox{RSOS}&\mbox{Dilute RSOS models~\cite{Kuniba1991,WNS1992,R92,WPSN1994,BNW94,ZPG1995,Suzuki1998}}\\[2pt]
&\mbox{Loop}&\mbox{Dilute TL loop models~\cite{DJS2010,SAPR2012,G12,FeherNien2015,GarbNien2017}}\\[4pt]
\hline\hline\vspace{-22pt}
\end{array}}
\nonumber
\end{eqnarray}
\caption{The different realisations of the $A_1^{(1)}$, $A_2^{(1)}$ and $A_2^{(2)}$ models. Some key references are given for studies of these models including the $T$- and $Y$-systems, $T$-$Q$ equations and Bethe ansatz.}
\end{table}

Somewhat remarkably, the algebraic structure of the local face operators of the critical $A_1^{(1)}$, $A_2^{(1)}$ and $A_2^{(2)}$ lattice models can all be expressed using variations of the planar Temperley-Lieb algebra~\cite{TempLieb1971,planarJones1999}. Indeed, the local relations, including the Yang-Baxter equation, the inversion relation and boundary Yang-Baxter equations, all follow from diagrammatic arguments in the (ordinary or dilute) planar Temperley-Lieb algebra. 
The representations of these algebras are of three distinct types: (i)~vertex, (ii)~RSOS and (iii)~loop. These correspond to physical degrees of freedom which are (i)~spins, (ii)~heights or (iii)~loop segments in the form of nonlocal connectivities or polymer segments. The face weights of these models admit the crossing parameter $\lambda\in{\Bbb R}$ as an additional parameter. 
We distinguish the {\em root of unity} case for which $\frac{\lambda}{\pi}\in {\Bbb Q}$ is rational from the {\em generic} case for which $\frac{\lambda}{\pi}\notin {\Bbb Q}$. The physical properties of the two cases are dramatically different. This is manifest in the continuum scaling limit. In this limit, the models are described by Conformal Field Theories (CFTs). The generic $A_1^{(1)}$, $A_2^{(1)}$, $A_2^{(2)}$ vertex and loop models with $\lambda\notin {\Bbb Q}$ are described by {\em irrational} CFTs~\cite{HKOC96}. For the RSOS models, $\frac{\lambda}{\pi}$ is rational and the associated CFTs are {\em rational} CFTs~\cite{MooreSeiberg}. Lastly, the vertex and loop models with $\frac{\lambda}{\pi}\in {\Bbb Q}$ are described by {\em logarithmic} CFTs~\cite{LCFT}.

\goodbreak
Rodney Baxter~\cite{BaxterBook1982} pioneered the use of functional equations to calculate various physically relevant lattice quantities for solvable lattice models. Nowadays, it is well understood that on an extended lattice, the integrability structures are embodied in the fusion hierarchy~\cite{BazhResh1989}, $T$- and $Y$-systems~\cite{Zam1991a,Zam1991b,KunibaNS9310,KNS2011,KP92,KNS94} and $T$-$Q$ functional equations~\cite{Baxter1972,Baxter1973,ZP95,BazhMang2007,FrahmMDP2019} satisfied by the transfer matrices. 
The $T$-system is well suited to the calculation of {\em non-universal\/} lattice quantities such as free energies. The $Y$-system is {\em universal\/}~\cite{CMP2001} in the sense that it is independent of the boundary conditions and topology. The $Y$-system and \mbox{$T$-$Q$} equations are the starting point to derive Thermodynamic Bethe Ansatz (TBA) and Non-Linear Integral Equations (NLIE). The TBA and NLIE are well suited to the calculation of the conformal data and conformal spectra from finite-size corrections. 
The $T$-$Q$ equations incorporate the usual Bethe Ansatz equations~\cite{Bethe1931}. 

Two dimensional loop models~\cite{Nienhuis90,NW93}, describing the statistical mechanics of systems with extended nonlocal degrees of freedom have been of growing interest since the early 1990's. The $T$- and $Y$-systems for the $A_1^{(1)}$ loop models have been known for a long time   and recently this was extended~\cite{MDPR2018} to the considerably more complicated $A_2^{(1)}$ loop models. 
In this paper, we derive the $T$- and $Y$-systems for the dilute $A_2^{(2)}$ loop models.  In particular, we derive closed finite $T$- and $Y$-systems at roots of unity. Due to the underlying $s\ell(3)$ structure, there are many similarities with the $T$- and $Y$-systems~\cite{MDPR2018} for the $A_2^{(1)}$ loop models. Some simplifications arise, however, due to the fact that the $A_2^{(2)}$ loop models are crossing symmetric whereas the $A_2^{(1)}$ loop models are not. 

The layout of the paper is as follows. After the introduction, in \cref{sec:A22.def}, we define the $A_2^{(2)}$ loop models on the square lattice. For the special case $\lambda=\frac{\pi}{3}$, we give the mapping between the $A_2^{(2)}$ loop model and site percolation on the triangular lattice. In \cref{sec:pdTL}, we introduce the periodic dilute Temperley-Lieb algebra and the standard modules generated by the action on link states. In \cref{sec:diag.calculus}, the local relations including the initial condition, crossing symmetry, inversion relation and Yang-Baxter equation are presented. Additionally, triangle operators are defined through the factorization of the face operators at the degeneration points $u= 2\lambda, 3\lambda$ and these satisfy local push-through properties. The $T$- and $Y$-systems are derived for generic $\lambda$ in \cref{sec:Ts.and.FH}. Fused transfer matrices, labelled by the dominant integral weights of the $s\ell(3)$, are constructed from the fundamental transfer matrix and its conjugate. The polynomial properties and braid limits of the fused transfer matrices are also discussed. The closure relations of the $T$- and $Y$-systems at roots of unity are presented in \cref{sec:closures}. The main result of the paper, namely the $Y$-system for $\lambda$ a root of unity, is given in \eqref{eq:Ysys} and \eqref{rootY}. The associated TBA diagram appears in Figure~\ref{fig:DynkinY}. Some open problems and final remarks are given in the conclusion. The technical details of the construction of some Wenzl-Jones projectors and the derivation of the $T$-system and closure relations are relegated to the appendices.

\section{Definition of the $\boldsymbol{A_2^{(2)}}$ loop models}\label{sec:A22.def}
The elementary face operator of the dilute $A_2^{(2)}$ loop model is defined as a linear combination of nine elementary tiles
\begin{alignat}{2}
\label{eq:face.op}
\begin{pspicture}[shift=-.40](0,0)(1,1)
\facegrid{(0,0)}{(1,1)}
\psarc[linewidth=0.025]{-}(0,0){0.16}{0}{90}
\rput(.5,.5){$u$}
\end{pspicture}
\ = \ &\rho_1(u)\ 
\begin{pspicture}[shift=-.40](0,0)(1,1)
\facegrid{(0,0)}{(1,1)}
\rput[bl](0,0){\loopa}
\end{pspicture}
\ + \rho_2(u)\ 
\begin{pspicture}[shift=-.40](0,0)(1,1)
\facegrid{(0,0)}{(1,1)}
\rput[bl](0,0){\loopb}
\end{pspicture}
\ + \rho_3(u)\ 
\begin{pspicture}[shift=-.40](0,0)(1,1)
\facegrid{(0,0)}{(1,1)}
\rput[bl](0,0){\loopc}
\end{pspicture}
\ + \rho_4(u)\ 
\begin{pspicture}[shift=-.40](0,0)(1,1)
\facegrid{(0,0)}{(1,1)}
\rput[bl](0,0){\loopd}
\end{pspicture}
\ + \rho_5(u)\ 
\begin{pspicture}[shift=-.40](0,0)(1,1)
\facegrid{(0,0)}{(1,1)}
\rput[bl](0,0){\loope}
\end{pspicture}
\nonumber\\[0.2cm] \ &\hspace{0.2cm}+ \rho_6(u)\ 
\begin{pspicture}[shift=-.40](0,0)(1,1)
\facegrid{(0,0)}{(1,1)}
\rput[bl](0,0){\loopf}
\end{pspicture}
\ + \rho_7(u)\ 
\begin{pspicture}[shift=-.40](0,0)(1,1)
\facegrid{(0,0)}{(1,1)}
\rput[bl](0,0){\loopg}
\end{pspicture}
\ + \rho_8(u)\ 
\begin{pspicture}[shift=-.40](0,0)(1,1)
\facegrid{(0,0)}{(1,1)}
\rput[bl](0,0){\looph}
\end{pspicture}
\ + \rho_9(u)\ 
\begin{pspicture}[shift=-.40](0,0)(1,1)
\facegrid{(0,0)}{(1,1)}
\rput[bl](0,0){\loopi}
\end{pspicture}\ ,
\end{alignat}
where the local Boltzmann weights are
\begin{subequations}
\label{eq:weights}
\begin{alignat}{4}
&\rho_1(u)=1+\frac{\sin u\sin(3\lambda-u)}{\sin 2\lambda \sin 3\lambda}, \quad
&&\rho_{2,3}(u)= \frac{\sin(3\lambda-u)}{\sin 3\lambda}, \quad
&&\rho_{4,5}(u)=\frac{\sin u}{\sin3\lambda},\\[0.15cm]
&\rho_{6,7}(u)= \frac{\sin u\sin(3\lambda-u)}{\sin2\lambda\sin3\lambda}, \quad
&&\rho_8(u)= \frac{\sin(2\lambda-u)\sin(3\lambda-u)}{\sin2\lambda\sin3\lambda}, \quad
&&\rho_9(u)= -\frac{\sin u\sin(\lambda-u)}{\sin2\lambda\sin3\lambda}.
\end{alignat}
\end{subequations}
The parameters $u$ and $\lambda$ are respectively the spectral and crossing parameters. The Boltzmann face weights are real for $u,\lambda \in \mathbb R$ but they can be negative.

The values of $\frac{\lambda}\pi \notin \mathbb Q$ are referred to as generic.
We are particularly interested in the $A_2^{(2)}$ loop models at {\it roots of unity}, that is, $\frac{\lambda}{\pi}\in{\Bbb Q}$. 
In these cases, we suppose $p$ and $p'$ are coprime integers 
and define
\begin{eqnarray}
\lambda=\frac{(2p'-p)\pi}{4p'},\qquad\quad\bar\lambda=4\lambda-\pi=\frac{(p'-p)\pi}{p'},\qquad\quad \mbox{gcd}(p,p')=1,\label{eq:lambda.pp'}
\end{eqnarray}
so that $\bar x=\eE^{\ir\bar\lambda}$ is a root of unity with $\bar x^{2p'}=1$. 
We refer to the $A_2^{(2)}$ loop model at this root of unity as the {\em dilute logarithmic minimal model} ${\cal DLM}(p,p')$. 
The fugacity of contractible loops is
\begin{eqnarray}
\beta=-2\cos 4\lambda=2\cos\bar\lambda \in [-2,2].
\end{eqnarray}
Taking $0<u<3\lambda$, the critical manifold of this model admits a number of branches subdividing the interval $\lambda\in [0,\pi]$. 
Two branches of particular interest are
\begin{eqnarray}
\label{eq:branches}
{\cal DLM}(p,p')=\begin{cases}
\mbox{dilute},&\mbox{$0<\lambda<\frac{\pi}{4}$ and $1<\frac{p}{p'}<2$},\\[2pt]
\mbox{dense},&\mbox{$\frac{\pi}{4}<\lambda<\frac{\pi}{2}$ and $0<\frac{p}{p'}<1$},
\end{cases}
\end{eqnarray}
with central charges
\begin{eqnarray}
c=1-\frac{6(p-p')^2}{pp'}.
\end{eqnarray}
The integrable models ${\cal DLM}(p,p')$ represent different universality classes of 2d logarithmic critical behaviour.
Some prototypical examples include
\begin{flalign}
\hspace{-18pt}{\cal DLM}(1,2)&=\mbox{critical dense polymers with $c=-2$, $\lambda=\frac{3\pi}{8}$, $\bar\lambda=\frac{\pi}{2}$, $\beta=0$},\nonumber\\
\hspace{-18pt}{\cal DLM}(2,3)&=\mbox{critical site percolation on the triangular lattice with $c=0$, $\lambda=\bar\lambda=\frac{\pi}{3}$, $\beta=1$},\\
\hspace{-10pt}{\cal DLM}(3,4)&=\mbox{dilute logarithmic Ising model with $c=\frac{1}{2}$, $\lambda=\frac{5\pi}{16}$, $\bar\lambda=\frac{\pi}{4}$, $\beta=\sqrt{2}$}.\nonumber 
\end{flalign}

The boundary points $\lambda=0, \frac \pi 4, \frac \pi 2$ in \eqref{eq:branches} are special.
The point $\lambda=\frac{\pi}{4}$ corresponds to $\beta = 2$. For the dilute $A_2^{(2)}$ models, the face operator is well defined at $\lambda = \frac{\pi}{4}$. This is in contrast to the dense  $A_1^{(1)}$ model at $\beta=2$ where the trigonometric functions are replaced with linear functions of $u$. It is unclear what exactly the $c=1$ CFT  is for this $A_2^{(2)}$ model. Separately, in the limit $\lambda\to 0$, we see that $c\to -2$, $\bar\lambda\to -\pi$ and $\beta\to -2$. Lastly, in the limit $\lambda\to\frac{\pi}{2}$, we see that $c\to-\infty$, $\bar\lambda\to \pi$ and $\beta\to -2$. 

\begin{figure}[ht]
\begin{center}
\psset{unit=0.35cm}
\begin{pspicture}[shift=-6.4](-5,-1)(9,13)
\rput(-6.062,10.5){\hexab}\rput(-4.33,10.5){\hexab}\rput(-2.598,10.5){\hexab}\rput(-0.866,10.5){\hexaw}\rput(0.866,10.5){\hexaw}\rput(2.598,10.5){\hexaw}
\rput(-5.196,9){\hexab}\rput(-3.464,9){\hexaw}\rput(-1.732,9){\hexab}\rput(0,9){\hexaw}\rput(1.732,9){\hexab}\rput(3.464,9){\hexaw}
\rput(-4.33,7.5){\hexab}\rput(-2.598,7.5){\hexaw}\rput(-0.866,7.5){\hexab}\rput(0.866,7.5){\hexab}\rput(2.598,7.5){\hexaw}\rput(4.33,7.5){\hexaw}
\rput(-3.464,6){\hexab}\rput(-1.732,6){\hexab}\rput(0,6){\hexaw}\rput(1.732,6){\hexab}\rput(3.464,6){\hexab}\rput(5.196,6){\hexaw}
\rput(-2.598,4.5){\hexab}\rput(-0.866,4.5){\hexab}\rput(0.866,4.5){\hexaw}\rput(2.598,4.5){\hexaw}\rput(4.33,4.5){\hexab}\rput(6.062,4.5){\hexaw}
\rput(-1.732,3){\hexab}\rput(0,3){\hexab}\rput(1.732,3){\hexaw}\rput(3.464,3){\hexab}\rput(5.196,3){\hexaw}\rput(6.928,3){\hexaw}
\rput(-0.866,1.5){\hexab}\rput(0.866,1.5){\hexaw}\rput(2.598,1.5){\hexab}\rput(4.33,1.5){\hexaw}\rput(6.062,1.5){\hexab}\rput(7.794,1.5){\hexaw}
\rput(0,0){\hexab}\rput(1.732,0){\hexab}\rput(3.464,0){\hexab}\rput(5.196,0){\hexaw}\rput(6.928,0){\hexaw}\rput(8.66,0){\hexaw}
\end{pspicture}
\quad$\longleftrightarrow$\quad
\begin{pspicture}[shift=-6.4](-5,-1)(9,12)
\rput(-6.062,10.5){\hexab}\rput(-4.33,10.5){\hexab}\rput(-2.598,10.5){\hexab}\rput(-0.866,10.5){\hexaw}\rput(0.866,10.5){\hexaw}\rput(2.598,10.5){\hexaw}
\rput(-5.196,9){\hexab}\rput(-3.464,9){\hexaw}\rput(-1.732,9){\hexab}\rput(0,9){\hexaw}\rput(1.732,9){\hexab}\rput(3.464,9){\hexaw}
\rput(-4.33,7.5){\hexab}\rput(-2.598,7.5){\hexaw}\rput(-0.866,7.5){\hexab}\rput(0.866,7.5){\hexab}\rput(2.598,7.5){\hexaw}\rput(4.33,7.5){\hexaw}
\rput(-3.464,6){\hexab}\rput(-1.732,6){\hexab}\rput(0,6){\hexaw}\rput(1.732,6){\hexab}\rput(3.464,6){\hexab}\rput(5.196,6){\hexaw}
\rput(-2.598,4.5){\hexab}\rput(-0.866,4.5){\hexab}\rput(0.866,4.5){\hexaw}\rput(2.598,4.5){\hexaw}\rput(4.33,4.5){\hexab}\rput(6.062,4.5){\hexaw}
\rput(-1.732,3){\hexab}\rput(0,3){\hexab}\rput(1.732,3){\hexaw}\rput(3.464,3){\hexab}\rput(5.196,3){\hexaw}\rput(6.928,3){\hexaw}
\rput(-0.866,1.5){\hexab}\rput(0.866,1.5){\hexaw}\rput(2.598,1.5){\hexab}\rput(4.33,1.5){\hexaw}\rput(6.062,1.5){\hexab}\rput(7.794,1.5){\hexaw}
\rput(0,0){\hexab}\rput(1.732,0){\hexab}\rput(3.464,0){\hexab}\rput(5.196,0){\hexaw}\rput(6.928,0){\hexaw}\rput(8.66,0){\hexaw}
\rput(-4.33,10.5){\trida}\rput(-2.598,10.5){\tridb}\rput(-0.866,10.5){\tridd}\rput(0.866,10.5){\trida}\rput(2.598,10.5){\tridb}
\rput(-3.464,10.5){\triud}\rput(-1.732,10.5){\triuc}\rput(0,10.5){\triuc}\rput(1.732,10.5){\triud}\rput(3.464,10.5){\triuc}
\rput(-3.464,9){\tridd}\rput(-1.732,9){\tridd}\rput(0,9){\tridd}\rput(1.732,9){\tridc}\rput(3.464,9){\tridc}
\rput(-2.598,9){\triuc}\rput(-0.866,9){\triuc}\rput(0.866,9){\triub}\rput(2.598,9){\triud}\rput(4.33,9){\triua}
\rput(-2.598,7.5){\tridd}\rput(-0.866,7.5){\tridc}\rput(0.866,7.5){\tridb}\rput(2.598,7.5){\tridd}\rput(4.33,7.5){\tridb}
\rput(-1.732,7.5){\triub}\rput(0,7.5){\triud}\rput(1.732,7.5){\triuc}\rput(3.464,7.5){\triub}\rput(5.196,7.5){\triuc}
\rput(-1.732,6){\trida}\rput(0,6){\tridd}\rput(1.732,6){\tridd}\rput(3.464,6){\tridb}\rput(5.196,6){\tridd}
\rput(-0.866,6){\triua}\rput(0.866,6){\triuc}\rput(2.598,6){\triub}\rput(4.33,6){\triuc}\rput(6.062,6){\triuc}
\rput(-0.866,4.5){\trida}\rput(0.866,4.5){\tridd}\rput(2.598,4.5){\trida}\rput(4.33,4.5){\tridc}\rput(6.062,4.5){\tridc}
\rput(0,4.5){\triua}\rput(1.732,4.5){\triuc}\rput(3.464,4.5){\triud}\rput(5.196,4.5){\triud}\rput(6.928,4.5){\triua}
\rput(0,3){\trida}\rput(1.732,3){\tridc}\rput(3.464,3){\tridc}\rput(5.196,3){\tridc}\rput(6.928,3){\tridb}
\rput(0.866,3){\triud}\rput(2.598,3){\triud}\rput(4.33,3){\triud}\rput(6.062,3){\triud}\rput(7.794,3){\triuc}
\rput(0.866,1.5){\tridd}\rput(2.598,1.5){\tridc}\rput(4.33,1.5){\tridd}\rput(6.062,1.5){\tridd}\rput(7.794,1.5){\tridc}
\rput(1.732,1.5){\triub}\rput(3.464,1.5){\triua}\rput(5.196,1.5){\triuc}\rput(6.928,1.5){\triub}\rput(8.66,1.5){\triua}
\end{pspicture}
\quad$\longleftrightarrow$\qquad
\psset{unit=0.5cm}
\begin{pspicture}[shift=-3.9](0,-0.5)(5,7.5)
\facegrid{(0,0)}{(5,7)}
\rput(0,6){\dloopthr}\rput(1,6){\dloopfou}\rput(2,6){\dloopsev}\rput(3,6){\dloopthr}\rput(4,6){\dloopfou}
\rput(0,5){\dloopsev}\rput(1,5){\dloopsev}\rput(2,5){\dloopfiv}\rput(3,5){\dloopeig}\rput(4,5){\dlooptwo}
\rput(0,4){\dloopfiv}\rput(1,4){\dloopeig}\rput(2,4){\dloopfou}\rput(3,4){\dloopfiv}\rput(4,4){\dloopfou}
\rput(0,3){\dloopone}\rput(1,3){\dloopsev}\rput(2,3){\dloopfiv}\rput(3,3){\dloopfou}\rput(4,3){\dloopsev}
\rput(0,2){\dloopone}\rput(1,2){\dloopsev}\rput(2,2){\dloopthr}\rput(3,2){\dloopeig}\rput(4,2){\dlooptwo}
\rput(0,1){\dloopthr}\rput(1,1){\dloopeig}\rput(2,1){\dloopeig}\rput(3,1){\dloopeig}\rput(4,1){\dloopfou}
\rput(0,0){\dloopfiv}\rput(1,0){\dlooptwo}\rput(2,0){\dloopsev}\rput(3,0){\dloopfiv}\rput(4,0){\dlooptwo}
\end{pspicture}
\caption{The bijection~\cite{FeherNien2015} between a configuration of site percolation on the triangular lattice and a configuration of the dilute $A^{(2)}_2$ loop model. At $u=\lambda=\frac{\pi}{3}$, the corresponding renormalized weights are 
$\rho_1=\rho_{2,3}=\rho_{4,5}=\rho_{6,7}=\rho_8=1$ and $\rho_9=0$.}\label{fig:perco.honey}
\end{center}
\end{figure}
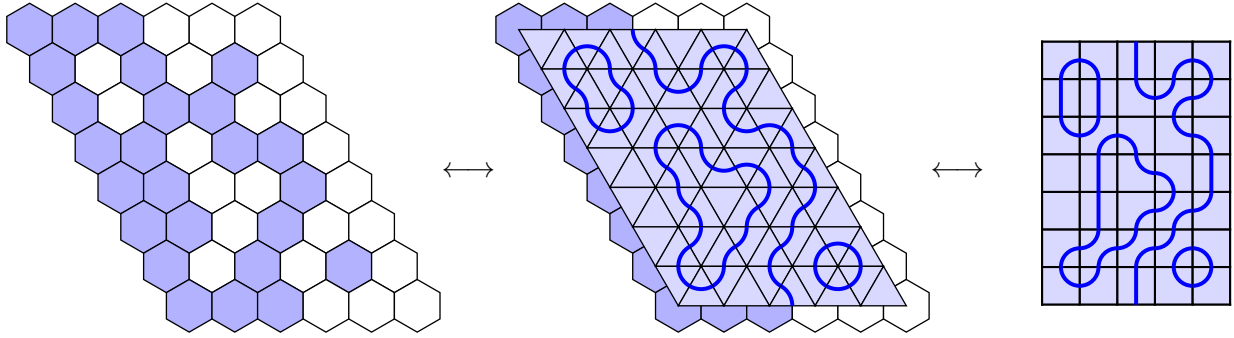

There are singular points in the Boltzmann weights \eqref{eq:weights} for $\sin 2 \lambda \sin 3 \lambda = 0$, that is, when $\lambda = \frac \pi 3, \frac \pi 2, \frac{2\pi}3$. These are artifacts of the choice of normalisation and correspond to removable singularities. The derivations of the functional equations given in this paper do not strictly cover these cases. However, it is a straightforward exercise to obtain similar results for these three values by renormalizing the face operators and taking the proper limits. We note that, for the important case $\lambda=\frac{\pi}{3}$ which is ${\cal DLM}(2,3)$, there is a mapping~\cite{FeherNien2015} between configurations of critical site percolation on the triangular lattice and loop configurations of the $A_2^{(2)}$ loop model with $u = \lambda = \frac \pi 3$, as shown in \cref{fig:perco.honey}.

\section{The periodic dilute Temperley-Lieb algebra}\label{sec:pdTL}

\subsection{Definition of the algebra} 

The dilute $A^{(2)}_2$ loop models are naturally described in the framework of the dilute Temperley-Lieb algebra. This algebra was discussed in \cite{GP93,P94,U96} and is of interest to physicists because of its role in the Izergin-Korepin 19-vertex model \cite{IK1981}, the critical $O(n)$ model \cite{N90} and integrable dilute RSOS lattice models \cite{WNS1992,R92,WPSN1994}. Its representation theory is studied in \cite{BSA14}. Here, we are interested in lattices with periodic boundary conditions. We therefore work with the periodic incarnation of this algebra which we denote by $\mathsf{pdTL}_N(\alpha,\beta)$. It was previously considered in \cite{G12,MDPR2018}. 

The algebra $\mathsf{pdTL}_N(\alpha,\beta)$ is unital and associative and is the linear span of connectivity diagrams drawn on a slice of the cylinder. We draw this slice in the plane as a rectangle with periodic boundary conditions in the horizontal direction. The rectangle has $N$ marked nodes on its top segment and likewise $N$ nodes on its bottom segment. A connectivity diagram consists of a set of non-intersecting planar loop segments that connect a subset of these nodes in pairs. These may or may not travel via the back of the cylinder. If a node is not attached to another node via a loop segment, we say that it is {\it vacant}. We usually depict vacant sites as solid black circles but, in some diagrams, we will simply not draw vacant sites. Some examples of connectivity diagrams for $N=5$ are
\be
\label{eq:conn.ex}
a_1 = \ 
\begin{pspicture}[shift=-0.3](-0.0,0)(2.0,0.8)
\pspolygon[fillstyle=solid,fillcolor=lightlightblue,linecolor=black,linewidth=0pt](0,0)(0,0.8)(2.0,0.8)(2.0,0)(0,0)
\psarc[linecolor=blue,linewidth=\elegant]{-}(0.8,0.8){0.2}{180}{0}
\psbezier[linecolor=blue,linewidth=\elegant]{-}(0.2,0.8)(0.2,0.4)(1.0,0.4)(1.0,0)
\psbezier[linecolor=blue,linewidth=\elegant]{-}(1.4,0.8)(1.4,0.5)(1.92,0.4)(2.02,0.26)
\psbezier[linecolor=blue,linewidth=\elegant]{-}(0.2,0)(0.2,0.15)(0.08,0.24)(-0.02,0.3)
\pscircle[fillstyle=solid,fillcolor=black](1.8,0.8){0.035}
\pscircle[fillstyle=solid,fillcolor=black](0.6,0){0.035}
\pscircle[fillstyle=solid,fillcolor=black](1.4,0){0.035}
\pscircle[fillstyle=solid,fillcolor=black](1.8,0){0.035}
\psframe[fillstyle=solid,linecolor=white,linewidth=0pt](-0.04,0)(0,0.8)
\psframe[fillstyle=solid,linecolor=white,linewidth=0pt](2.0,0)(2.04,0.8)
\end{pspicture}\ \ ,
\qquad
a_2 = \  
\begin{pspicture}[shift=-0.3](-0.0,0)(2.0,0.8)
\pspolygon[fillstyle=solid,fillcolor=lightlightblue,linecolor=black,linewidth=0pt](0,0)(0,0.8)(2.0,0.8)(2.0,0)(0,0)
\psarc[linecolor=blue,linewidth=\elegant]{-}(1.6,0.8){0.2}{180}{0}
\psarc[linecolor=blue,linewidth=\elegant]{-}(0.8,0){0.2}{0}{180}
\psbezier[linecolor=blue,linewidth=\elegant]{-}(0.2,0)(0.2,0.5)(1.4,0.5)(1.4,0)
\psbezier[linecolor=blue,linewidth=\elegant]{-}(1.0,0.8)(1.0,0.33)(1.92,0.39)(2.02,0.53)
\psbezier[linecolor=blue,linewidth=\elegant]{-}(0.2,0.8)(0.2,0.65)(0.08,0.6)(-0.02,0.56)
\pscircle[fillstyle=solid,fillcolor=black](0.6,0.8){0.035}
\pscircle[fillstyle=solid,fillcolor=black](1.8,0){0.035}
\psframe[fillstyle=solid,linecolor=white,linewidth=0pt](-0.04,0)(0,0.8)
\psframe[fillstyle=solid,linecolor=white,linewidth=0pt](2.0,0)(2.04,0.8)
\end{pspicture}\ \ ,
\qquad
a_3 = \  
\begin{pspicture}[shift=-0.3](-0.0,0)(2.0,0.8)
\pspolygon[fillstyle=solid,fillcolor=lightlightblue,linecolor=black,linewidth=0pt](0,0)(0,0.8)(2.0,0.8)(2.0,0)(0,0)
\pscircle[fillstyle=solid,fillcolor=black](0.6,0.8){0.035}
\pscircle[fillstyle=solid,fillcolor=black](1.4,0.8){0.035}
\pscircle[fillstyle=solid,fillcolor=black](1.8,0.8){0.035}
\pscircle[fillstyle=solid,fillcolor=black](0.6,0){0.035}
\psbezier[linecolor=blue,linewidth=\elegant]{-}(0.2,0)(0.2,0.4)(1.0,0.4)(1.0,0)
\psbezier[linecolor=blue,linewidth=\elegant]{-}(1.4,0)(1.4,0.3)(1.82,0.38)(2.02,0.52)
\psbezier[linecolor=blue,linewidth=\elegant]{-}(0.2,0.8)(0.2,0.65)(0.08,0.55)(-0.02,0.50)
\psbezier[linecolor=blue,linewidth=\elegant]{-}(1.8,0)(1.8,0.14)(1.92,0.23)(2.02,0.28)
\psbezier[linecolor=blue,linewidth=\elegant]{-}(1.0,0.8)(1.0,0.56)(0.08,0.44)(-0.02,0.24)
\psframe[fillstyle=solid,linecolor=white,linewidth=0pt](-0.04,0)(0,0.8)
\psframe[fillstyle=solid,linecolor=white,linewidth=0pt](2.0,0)(2.04,0.8)
\end{pspicture}\ \ .
\ee

The sum of connectivity diagrams is the simple commutative sum: $a_1+a_2 = a_2 + a_1$ for all $a_1, a_2 \in \pdTL_N(\alpha,\beta)$. The product is non-commutative and defined via concatenation. To compute the product $a_1a_2$ of two connectivities, we diagrammatically stack $a_2$ above $a_1$ and connect their nodes on the intermediate segment. If on this segment, one or more loop segments connect to vacancies, the result is set to zero. Otherwise, we contract the loop segments where possible, read the connectivity between the nodes of the top and bottom segments, and set $a_1a_2$ equal to this connectivity diagram times a scalar prefactor. This prefactor is $\alpha^{n_\alpha}\beta^{n_\beta}$, where $\alpha$ and $\beta$ are the fugacities of the non-contractible and contractible loops appearing in the diagram, and $n_\alpha$ and $n_\beta$ are the respective numbers of these two kinds of closed loops. Some examples of products are 
{\allowdisplaybreaks
\begin{subequations}
\begin{alignat}{2}
a_1 a_2&= \ 
\begin{pspicture}[shift=-0.7](-0.0,0)(2.0,1.6)
\pspolygon[fillstyle=solid,fillcolor=lightlightblue,linecolor=black,linewidth=0pt](0,0)(0,0.8)(2.0,0.8)(2.0,0)(0,0)
\psarc[linecolor=blue,linewidth=\elegant]{-}(0.8,0.8){0.2}{180}{0}
\psbezier[linecolor=blue,linewidth=\elegant]{-}(0.2,0.8)(0.2,0.4)(1.0,0.4)(1.0,0)
\psbezier[linecolor=blue,linewidth=\elegant]{-}(1.4,0.8)(1.4,0.5)(1.92,0.4)(2.02,0.26)
\psbezier[linecolor=blue,linewidth=\elegant]{-}(0.2,0)(0.2,0.15)(0.08,0.24)(-0.02,0.3)
\pscircle[fillstyle=solid,fillcolor=black](1.8,0.8){0.035}
\pscircle[fillstyle=solid,fillcolor=black](0.6,0){0.035}
\pscircle[fillstyle=solid,fillcolor=black](1.4,0){0.035}
\pscircle[fillstyle=solid,fillcolor=black](1.8,0){0.035}
\psframe[fillstyle=solid,linecolor=white,linewidth=0pt](-0.04,0)(0,0.8)
\psframe[fillstyle=solid,linecolor=white,linewidth=0pt](2.0,0)(2.04,0.8)
\rput(0,0.8)
{
\pspolygon[fillstyle=solid,fillcolor=lightlightblue,linecolor=black,linewidth=0pt](0,0)(0,0.8)(2.0,0.8)(2.0,0)(0,0)
\psarc[linecolor=blue,linewidth=\elegant]{-}(1.6,0.8){0.2}{180}{0}
\psarc[linecolor=blue,linewidth=\elegant]{-}(0.8,0){0.2}{0}{180}
\psbezier[linecolor=blue,linewidth=\elegant]{-}(0.2,0)(0.2,0.5)(1.4,0.5)(1.4,0)
\psbezier[linecolor=blue,linewidth=\elegant]{-}(1.0,0.8)(1.0,0.33)(1.92,0.39)(2.02,0.53)
\psbezier[linecolor=blue,linewidth=\elegant]{-}(0.2,0.8)(0.2,0.65)(0.08,0.6)(-0.02,0.56)
\pscircle[fillstyle=solid,fillcolor=black](0.6,0.8){0.035}
\pscircle[fillstyle=solid,fillcolor=black](1.8,0){0.035}
\psframe[fillstyle=solid,linecolor=white,linewidth=0pt](-0.04,0)(0,0.8)
\psframe[fillstyle=solid,linecolor=white,linewidth=0pt](2.0,0)(2.04,0.8)
}
\end{pspicture}
\ = \beta\ \,
\begin{pspicture}[shift=-0.4](-0.0,0)(2.0,1.0)
\pspolygon[fillstyle=solid,fillcolor=lightlightblue,linecolor=black,linewidth=0pt](0,0)(0,1.0)(2.0,1.0)(2.0,0)(0,0)
\psarc[linecolor=blue,linewidth=\elegant]{-}(1.6,1){0.2}{180}{0}
\psbezier[linecolor=blue,linewidth=\elegant]{-}(1.0,1)(1.0,0.53)(1.92,0.59)(2.02,0.73)
\psbezier[linecolor=blue,linewidth=\elegant]{-}(0.2,1)(0.2,0.85)(0.08,0.8)(-0.02,0.76)
\psbezier[linecolor=blue,linewidth=\elegant]{-}(1.0,0)(1.0,0.47)(1.92,0.41)(2.02,0.27)
\psbezier[linecolor=blue,linewidth=\elegant]{-}(0.2,0)(0.2,0.15)(0.08,0.2)(-0.02,0.24)
\pscircle[fillstyle=solid,fillcolor=black](0.6,1){0.035}
\pscircle[fillstyle=solid,fillcolor=black](0.6,0){0.035}
\pscircle[fillstyle=solid,fillcolor=black](1.4,0){0.035}
\pscircle[fillstyle=solid,fillcolor=black](1.8,0){0.035}
\psframe[fillstyle=solid,linecolor=white,linewidth=0pt](-0.04,0)(0,1)
\psframe[fillstyle=solid,linecolor=white,linewidth=0pt](2.0,0)(2.04,1)
\end{pspicture}\ \ ,
\\[0.3cm]
a_2 a_3&= \ 
\begin{pspicture}[shift=-0.7](-0.0,0)(2.0,1.6)
\pspolygon[fillstyle=solid,fillcolor=lightlightblue,linecolor=black,linewidth=0pt](0,0)(0,0.8)(2.0,0.8)(2.0,0)(0,0)
\psarc[linecolor=blue,linewidth=\elegant]{-}(1.6,0.8){0.2}{180}{0}
\psarc[linecolor=blue,linewidth=\elegant]{-}(0.8,0){0.2}{0}{180}
\psbezier[linecolor=blue,linewidth=\elegant]{-}(0.2,0)(0.2,0.5)(1.4,0.5)(1.4,0)
\psbezier[linecolor=blue,linewidth=\elegant]{-}(1.0,0.8)(1.0,0.33)(1.92,0.39)(2.02,0.53)
\psbezier[linecolor=blue,linewidth=\elegant]{-}(0.2,0.8)(0.2,0.65)(0.08,0.6)(-0.02,0.56)
\pscircle[fillstyle=solid,fillcolor=black](0.6,0.8){0.035}
\pscircle[fillstyle=solid,fillcolor=black](1.8,0){0.035}
\psframe[fillstyle=solid,linecolor=white,linewidth=0pt](-0.04,0)(0,0.8)
\psframe[fillstyle=solid,linecolor=white,linewidth=0pt](2.0,0)(2.04,0.8)
\rput(0,0.8)
{
\pspolygon[fillstyle=solid,fillcolor=lightlightblue,linecolor=black,linewidth=0pt](0,0)(0,0.8)(2.0,0.8)(2.0,0)(0,0)
\pscircle[fillstyle=solid,fillcolor=black](0.6,0.8){0.035}
\pscircle[fillstyle=solid,fillcolor=black](1.4,0.8){0.035}
\pscircle[fillstyle=solid,fillcolor=black](1.8,0.8){0.035}
\pscircle[fillstyle=solid,fillcolor=black](0.6,0){0.035}
\psbezier[linecolor=blue,linewidth=\elegant]{-}(0.2,0)(0.2,0.4)(1.0,0.4)(1.0,0)
\psbezier[linecolor=blue,linewidth=\elegant]{-}(1.4,0)(1.4,0.3)(1.82,0.38)(2.02,0.52)
\psbezier[linecolor=blue,linewidth=\elegant]{-}(0.2,0.8)(0.2,0.65)(0.08,0.55)(-0.02,0.50)
\psbezier[linecolor=blue,linewidth=\elegant]{-}(1.8,0)(1.8,0.14)(1.92,0.23)(2.02,0.28)
\psbezier[linecolor=blue,linewidth=\elegant]{-}(1.0,0.8)(1.0,0.56)(0.08,0.44)(-0.02,0.24)
\psframe[fillstyle=solid,linecolor=white,linewidth=0pt](-0.04,0)(0,0.8)
\psframe[fillstyle=solid,linecolor=white,linewidth=0pt](2.0,0)(2.04,0.8)
}
\end{pspicture}
\ = \alpha\ \,
\begin{pspicture}[shift=-0.3](-0.0,0)(2.0,0.8)
\pspolygon[fillstyle=solid,fillcolor=lightlightblue,linecolor=black,linewidth=0pt](0,0)(0,0.8)(2.0,0.8)(2.0,0)(0,0)
\psarc[linecolor=blue,linewidth=\elegant]{-}(0.8,0){0.2}{0}{180}
\psbezier[linecolor=blue,linewidth=\elegant]{-}(0.2,0)(0.2,0.5)(1.4,0.5)(1.4,0)
\psbezier[linecolor=blue,linewidth=\elegant]{-}(0.2,0.8)(0.2,0.4)(1.0,0.4)(1.0,0.8)
\pscircle[fillstyle=solid,fillcolor=black](0.6,0.8){0.035}
\pscircle[fillstyle=solid,fillcolor=black](1.4,0.8){0.035}
\pscircle[fillstyle=solid,fillcolor=black](1.8,0.8){0.035}
\pscircle[fillstyle=solid,fillcolor=black](1.8,0){0.035}
\psframe[fillstyle=solid,linecolor=white,linewidth=0pt](-0.04,0)(0,0.8)
\psframe[fillstyle=solid,linecolor=white,linewidth=0pt](2.0,0)(2.04,0.8)
\end{pspicture}\ \ ,
\\[0.3cm]
a_1 a_3&= \ 
\begin{pspicture}[shift=-0.7](-0.0,0)(2.0,1.6)
\pspolygon[fillstyle=solid,fillcolor=lightlightblue,linecolor=black,linewidth=0pt](0,0)(0,0.8)(2.0,0.8)(2.0,0)(0,0)
\psarc[linecolor=blue,linewidth=\elegant]{-}(0.8,0.8){0.2}{180}{0}
\psbezier[linecolor=blue,linewidth=\elegant]{-}(0.2,0.8)(0.2,0.4)(1.0,0.4)(1.0,0)
\psbezier[linecolor=blue,linewidth=\elegant]{-}(1.4,0.8)(1.4,0.5)(1.92,0.4)(2.02,0.26)
\psbezier[linecolor=blue,linewidth=\elegant]{-}(0.2,0)(0.2,0.15)(0.08,0.24)(-0.02,0.3)
\pscircle[fillstyle=solid,fillcolor=black](1.8,0.8){0.035}
\pscircle[fillstyle=solid,fillcolor=black](0.6,0){0.035}
\pscircle[fillstyle=solid,fillcolor=black](1.4,0){0.035}
\pscircle[fillstyle=solid,fillcolor=black](1.8,0){0.035}
\psframe[fillstyle=solid,linecolor=white,linewidth=0pt](-0.04,0)(0,0.8)
\psframe[fillstyle=solid,linecolor=white,linewidth=0pt](2.0,0)(2.04,0.8)
\rput(0,0.8)
{
\pspolygon[fillstyle=solid,fillcolor=lightlightblue,linecolor=black,linewidth=0pt](0,0)(0,0.8)(2.0,0.8)(2.0,0)(0,0)
\pscircle[fillstyle=solid,fillcolor=black](0.6,0.8){0.035}
\pscircle[fillstyle=solid,fillcolor=black](1.4,0.8){0.035}
\pscircle[fillstyle=solid,fillcolor=black](1.8,0.8){0.035}
\pscircle[fillstyle=solid,fillcolor=black](0.6,0){0.035}
\psbezier[linecolor=blue,linewidth=\elegant]{-}(0.2,0)(0.2,0.4)(1.0,0.4)(1.0,0)
\psbezier[linecolor=blue,linewidth=\elegant]{-}(1.4,0)(1.4,0.3)(1.82,0.38)(2.02,0.52)
\psbezier[linecolor=blue,linewidth=\elegant]{-}(0.2,0.8)(0.2,0.65)(0.08,0.55)(-0.02,0.50)
\psbezier[linecolor=blue,linewidth=\elegant]{-}(1.8,0)(1.8,0.14)(1.92,0.23)(2.02,0.28)
\psbezier[linecolor=blue,linewidth=\elegant]{-}(1.0,0.8)(1.0,0.56)(0.08,0.44)(-0.02,0.24)
\psframe[fillstyle=solid,linecolor=white,linewidth=0pt](-0.04,0)(0,0.8)
\psframe[fillstyle=solid,linecolor=white,linewidth=0pt](2.0,0)(2.04,0.8)
}
\end{pspicture}
\ = 0.
\end{alignat}
\end{subequations}
}
The identity element, denoted $\Ib$, is a sum of $2^N$ connectivity diagrams:
\be
\label{eq:id}
\Ib = \
\begin{pspicture}[shift=-0.525](0,-0.25)(2.4,0.8)
\pspolygon[fillstyle=solid,fillcolor=lightlightblue,linecolor=black,linewidth=0pt](0,0)(0,0.8)(2.4,0.8)(2.4,0)(0,0)
\psline[linecolor=blue,linewidth=\elegant,linestyle=dashed,dash=2pt 2pt]{-}(0.2,0)(0.2,0.8)
\psline[linecolor=blue,linewidth=\elegant,linestyle=dashed,dash=2pt 2pt]{-}(0.6,0)(0.6,0.8)
\psline[linecolor=blue,linewidth=\elegant,linestyle=dashed,dash=2pt 2pt]{-}(1.0,0)(1.0,0.8)
\rput(1.4,0.4){$...$}
\psline[linecolor=blue,linewidth=\elegant,linestyle=dashed,dash=2pt 2pt]{-}(1.8,0)(1.8,0.8)
\psline[linecolor=blue,linewidth=\elegant,linestyle=dashed,dash=2pt 2pt]{-}(2.2,0)(2.2,0.8)
\rput(0.2,-0.25){$_1$}
\rput(0.6,-0.25){$_2$}
\rput(1.0,-0.25){$_3$}
\rput(2.2,-0.25){$_N$}
\end{pspicture}\ \ ,
\qquad
\begin{pspicture}[shift=-0.525](0,-0.25)(0.4,0.8)
\pspolygon[fillstyle=solid,fillcolor=lightlightblue,linecolor=black,linewidth=0pt](0,0)(0,0.8)(0.4,0.8)(0.4,0)(0,0)
\psline[linecolor=blue,linewidth=\elegant,linestyle=dashed,dash=2pt 2pt]{-}(0.2,0)(0.2,0.8)
\end{pspicture} 
\ = \ 
\begin{pspicture}[shift=-0.525](0,-0.25)(0.4,0.8)
\pspolygon[fillstyle=solid,fillcolor=lightlightblue,linecolor=black,linewidth=0pt](0,0)(0,0.8)(0.4,0.8)(0.4,0)(0,0)
\psline[linecolor=blue,linewidth=\elegant]{-}(0.2,0)(0.2,0.8)
\end{pspicture} 
\ + \ 
\begin{pspicture}[shift=-0.525](0,-0.25)(0.4,0.8)
\pspolygon[fillstyle=solid,fillcolor=lightlightblue,linecolor=black,linewidth=0pt](0,0)(0,0.8)(0.4,0.8)(0.4,0)(0,0)
\pscircle[fillstyle=solid,fillcolor=black](0.2,0){0.035}
\pscircle[fillstyle=solid,fillcolor=black](0.2,0.8){0.035}
\end{pspicture} \ .
\ee
Because one can draw connectivity diagrams with loop segments that wrap around the cylinder an arbitrary number of times, $\pdTL_N(\alpha,\beta)$ is an infinite-dimensional algebra.

\subsection{Standard modules}
 
The algebra $\pdTL_N(\alpha,\beta)$ has a set of standard modules $\stan_{N,d}$. These are defined in terms of the action on a vector space of diagrams, called link states, that are drawn above a horizontal segment with identified endpoints and $N$ marked nodes. Each node is either left vacant, paired to another node by a loop segment or connected above to infinity by a loop segment. In this last case, we say that the node is occupied by a defect. The loop segments are non-intersecting. The boundary conditions are cylindric, so that two nodes may be connected by a loop segment via the front or back of the cylinder. The standard module $\stan_{N,d}$ is then defined on the span of link states with exactly $d$ defects, with $0 \le d \le N$. The link states for $N=3$ are
\be
\begin{array}{l}
\stan_{3,3}:\qquad 
\begin{pspicture}[shift=0](0.0,0)(1.2,0.5)
\psline[linewidth=\mince](0,0)(1.2,0)
\psline[linecolor=blue,linewidth=\elegant]{-}(0.2,0)(0.2,0.4)
\psline[linecolor=blue,linewidth=\elegant]{-}(0.6,0)(0.6,0.4)
\psline[linecolor=blue,linewidth=\elegant]{-}(1.0,0)(1.0,0.4)
\end{pspicture}\ \ ,
\qquad \qquad
\stan_{3,2}:\qquad 
\begin{pspicture}[shift=0](0.0,0)(1.2,0.5)
\psline[linewidth=\mince](0,0)(1.2,0)
\psline[linecolor=blue,linewidth=\elegant]{-}(0.2,0)(0.2,0.4)
\psline[linecolor=blue,linewidth=\elegant]{-}(0.6,0)(0.6,0.4)
\pscircle[fillstyle=solid,fillcolor=black](1.0,0){0.04}
\end{pspicture}
\quad
\begin{pspicture}[shift=0](0.0,0)(1.2,0.5)
\psline[linewidth=\mince](0,0)(1.2,0)
\psline[linecolor=blue,linewidth=\elegant]{-}(0.2,0)(0.2,0.4)
\psline[linecolor=blue,linewidth=\elegant]{-}(1.0,0)(1.0,0.4)
\pscircle[fillstyle=solid,fillcolor=black](0.6,0){0.04}
\end{pspicture}
\quad
\begin{pspicture}[shift=0](0.0,0)(1.2,0.5)
\psline[linewidth=\mince](0,0)(1.2,0)
\psline[linecolor=blue,linewidth=\elegant]{-}(0.6,0)(0.6,0.4)
\psline[linecolor=blue,linewidth=\elegant]{-}(1.0,0)(1.0,0.4)
\pscircle[fillstyle=solid,fillcolor=black](0.2,0){0.04}
\end{pspicture}\ \ ,
\\[0.2cm]
\stan_{3,1}:\qquad 
\begin{pspicture}[shift=0](0.0,0)(1.2,0.5)
\psline[linewidth=\mince](0,0)(1.2,0)
\psarc[linecolor=blue,linewidth=\elegant]{-}(0.8,0){0.2}{0}{180}
\psline[linecolor=blue,linewidth=\elegant]{-}(0.2,0)(0.2,0.4)
\end{pspicture}
\quad
\begin{pspicture}[shift=0](0.0,0)(1.2,0.5)
\psline[linewidth=\mince](0,0)(1.2,0)
\psarc[linecolor=blue,linewidth=\elegant]{-}(0.4,0){0.2}{0}{180}
\psline[linecolor=blue,linewidth=\elegant]{-}(1.0,0)(1.0,0.4)
\end{pspicture}
\quad
\begin{pspicture}[shift=0](0.0,0)(1.2,0.5)
\psline[linewidth=\mince](0,0)(1.2,0)
\psarc[linecolor=blue,linewidth=\elegant]{-}(0,0){0.2}{0}{90}
\psarc[linecolor=blue,linewidth=\elegant]{-}(1.2,0){0.2}{90}{180}
\psline[linecolor=blue,linewidth=\elegant]{-}(0.6,0)(0.6,0.4)
\end{pspicture}
\quad
\begin{pspicture}[shift=0](0.0,0)(1.2,0.5)
\psline[linewidth=\mince](0,0)(1.2,0)
\pscircle[fillstyle=solid,fillcolor=black](0.6,0){0.04}
\pscircle[fillstyle=solid,fillcolor=black](1,0){0.04}
\psline[linecolor=blue,linewidth=\elegant]{-}(0.2,0)(0.2,0.4)
\end{pspicture}
\quad
\begin{pspicture}[shift=0](0.0,0)(1.2,0.5)
\psline[linewidth=\mince](0,0)(1.2,0)
\pscircle[fillstyle=solid,fillcolor=black](0.2,0){0.04}
\pscircle[fillstyle=solid,fillcolor=black](1.0,0){0.04}
\psline[linecolor=blue,linewidth=\elegant]{-}(0.6,0)(0.6,0.4)
\end{pspicture}
\quad
\begin{pspicture}[shift=0](0.0,0)(1.2,0.5)
\psline[linewidth=\mince](0,0)(1.2,0)
\pscircle[fillstyle=solid,fillcolor=black](0.2,0){0.04}
\pscircle[fillstyle=solid,fillcolor=black](0.6,0){0.04}
\psline[linecolor=blue,linewidth=\elegant]{-}(1.0,0)(1.0,0.4)
\end{pspicture}
\ \ ,
\\[0.2cm]%
\stan_{3,0}:\qquad
\begin{pspicture}[shift=0](0.0,0)(1.2,0.5)
\psline[linewidth=\mince](0,0)(1.2,0)
\pscircle[fillstyle=solid,fillcolor=black](0.2,0){0.04}
\psarc[linecolor=blue,linewidth=\elegant]{-}(0.8,0){0.2}{0}{180}
\end{pspicture}
\quad
\begin{pspicture}[shift=0](0.0,0)(1.2,0.5)
\psline[linewidth=\mince](0,0)(1.2,0)
\pscircle[fillstyle=solid,fillcolor=black](1.0,0){0.04}
\psarc[linecolor=blue,linewidth=\elegant]{-}(0.4,0){0.2}{0}{180}
\end{pspicture}
\quad
\begin{pspicture}[shift=0](0.0,0)(1.2,0.5)
\psline[linewidth=\mince](0,0)(1.2,0)
\pscircle[fillstyle=solid,fillcolor=black](0.6,0){0.04}
\psbezier[linecolor=blue,linewidth=\elegant](0.2,0)(0.2,0.4)(1.0,0.4)(1.0,0)
\end{pspicture}
\quad
\begin{pspicture}[shift=0](0.0,0)(1.2,0.5)
\psline[linewidth=\mince](0,0)(1.2,0)
\pscircle[fillstyle=solid,fillcolor=black](0.2,0){0.04}
\psbezier[linecolor=blue,linewidth=\elegant](1.0,0)(1.0,0.4)(1.8,0.4)(1.8,0)\rput(-1.2,0){\psbezier[linecolor=blue,linewidth=\elegant](1.0,0)(1.0,0.4)(1.8,0.4)(1.8,0)}
\psframe[fillstyle=solid,linecolor=white,linewidth=0pt](1.2,0)(2.5,0.4)
\psframe[fillstyle=solid,linecolor=white,linewidth=0pt](0,0)(-0.5,0.4)
\end{pspicture}
\quad
\begin{pspicture}[shift=0](0.0,0)(1.2,0.5)
\psline[linewidth=\mince](0,0)(1.2,0)
\pscircle[fillstyle=solid,fillcolor=black](1.0,0){0.04}
\psbezier[linecolor=blue,linewidth=\elegant](0.6,0)(0.6,0.4)(1.4,0.4)(1.4,0)\psbezier[linecolor=blue,linewidth=\elegant](-0.1,0.30)(0.1,0.23)(0.2,0.16)(0.2,0)
\psframe[fillstyle=solid,linecolor=white,linewidth=0pt](1.2,0)(2.5,0.4)
\psframe[fillstyle=solid,linecolor=white,linewidth=0pt](0,0)(-0.2,0.4)
\end{pspicture}
\quad
\begin{pspicture}[shift=0](0.0,0)(1.2,0.5)
\psline[linewidth=\mince](0,0)(1.2,0)
\pscircle[fillstyle=solid,fillcolor=black](0.6,0){0.04}
\psarc[linecolor=blue,linewidth=\elegant]{-}(0,0){0.2}{0}{90}
\psarc[linecolor=blue,linewidth=\elegant]{-}(1.2,0){0.2}{90}{180}
\end{pspicture}
\quad
\begin{pspicture}[shift=0](0.0,0)(1.2,0.5)
\psline[linewidth=\mince](0,0)(1.2,0)
\pscircle[fillstyle=solid,fillcolor=black](0.2,0){0.04}
\pscircle[fillstyle=solid,fillcolor=black](0.6,0){0.04}
\pscircle[fillstyle=solid,fillcolor=black](1.0,0){0.04}
\end{pspicture}\ \ .
\end{array}
\ee

The standard action of a connectivity $a \in \pdTL_N(\alpha,\beta)$ on a link state $w\in \stan_{N,d}$ is denoted $a w$ and defined as follows. The link state $w$ is placed on top of $a$ and the $N$ nodes are connected together. If two defects become connected, or if a loop segment becomes connected to a vacancy, the result is set to zero. Otherwise, the result is the link state read from the bottom segment of the rectangle, times the scalar $\alpha^{n_\alpha}\beta^{n_\beta}\omega^{n_\omega}$. As before, $n_\alpha$ and $n_\beta$ count the non-contractible and contractible loops appearing in the diagram. Moreover $n_\omega$, where $\omega$ is the twist parameter, measures the winding of the defects around the cylinder. Each defect that crosses the vertical line where the cylinder is cut contributes $+1$ to $n_\omega$ if it travels towards the left, whereas it contributes $-1$ if it travels towards the right as it progresses down the cylinder. Clearly, powers of $\omega$ only appear in the standard modules $\stan_{N,d}$ for $d>0$. Likewise, factors of $\alpha$ only appear for $\stan_{N,d=0}$. Some examples of the standard action are
\begin{subequations}
\begin{alignat}{3}
&
\begin{pspicture}[shift=-0.3](-0.0,0)(2.0,1.2)
\pspolygon[fillstyle=solid,fillcolor=lightlightblue,linecolor=black,linewidth=0pt](0,0)(0,0.8)(2.0,0.8)(2.0,0)(0,0)
\psarc[linecolor=blue,linewidth=\elegant]{-}(0.8,0.8){0.2}{180}{0}
\psbezier[linecolor=blue,linewidth=\elegant]{-}(0.2,0.8)(0.2,0.4)(1.0,0.4)(1.0,0)
\psbezier[linecolor=blue,linewidth=\elegant]{-}(1.4,0.8)(1.4,0.5)(1.92,0.4)(2.02,0.26)
\psbezier[linecolor=blue,linewidth=\elegant]{-}(0.2,0)(0.2,0.15)(0.08,0.24)(-0.02,0.3)
\pscircle[fillstyle=solid,fillcolor=black](1.8,0.8){0.035}
\pscircle[fillstyle=solid,fillcolor=black](0.6,0){0.035}
\pscircle[fillstyle=solid,fillcolor=black](1.4,0){0.035}
\pscircle[fillstyle=solid,fillcolor=black](1.8,0){0.035}
\psframe[fillstyle=solid,linecolor=white,linewidth=0pt](-0.04,0)(0,0.8)
\psframe[fillstyle=solid,linecolor=white,linewidth=0pt](2.0,0)(2.04,0.8)
\rput(0,0.8){
\psarc[linecolor=blue,linewidth=\elegant]{-}(0.4,0){0.2}{0}{180}
\pscircle[fillstyle=solid,fillcolor=black](1.8,0){0.04}
\psline[linecolor=blue,linewidth=\elegant]{-}(1.0,0)(1.0,0.4)
\psline[linecolor=blue,linewidth=\elegant]{-}(1.4,0)(1.4,0.4)
}
\end{pspicture} \ = \omega^{-1} \ 
\begin{pspicture}[shift=0](0.0,0)(2.0,0.5)
\psline[linewidth=\mince](0,0)(2.0,0)
\pscircle[fillstyle=solid,fillcolor=black](0.6,0){0.04}
\pscircle[fillstyle=solid,fillcolor=black](1.4,0){0.04}
\pscircle[fillstyle=solid,fillcolor=black](1.8,0){0.04}
\psline[linecolor=blue,linewidth=\elegant]{-}(0.2,0)(0.2,0.4)
\psline[linecolor=blue,linewidth=\elegant]{-}(1.0,0)(1.0,0.4)
\end{pspicture}\ ,
\qquad\qquad&
\begin{pspicture}[shift=-0.3](-0.0,0)(2.0,1.2)
\pspolygon[fillstyle=solid,fillcolor=lightlightblue,linecolor=black,linewidth=0pt](0,0)(0,0.8)(2.0,0.8)(2.0,0)(0,0)
\psarc[linecolor=blue,linewidth=\elegant]{-}(1.6,0.8){0.2}{180}{0}
\psarc[linecolor=blue,linewidth=\elegant]{-}(0.8,0){0.2}{0}{180}
\psbezier[linecolor=blue,linewidth=\elegant]{-}(0.2,0)(0.2,0.5)(1.4,0.5)(1.4,0)
\psbezier[linecolor=blue,linewidth=\elegant]{-}(1.0,0.8)(1.0,0.33)(1.92,0.39)(2.02,0.53)
\psbezier[linecolor=blue,linewidth=\elegant]{-}(0.2,0.8)(0.2,0.65)(0.08,0.6)(-0.02,0.56)
\pscircle[fillstyle=solid,fillcolor=black](0.6,0.8){0.035}
\pscircle[fillstyle=solid,fillcolor=black](1.8,0){0.035}
\psframe[fillstyle=solid,linecolor=white,linewidth=0pt](-0.04,0)(0,0.8)
\psframe[fillstyle=solid,linecolor=white,linewidth=0pt](2.0,0)(2.04,0.8)
\rput(0,0.8){
\psarc[linecolor=blue,linewidth=\elegant]{-}(1.2,0){0.2}{0}{180}
\pscircle[fillstyle=solid,fillcolor=black](0.6,0){0.04}
\psline[linecolor=blue,linewidth=\elegant]{-}(0.2,0)(0.2,0.4)
\psline[linecolor=blue,linewidth=\elegant]{-}(1.8,0)(1.8,0.4)
}
\end{pspicture}\ = 0,
\\[0.3cm]
&
\begin{pspicture}[shift=-0.3](-0.0,0)(2.0,1.2)
\pspolygon[fillstyle=solid,fillcolor=lightlightblue,linecolor=black,linewidth=0pt](0,0)(0,0.8)(2.0,0.8)(2.0,0)(0,0)
\psarc[linecolor=blue,linewidth=\elegant]{-}(1.6,0.8){0.2}{180}{0}
\psarc[linecolor=blue,linewidth=\elegant]{-}(0.8,0){0.2}{0}{180}
\psbezier[linecolor=blue,linewidth=\elegant]{-}(0.2,0)(0.2,0.5)(1.4,0.5)(1.4,0)
\psbezier[linecolor=blue,linewidth=\elegant]{-}(1.0,0.8)(1.0,0.33)(1.92,0.39)(2.02,0.53)
\psbezier[linecolor=blue,linewidth=\elegant]{-}(0.2,0.8)(0.2,0.65)(0.08,0.6)(-0.02,0.56)
\pscircle[fillstyle=solid,fillcolor=black](0.6,0.8){0.035}
\pscircle[fillstyle=solid,fillcolor=black](1.8,0){0.035}
\psframe[fillstyle=solid,linecolor=white,linewidth=0pt](-0.04,0)(0,0.8)
\psframe[fillstyle=solid,linecolor=white,linewidth=0pt](2.0,0)(2.04,0.8)
\rput(0,0.8){
\psarc[linecolor=blue,linewidth=\elegant]{-}(1.6,0){0.2}{0}{180}
\pscircle[fillstyle=solid,fillcolor=black](0.6,0){0.04}
\psbezier[linecolor=blue,linewidth=\elegant]{-}(0.2,0)(0.2,0.4)(1.0,0.4)(1.0,0)
}
\end{pspicture} \ = \alpha\beta \ 
\begin{pspicture}[shift=0](0.0,0)(2.0,0.5)
\psline[linewidth=\mince](0,0)(2.0,0)
\pscircle[fillstyle=solid,fillcolor=black](1.8,0){0.04}
\psbezier[linecolor=blue,linewidth=\elegant]{-}(0.2,0)(0.2,0.5)(1.4,0.5)(1.4,0)
\psarc[linecolor=blue,linewidth=\elegant]{-}(0.8,0){0.2}{0}{180}
\end{pspicture}\ ,
\qquad\qquad&
\begin{pspicture}[shift=-0.3](-0.0,0)(2.0,1.2)
\pspolygon[fillstyle=solid,fillcolor=lightlightblue,linecolor=black,linewidth=0pt](0,0)(0,0.8)(2.0,0.8)(2.0,0)(0,0)
\pscircle[fillstyle=solid,fillcolor=black](0.6,0.8){0.035}
\pscircle[fillstyle=solid,fillcolor=black](1.4,0.8){0.035}
\pscircle[fillstyle=solid,fillcolor=black](1.8,0.8){0.035}
\pscircle[fillstyle=solid,fillcolor=black](0.6,0){0.035}
\psbezier[linecolor=blue,linewidth=\elegant]{-}(0.2,0)(0.2,0.4)(1.0,0.4)(1.0,0)
\psbezier[linecolor=blue,linewidth=\elegant]{-}(1.4,0)(1.4,0.3)(1.82,0.38)(2.02,0.52)
\psbezier[linecolor=blue,linewidth=\elegant]{-}(0.2,0.8)(0.2,0.65)(0.08,0.55)(-0.02,0.50)
\psbezier[linecolor=blue,linewidth=\elegant]{-}(1.8,0)(1.8,0.14)(1.92,0.23)(2.02,0.28)
\psbezier[linecolor=blue,linewidth=\elegant]{-}(1.0,0.8)(1.0,0.56)(0.08,0.44)(-0.02,0.24)
\psframe[fillstyle=solid,linecolor=white,linewidth=0pt](-0.04,0)(0,0.8)
\psframe[fillstyle=solid,linecolor=white,linewidth=0pt](2.0,0)(2.04,0.8)
\rput(0,0.8){
\psarc[linecolor=blue,linewidth=\elegant]{-}(1.2,0){0.2}{0}{180}
\pscircle[fillstyle=solid,fillcolor=black](1.8,0){0.04}
\psline[linecolor=blue,linewidth=\elegant]{-}(0.2,0)(0.2,0.4)
\psline[linecolor=blue,linewidth=\elegant]{-}(0.6,0)(0.6,0.4)
}
\end{pspicture}\ = 0.
\end{alignat}
\end{subequations}
The standard module $\stan_{N,d}$ has dimension $\binom{N}{d}_2$, where the trinomial coefficient is defined by
\be
(x+1+x^{-1})^N = \sum_{k=-N}^N \binom{N}{k}_{\!\!2}\, x^k.
\ee

\section{Diagrammatic calculus}\label{sec:diag.calculus}

\subsection{Local relations}

The diagrams appearing in the decomposition \eqref{eq:face.op} of the face operator have a free node on each of the four sides, so it is not a priori an element of $\pdTL_N(\alpha,\beta)$. It is instead~\cite{planarJones1999} an element of the planar dilute Temperley-Lieb algebra. In the planar setting, the algebraic relations are implemented in terms of the natural diagrammatic products of the nine tiles in \eqref{eq:face.op}. We refer to objects in the planar algebra as {\it tangles}. Still, the face operator can be seen as an element of $\pdTL_2(\alpha,\beta)$ by fixing a direction of action. One can, for instance, decide that the top and right nodes live on the top segment of the rectangle, and that the left and bottom nodes live on the bottom segment. 

The face operator satisfies a number of local relations. These express the equality of tangles in the planar algebra and can be converted to identities in $\pdTL_N(\alpha,\beta)$ by fixing a direction of action. Recalling the definition \eqref{eq:id} of the dashed loop segment, we find the initial condition
\be
\label{eq:face.u=0}
\begin{pspicture}[shift=-.40](0,0)(1,1)
\facegrid{(0,0)}{(1,1)}
\psarc[linewidth=0.025]{-}(0,0){0.16}{0}{90}
\rput(.5,.5){\small $u=0$}
\end{pspicture}\ = \
\begin{pspicture}[shift=-.40](0,0)(1,1)
\facegrid{(0,0)}{(1,1)}
\rput[bl](0,0){\loopa}
\end{pspicture}
\ + \
\begin{pspicture}[shift=-.40](0,0)(1,1)
\facegrid{(0,0)}{(1,1)}
\rput[bl](0,0){\loopb}
\end{pspicture} 
\ + \
\begin{pspicture}[shift=-.40](0,0)(1,1)
\facegrid{(0,0)}{(1,1)}
\rput[bl](0,0){\loopc}
\end{pspicture} 
\ + \
\begin{pspicture}[shift=-.40](0,0)(1,1)
\facegrid{(0,0)}{(1,1)}
\rput[bl](0,0){\looph}
\end{pspicture} 
\ = \
\begin{pspicture}[shift=-.40](0,0)(1,1)
\facegrid{(0,0)}{(1,1)}
\rput[bl](0,0){\loopid}
\end{pspicture} \ \ .
\ee
With the direction of action fixed to be up and to the right, 
this tangle corresponds to the identity element of $\pdTL_2(\alpha,\beta)$.
The face operators also satisfy the crossing and the inversion relations
\be
\label{eq:cr+id}
\begin{pspicture}[shift=-.40](0,0)(1,1)
\facegrid{(0,0)}{(1,1)}
\psarc[linewidth=0.025]{-}(0,0){0.16}{0}{90}
\rput(.5,.5){$u$}
\end{pspicture}\ = \
\begin{pspicture}[shift=-.40](0,0)(1,1)
\facegrid{(0,0)}{(1,1)}
\psarc[linewidth=0.025]{-}(1,0){0.16}{90}{180}
\rput(.5,.5){\scriptsize$3\lambda-u$}
\end{pspicture}\ \ , \qquad\quad
\psset{unit=0.7364cm}
\begin{pspicture}[shift=-0.95](-0.1,-1.1)(4.3,1.1)
\pspolygon[fillstyle=solid,fillcolor=lightlightblue](0,0)(1,1)(3,-1)(4,0)(3,1)(1,-1)
\psarc[linewidth=1.5pt,linecolor=blue,linestyle=dashed,dash=2pt 2pt](2,0){.707}{45}{135}
\psarc[linewidth=1.5pt,linecolor=blue,linestyle=dashed,dash=2pt 2pt](2,0){.707}{-135}{-45}
\psarc[linewidth=0.025]{-}(0,0){0.21}{-45}{45}
\psarc[linewidth=0.025]{-}(2,0){0.21}{-45}{45}
\rput(1,0){$u$}
\rput(3,0){$-u$}
\end{pspicture}
= \ \rho_8(u)\rho_8(-u) \, \, \,
\begin{pspicture}[shift=-0.95](-0.1,-1.1)(2.3,1.1)
\pspolygon[fillstyle=solid,fillcolor=lightlightblue](0,0)(1,1)(2,0)(1,-1)(0,0)
\psarc[linewidth=1.5pt,linecolor=blue,linestyle=dashed,dash=2pt 2pt](1,-1){.707}{45}{135}
\psarc[linewidth=1.5pt,linecolor=blue,linestyle=dashed,dash=2pt 2pt](1,1){.707}{-135}{-45}
\end{pspicture}\, .
\ee
The inversion relation is proved by expanding the face operators in terms of the nine tiles on the right side of \eqref{eq:face.op}, removing all terms in which vacant sites are connected to loop segments, collecting the remaining terms according to the connectivity of the nodes, simplifying the resulting coefficients, and checking that this reproduces the right side of the equation. 
The explicit diagram decomposition of a similar inversion relation for the $A_2^{(1)}$ loop model is given in (3.5) of \cite{MDPR2018}.

An alternative proof of the inversion relation, more in the spirit of methods used later in this paper, is as follows. 
Let us call a Laurent polynomial with maximal and minimal powers $z^n$ and $z^{-n}$ a {\em centered Laurent polynomial\/} of degree $n$.
We then note that both the left and right sides of the equality are centered Laurent polynomials in $z = \eE^{\ir u}$ of degree $4$. 
It therefore suffices to prove the polynomial equality at nine distinct values of $z$, for instance, at $u = 0, \pm 2 \lambda, \pm 3 \lambda$ and at the same values shifted by $\pi$. In this instance, the alternative derivation is more tedious but this idea to use the polynomial properties of tangles in $z$ to establish an equality will prove useful subsequently.

The face operators also satisfy the Yang-Baxter equation
\be
\begin{pspicture}[shift=-0.9](0,0)(3,2.3)
\facegrid{(2,0)}{(3,2)}
\pspolygon[fillstyle=solid,fillcolor=lightlightblue](0,1)(1,2)(2,1)(1,0)(0,1)
\psarc[linewidth=0.025]{-}(2,0){0.16}{0}{90}
\psarc[linewidth=0.025]{-}(2,1){0.16}{0}{90}
\psarc[linewidth=0.025]{-}(0,1){0.21}{-45}{45}
\rput(2.5,.5){$u$}
\rput(2.5,1.5){$v$}
\rput(1,1){$u-v$}
\psline[linecolor=blue,linewidth=1.5pt,linestyle=dashed,dash=2pt 2pt]{-}(2,0.5)(1.5,0.5)
\psline[linecolor=blue,linewidth=1.5pt,linestyle=dashed,dash=2pt 2pt]{-}(2,1.5)(1.5,1.5)
\end{pspicture}
 \ =\ 
\begin{pspicture}[shift=-0.9](0,0)(3,2)
\facegrid{(0,0)}{(1,2)}
\pspolygon[fillstyle=solid,fillcolor=lightlightblue](1,1)(2,2)(3,1)(2,0)(1,1)
\psarc[linewidth=0.025]{-}(0,0){0.16}{0}{90}
\psarc[linewidth=0.025]{-}(0,1){0.16}{0}{90}
\psarc[linewidth=0.025]{-}(1,1){0.21}{-45}{45}
\rput(.5,.5){$v$}
\rput(.5,1.5){$u$}
\rput(2,1){$u-v$}
\psline[linecolor=blue,linewidth=1.5pt,linestyle=dashed,dash=2pt 2pt]{-}(1,0.5)(1.5,0.5)
\psline[linecolor=blue,linewidth=1.5pt,linestyle=dashed,dash=2pt 2pt]{-}(1,1.5)(1.5,1.5)
\end{pspicture}\ \ .
\label{eq:YBE}
\ee
This is the key relation ensuring that the $A_2^{(2)}$ loop models are integrable. It can be proved using either of the two methods described above.

\subsection{Degeneration points and push-through properties}

The face operator is non-invertible or singular at certain degeneration points. They coincide with the values of $u$ where the right side of the inversion relation vanishes, namely $u = \pm 2 \lambda, \pm 3 \lambda$. 
At such degeneration points, the face operator can factorise into triangle face operators. 
For $u = 3\lambda$, we find
\be
\begin{pspicture}[shift=-.40](0,0)(1,1)
\facegrid{(0,0)}{(1,1)}
\psarc[linewidth=0.025]{-}(0,0){0.16}{0}{90}
\rput(.5,.5){$3\lambda$}
\end{pspicture}\ = \
\begin{pspicture}[shift=-.40](0,0)(1,1)
\facegrid{(0,0)}{(1,1)}
\psline{-}(1,0)(0,1)
\rput[bl](0,0){
\psarc[linewidth=1.5pt,linecolor=blue,linestyle=dashed,dash=2pt 2pt](0,0){.5}{0}{90}
\psarc[linewidth=1.5pt,linecolor=blue,linestyle=dashed,dash=2pt 2pt](1,1){.5}{180}{-90}
}
\end{pspicture}
\ee
which follows directly from \eqref{eq:face.u=0} and the crossing relation. The diagonal line on the right face is to emphasise the fact that
the action of this operator splits as a product of two triangle operators.
Similarly, for $u = 2\lambda$, we find
\be
\begin{pspicture}[shift=-.40](0,0)(1,1)
\facegrid{(0,0)}{(1,1)}
\psarc[linewidth=0.025]{-}(0,0){0.16}{0}{90}
\rput(.5,.5){$2\lambda$}
\end{pspicture}
\ = \
\frac{\sin \lambda}{\sin 3\lambda} \ \ 
\begin{pspicture}[shift=-.40](0,0)(1,1)
\facegrid{(0,0)}{(1,1)}
\psline{-}(1,0)(0,1)
\rput(0.3,0.3){\specialcircle{0.05}}
\rput(0.7,0.7){\specialcircle{0.05}}
\end{pspicture} 
\ee
where
\be
\psset{unit=0.8}
\begin{pspicture}[shift=-.30](0,0)(2,1)
\pspolygon[fillstyle=solid,fillcolor=lightlightblue](0,0)(1,1)(2,0)
\rput(1,0.4){\specialcircle{0.07}}
\end{pspicture} 
\ = 2 \cos \lambda \
\begin{pspicture}[shift=-.30](0,0)(2,1)
\pspolygon[fillstyle=solid,fillcolor=lightlightblue](0,0)(1,1)(2,0)
\end{pspicture} 
\ + \ 
\begin{pspicture}[shift=-.30](0,0)(2,1)
\pspolygon[fillstyle=solid,fillcolor=lightlightblue](0,0)(1,1)(2,0)
\psarc[linewidth=1.5pt,linecolor=blue](1,1){.707}{-135}{-45}
\end{pspicture} 
\ + \
\begin{pspicture}[shift=-.30](0,0)(2,1)
\pspolygon[fillstyle=solid,fillcolor=lightlightblue](0,0)(1,1)(2,0)
\psline[linewidth=1.5pt,linecolor=blue](0.5,0.5)(1,0)
\end{pspicture} 
\ + \
\begin{pspicture}[shift=-.30](0,0)(2,1)
\pspolygon[fillstyle=solid,fillcolor=lightlightblue](0,0)(1,1)(2,0)
\psline[linewidth=1.5pt,linecolor=blue](1.5,0.5)(1,0)
\end{pspicture} \ .
\ee
Such a decomposition was previously found in \cite{GarbNien2017}. 

These two triangle face operators play an important role for the $A^{(2)}_2$ models. They satisfy the push-through properties
\be
\begin{pspicture}[shift=-0.9](0,0.0)(2,2)
\facegrid{(0,0)}{(1,2)}
\psarc[linewidth=0.025]{-}(0,0){0.16}{0}{90}
\psarc[linewidth=0.025]{-}(0,1){0.16}{0}{90}
\psline[linewidth=1.5pt,linecolor=blue,linestyle=dashed,dash=2pt 2pt]{-}(1,0.5)(1.5,0.5)
\psline[linewidth=1.5pt,linecolor=blue,linestyle=dashed,dash=2pt 2pt]{-}(1,1.5)(1.5,1.5)
\pspolygon[fillstyle=solid,fillcolor=lightlightblue](1,1)(2,0)(2,2)
\rput(1.6,1){\specialcircle{0.06}}
\rput(0.5,.5){$u_0$}
\rput(0.5,1.5){$u_2$}
\end{pspicture} \ \ = s_2(u)s_{3}(-u) \ \
\begin{pspicture}[shift=-0.4](-0.5,0)(1,1)
\facegrid{(0,0)}{(1,1)}
\pspolygon[fillstyle=solid,fillcolor=lightlightblue](0,0)(0,1)(-0.5,0.5)\rput(-0.2,0.5){\specialcircle{0.05}}
\psarc[linewidth=0.025]{-}(0,0){0.16}{0}{90}
\rput(0.5,0.5){$u_1$}
\end{pspicture}\ \ ,\qquad
\begin{pspicture}[shift=-0.9](0,0)(1.5,2)
\facegrid{(0,0)}{(1,2)}
\psarc[linewidth=0.025]{-}(0,0){0.16}{0}{90}
\psarc[linewidth=0.025]{-}(0,1){0.16}{0}{90}
\psarc[linewidth=1.5pt,linecolor=blue,linestyle=dashed,dash=2pt 2pt]{-}(1,1){0.5}{-90}{90}
\rput(0.5,.5){$u_0$}
\rput(0.5,1.5){$u_3$}
\end{pspicture} \ \ = s_2(u)s_2(-u)s_3(u)s_3(-u)  \ \
\begin{pspicture}[shift=-0.9](0,0.0)(1.5,2)
\facegrid{(0,0)}{(1,2)}
\psarc[linewidth=1.5pt,linecolor=blue,linestyle=dashed,dash=2pt 2pt]{-}(1,1){0.5}{-90}{90}
\psarc[linewidth=1.5pt,linecolor=blue,linestyle=dashed,dash=2pt 2pt]{-}(1,0){0.5}{90}{180}
\psarc[linewidth=1.5pt,linecolor=blue,linestyle=dashed,dash=2pt 2pt]{-}(0,1){0.5}{-90}{90}
\psarc[linewidth=1.5pt,linecolor=blue,linestyle=dashed,dash=2pt 2pt]{-}(1,2){0.5}{180}{270}
\end{pspicture}\ \ ,
\label{eq:pushthru}
\ee
where we use the short-hand notations
\be
u_k = u+ k \lambda, \qquad s_k(u) = \frac{\sin(u+k\lambda)}{(\sin 2 \lambda \sin 3\lambda)^{1/2}}.
\ee
The normalization of the trigonometric functions is convenient but the square root can introduce factors of the imaginary unit $\ir$ when the argument is negative. In principle, this can be avoided by changing the normalization of the face weights and the function $s_k(u)$, so that the factors of $\sin 2 \lambda \sin 3\lambda$ are absent from the denominators. In fact, these renormalization factors cancel out of all the homogeneous functional equations that follow.

\section{Transfer tangles and functional equations}\label{sec:Ts.and.FH}

\subsection{Fundamental transfer tangles}
In this subsection and the next, we define mutually commuting families of transfer tangles $\Tb^{m,n}(u)$ whose labels $(m,n)$ live on the infinite dominant integral weight lattice of $s\ell(3)$ as shown in \cref{fig:su3weights} with
\be
\label{Young}
(m,n) \qquad \longleftrightarrow\qquad
\psset{unit=4}
\begin{pspicture}[shift=-0.18](0,-0.1)(0.7,0.2)
\multiput(0,0)(.1,0){4}{\psline[linewidth=0.02cm]{-}(0,0)(0,.1)(.1,.1)(.1,0)(0,0)}
\multiput(0,.1)(.1,0){7}{\psline[linewidth=0.02cm]{-}(0,0)(0,.1)(.1,.1)(.1,0)(0,0)}
\rput(0.2,-0.08){$\underbrace{\ \hspace{1.3cm} \ }_n$}
\rput(0.55,0.02){$\underbrace{\ \hspace{0.85cm} \ }_m$}
\end{pspicture} \ \ .
\ee
We start by defining the single-row fundamental transfer tangle $\Tb^{1,0}(u) = \Tb(u)$. It is an element of $\pdTL_N(\alpha,\beta)$ given by
\be
\psset{unit=1.1cm}
\Tb^{1,0}(u) = \ \
\begin{pspicture}[shift=-0.4](-0.3,0)(4.3,1.0)
\facegrid{(0,0)}{(4,1)}
\psarc[linewidth=0.025]{-}(0,0){0.16}{0}{90}
\psarc[linewidth=0.025]{-}(1,0){0.16}{0}{90}
\psarc[linewidth=0.025]{-}(3,0){0.16}{0}{90}
\psline[linewidth=1.5pt,linecolor=blue,linestyle=dashed,dash=2pt 2pt]{-}(0,0.5)(-0.3,0.5)
\psline[linewidth=1.5pt,linecolor=blue,linestyle=dashed,dash=2pt 2pt]{-}(4,0.5)(4.3,0.5)
\rput(2.5,0.5){$\ldots$}
\rput(0.5,.5){$u\!-\!\xi_1$}
\rput(1.5,.5){$u\!-\!\xi_2$}
\rput(3.5,.5){$u\!-\!\xi_N$}
\end{pspicture}\ \ =\Tb(u)
\ee
where $\xi_1, \dots, \xi_N$ are arbitrary column inhomogeneities. The transfer tangle or transfer matrix $\Tb(u)$ is a centered Laurent polynomial in $z = \eE^{\ir u}$ of degree $2N$ with coefficients that are elements of $\pdTL_N(\alpha,\beta)$. Two transfer tangles with the same column inhomogeneities but different spectral parameters commute: $[\Tb(u),\Tb(v)]=0$. The proof uses the inversion relation and the Yang-Baxter equation and follows the usual graphical arguments~\cite{BaxterBook1982}. The transfer tangle also satisfies the periodicity $\Tb(u+\pi) = \Tb(u)$. To prove this, one observes that the face operators at $u$ and  $u+\pi$ differ by a simple gauge transformation, as in the $A_2^{(1)}$ case, see (3.23) of \cite{MDPR2018}.

The conjugate fundamental transfer tangle $\Tb^{0,1}(u)$ is given by
\be
\psset{unit=1.1cm}
\Tb^{0,1}(u) = \ \
\begin{pspicture}[shift=-0.4](-0.3,0)(4.3,1.0)
\facegrid{(0,0)}{(4,1)}
\psarc[linewidth=0.025]{-}(1,0){0.16}{90}{180}
\psarc[linewidth=0.025]{-}(2,0){0.16}{90}{180}
\psarc[linewidth=0.025]{-}(4,0){0.16}{90}{180}
\psline[linewidth=1.5pt,linecolor=blue,linestyle=dashed,dash=2pt 2pt]{-}(0,0.5)(-0.3,0.5)
\psline[linewidth=1.5pt,linecolor=blue,linestyle=dashed,dash=2pt 2pt]{-}(4,0.5)(4.3,0.5)
\rput(2.5,0.5){$\ldots$}
\rput(0.5,.5){$u'\!\!+\!\xi_1$}
\rput(1.5,.5){$u'\!\!+\!\xi_2$}
\rput(3.5,.5){$u'\!\!+\!\xi_N$}
\end{pspicture}\ \ =\Tb(u+\lambda),\qquad u'=2\lambda-u,
\ee
and is equal, by the crossing symmetry, to the first fundamental transfer tangle up to a shift of the spectral parameter.
This is in marked contrast to the situation for $A_2^{(1)}$ loop models where, due to the lack of crossing symmetry, the two fundamental transfer tangles are inequivalent~\cite{MDPR2018}. 

\subsection{Fused transfer tangles}

The usual technique to define general $\Tb^{m,n}(u)$ fused transfer matrices  is to introduce Wenzl-Jones projectors. These are used to construct fused face operators that are Laurent polynomials in $z$ of finite degree. The fused transfer tangles are defined in terms of the fused face operators by
\be
\label{eq:fusedTmn}
\psset{unit=1.1cm}
\Tb^{m,n} (u)= \  \
\begin{pspicture}[shift=-0.4](-0.3,0)(4.3,1.0)
\facegrid{(0,0)}{(4,1)}
\psarc[linewidth=0.025]{-}(0,0){0.16}{0}{90}
\psarc[linewidth=0.025]{-}(1,0){0.16}{0}{90}
\psarc[linewidth=0.025]{-}(3,0){0.16}{0}{90}
\psline[linewidth=1.5pt,linecolor=blue,linestyle=dashed,dash=2pt 2pt]{-}(0,0.5)(-0.3,0.5)
\psline[linewidth=1.5pt,linecolor=blue,linestyle=dashed,dash=2pt 2pt]{-}(4,0.5)(4.3,0.5)
\rput(2.5,0.5){$\ldots$}
\rput(0.5,.4){\small$u\!-\!\xi_1$}\rput(.5,.7){\tiny $(m,n)$}
\rput(1.5,.4){\small$u\!-\!\xi_2$}\rput(1.5,.7){\tiny $(m,n)$}
\rput(3.5,.4){\small$u\!-\!\xi_N$}\rput(3.5,.7){\tiny $(m,n)$}
\end{pspicture}\ 
\ee
and are centered Laurent polynomials in $z$ with degree proportional to $N$. 

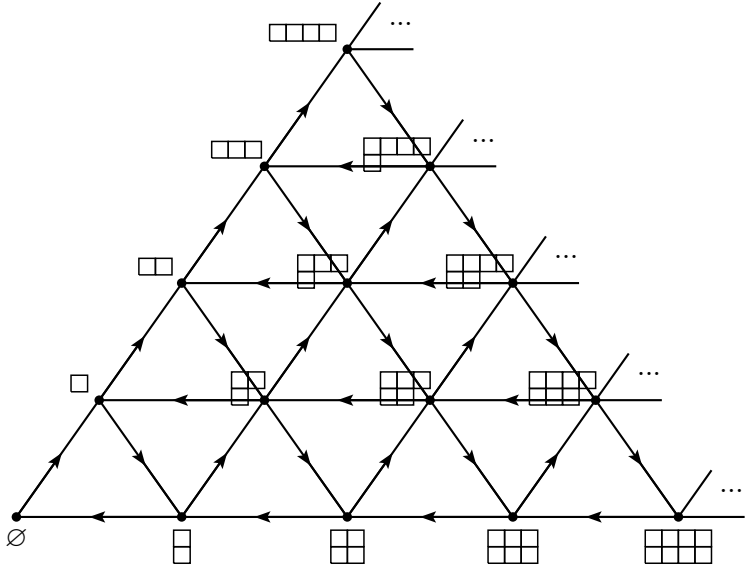
\begin{figure}
\begin{center}
$
\psset{unit=2.2}
\begin{pspicture}[shift=-0.9](0,-0.5)(4,3)
\multiput(0,0)(1,0){5}{\psdot(0,0)}
\multiput(0.5,0.707)(1,0){4}{\psdot(0,0)}
\multiput(1,1.414)(1,0){3}{\psdot(0,0)}
\multiput(1.5,2.121)(1,0){2}{\psdot(0,0)}
\multiput(2,2.828)(1,0){1}{\psdot(0,0)}
\multiput(0,0)(1,0){4}{\psline[arrowscale=1.4,arrowinset=0.2]{->}(0,0)(0.275, 0.3883)\psline{-}(0,0)(0.5,0.707)\psline[arrowscale=1.4,arrowinset=0.2]{>-}(0.725, 0.3883)(1,0)\psline{-}(0.5,0.707)(1,0)\psline[arrowscale=1.4,arrowinset=0.2]{<-}(0.44,0)(1,0)\psline{-}(0,0)(1,0)}
\multiput(0.5,0.707)(1,0){3}{\psline[arrowscale=1.4,arrowinset=0.2]{->}(0,0)(0.275, 0.3883)\psline{-}(0,0)(0.5,0.707)\psline[arrowscale=1.4,arrowinset=0.2]{>-}(0.725, 0.3883)(1,0)\psline{-}(0.5,0.707)(1,0)\psline[arrowscale=1.4,arrowinset=0.2]{<-}(0.44,0)(1,0)\psline{-}(0,0)(1,0)}
\multiput(1,1.414)(1,0){2}{\psline[arrowscale=1.4,arrowinset=0.2]{->}(0,0)(0.275, 0.3883)\psline{-}(0,0)(0.5,0.707)\psline[arrowscale=1.4,arrowinset=0.2]{>-}(0.725, 0.3883)(1,0)\psline{-}(0.5,0.707)(1,0)\psline[arrowscale=1.4,arrowinset=0.2]{<-}(0.44,0)(1,0)\psline{-}(0,0)(1,0)}
\multiput(1.5,2.121)(1,0){1}{\psline[arrowscale=1.4,arrowinset=0.2]{->}(0,0)(0.275, 0.3883)\psline{-}(0,0)(0.5,0.707)\psline[arrowscale=1.4,arrowinset=0.2]{>-}(0.725, 0.3883)(1,0)\psline{-}(0.5,0.707)(1,0)\psline[arrowscale=1.4,arrowinset=0.2]{<-}(0.44,0)(1,0)\psline{-}(0,0)(1,0)}
\multiput(4,0)(-0.5,0.707){5}{\psline{-}(0,0)(0.4,0)}
\multiput(4,0)(-0.5,0.707){5}{\psline{-}(0,0)(0.2, 0.2824)}
\rput(0,-0.13){$\varnothing$}
\rput(0.95,-0.23){\young{1}{1}}
\rput(1.9,-0.23){\young{2}{2}}
\rput(2.85,-0.23){\young{3}{3}}
\rput(3.8,-0.23){\young{4}{4}}
\rput(0.33,0.707){\young{1}{0}}
\rput(0.74,1.414){\young{2}{0}}
\rput(1.18,2.121){\young{3}{0}}
\rput(1.53,2.828){\young{4}{0}}
\rput(1.3,0.7271){\young21}
\rput(2.2,0.7271){\young32}
\rput(3.1,0.7271){\young43}
\rput(1.7,1.434){\young31}
\rput(2.6,1.434){\young42}
\rput(2.1,2.141){\young41}
\multiput(4.25,0.15)(-0.5,0.707){5}{...}
\end{pspicture}
$
\caption{The infinite dominant integral weight lattice of $s\ell(3)$ with nodes $(m,n)$ labelled by Young tableaux as in \eqref{Young}.}
\label{fig:su3weights}
\end{center}
\end{figure}

Normally, the recursive definitions of the projectors are used to derive the fusion hierarchy functional equations satisfied by the fused transfer tangles. This program has been successfully implemented for the $A_1^{(1)}$ loop models in \cite{MDPR2014} and the $A_2^{(1)}$ loop models in \cite{MDPR2018}. 
For the $A_2^{(2)}$ models, we have not succeeded to construct a full set of $(m,n)$ projectors. A construction of a partial set of projectors, namely those with labels $(m,0)$, $(m,1)$, $(0,n)$ and $(1,n)$, is given in \cref{sec:partial.WJ}. So instead, we proceed here by using a different method. 
We first write down fusion hierarchy relations directly and use them to define recursively the fused transfer tangles $\Tb^{m,n}(u)$ in terms of the two fundamentals. We separately prove that, as required, the resulting tangles are centered Laurent polynomials. It follows from our construction that $\Tb^{m,n}(u)$ is a centered Laurent polynomial in $z$ of degree $2N$ if $m = 0$ or $n=0$, but degree $3N$ otherwise.

We start by constructing $\Tb^{2,0}(u)$ and $\Tb^{0,2}(u)$ from the relations
\be
\label{eq:T20}
\Tb^{2,0}_0 = \frac{1}{f_{-1}f_0}\big(\Tb^{1,0}_0\Tb^{1,0}_2 - \sigma f_{-3}f_2 \Tb^{0,1}_0\big),
\qquad
\Tb^{0,2}_0 = \frac{1}{f_{0}f_1}\big(\Tb^{0,1}_0\Tb^{0,1}_2 - \sigma f_{-2}f_3 \Tb^{1,0}_2\big),
\ee
where we use the compact notations
\be
\Tb^{m,n}_k = \Tb^{m,n} (u+k \lambda), \qquad f_k = \prod_{j=1}^N s_k(u-\xi_j), \qquad \sigma = (-1)^{N}.
\ee
With these definitions, we have $\Tb^{0,2}(u) = \Tb^{2,0}(u+\lambda)$. It is not clear at this stage that $\Tb^{2,0}(u)$ is a Laurent polynomial, because the function $f_{-1}f_0$ in the denominator could produce poles. To show that this is not the case, we consider generic values for  crossing parameter $\lambda$ and inhomogeneities $\xi_j$. The polynomial properties we establish under this assumption then extend to the non-generic cases by continuity. For generic values, the functions $f_{-1}$ and $f_0$ are not equal and the zeros of the function $f_{-1}f_0$ are all simple. To investigate the presence of potential poles, let us evaluate $\Tb^{1,0}_0\Tb^{1,0}_2$ at $u = \xi_N$, a value for which $f_0=0$. We find
\be
\Tb^{1,0}_0\Tb^{1,0}_2\Big|_{u = \xi_N} = \ \
\psset{unit=1.1cm}
\begin{pspicture}[shift=-0.9](-0.3,0)(4.3,2.0)
\facegrid{(0,0)}{(4,2)}
\psarc[linewidth=0.025]{-}(0,0){0.16}{0}{90}
\psarc[linewidth=0.025]{-}(1,0){0.16}{0}{90}
\psarc[linewidth=0.025]{-}(3,0){0.16}{0}{90}
\psarc[linewidth=0.025]{-}(0,1){0.16}{0}{90}
\psarc[linewidth=0.025]{-}(1,1){0.16}{0}{90}
\psarc[linewidth=0.025]{-}(3,1){0.16}{0}{90}
\psline[linewidth=1.5pt,linecolor=blue,linestyle=dashed,dash=2pt 2pt]{-}(0,0.5)(-0.3,0.5)
\psline[linewidth=1.5pt,linecolor=blue,linestyle=dashed,dash=2pt 2pt]{-}(4,0.5)(4.3,0.5)
\psline[linewidth=1.5pt,linecolor=blue,linestyle=dashed,dash=2pt 2pt]{-}(0,1.5)(-0.3,1.5)
\psline[linewidth=1.5pt,linecolor=blue,linestyle=dashed,dash=2pt 2pt]{-}(4,1.5)(4.3,1.5)
\rput(2.5,0.5){$\ldots$}\rput(2.5,1.5){$\ldots$}
\rput(0.5,0.5){\footnotesize$\xi_{N,1}$}\rput(1.5,0.5){\footnotesize$\xi_{N,2}$}\rput(3.5,0.5){$0$}
\rput(0.5,1.5){\footnotesize$(\xi_{N,1})_2$}\rput(1.5,1.5){\footnotesize$(\xi_{N,2})_2$}\rput(3.5,1.5){$2\lambda$}
\end{pspicture}
\ \ = \ \ 
\begin{pspicture}[shift=-0.9](-0.3,0)(4.3,2.0)
\facegrid{(0,0)}{(4,2)}
\psarc[linewidth=0.025]{-}(0,0){0.16}{0}{90}
\psarc[linewidth=0.025]{-}(1,0){0.16}{0}{90}
\psarc[linewidth=0.025]{-}(0,1){0.16}{0}{90}
\psarc[linewidth=0.025]{-}(1,1){0.16}{0}{90}
\psline[linewidth=1.5pt,linecolor=blue,linestyle=dashed,dash=2pt 2pt]{-}(0,0.5)(-0.3,0.5)
\psline[linewidth=1.5pt,linecolor=blue,linestyle=dashed,dash=2pt 2pt]{-}(4,0.5)(4.3,0.5)
\psline[linewidth=1.5pt,linecolor=blue,linestyle=dashed,dash=2pt 2pt]{-}(0,1.5)(-0.3,1.5)
\psline[linewidth=1.5pt,linecolor=blue,linestyle=dashed,dash=2pt 2pt]{-}(4,1.5)(4.3,1.5)
\rput(2.5,0.5){$\ldots$}\rput(2.5,1.5){$\ldots$}
\rput(0.5,0.5){\footnotesize$\xi_{N,1}$}\rput(1.5,0.5){\footnotesize$\xi_{N,2}$}
\rput(0.5,1.5){\footnotesize$(\xi_{N,1})_2$}\rput(1.5,1.5){\footnotesize$(\xi_{N,2})_2$}
\rput(3,0){\loopid}
\rput(3,1){\psline{-}(1,0)(0,1)\rput(0.3,0.3){\specialcircle{0.05}}\rput(0.7,0.7){\specialcircle{0.05}}}
\end{pspicture}
\ee
where $\xi_{i,j}=\xi_i-\xi_j$ and $(\xi_{i,j})_k = \xi_i-\xi_j + k \lambda$. Using the push-through property for the triangle operator $N-1$ times, we find
\be
\Tb^{1,0}_0\Tb^{1,0}_2\Big|_{u = \xi_N} = \prod_{j=1}^{N-1}s_2(\xi_{N,j})s_3(-\xi_{N,j})\ \ 
\psset{unit=1.1cm}
\begin{pspicture}[shift=-0.4](-0.3,0)(4.3,1.0)
\facegrid{(0,0)}{(4,1)}
\psarc[linewidth=0.025]{-}(0,0){0.16}{0}{90}
\psarc[linewidth=0.025]{-}(1,0){0.16}{0}{90}
\psline[linewidth=1.5pt,linecolor=blue,linestyle=dashed,dash=2pt 2pt]{-}(0,0.5)(-0.3,0.5)
\psline[linewidth=1.5pt,linecolor=blue,linestyle=dashed,dash=2pt 2pt]{-}(4,0.5)(4.3,0.5)
\rput(2.5,0.5){$\ldots$}
\rput(0.5,0.5){\footnotesize$(\xi_{N,1})_1$}\rput(1.5,0.5){\footnotesize$(\xi_{N,2})_1$}
\rput(3,0){\psline{-}(0,0)(1,1)\rput(0.3,0.7){\specialcircle{0.05}}\rput(0.7,0.3){\specialcircle{0.05}}}
\end{pspicture} \ \ = \sigma f_{-3}f_2 \Tb^{0,1}_0\Big|_{u = \xi_N}.
\ee
It follows that the numerator in the leftmost equation in \eqref{eq:T20} vanishes at $u = \xi_N$. We conclude that $\Tb^{2,0}(u)$ is regular at $u = \xi_N$. The same argument holds for $u = \xi_j$ and $u = \lambda + \xi_j$ with $j = 1,2, \dots, N$, and shows that the numerator in the definition \eqref{eq:T20} of $\Tb^{2,0}(u)$ vanishes at all the simple zeros of $f_{-1}$ and $f_0$. The tangle $\Tb^{2,0}(u)$ is therefore a  Laurent polynomial. By subtracting the degrees from the numerator and denominator of \eqref{eq:T20}, we find that $\Tb^{2,0}(u)$ is a centered Laurent polynomial of degree $2N$ as previously claimed. The same holds for $\Tb^{0,2}(u)$.

Next, we construct $\Tb^{1,1}(u)$ from the relation 
\be
\label{eq:T11}
\Tb^{1,1}(u)= \frac1{f_0}\big(\Tb^{1,0}_0\Tb^{0,1}_2-f_{-3}f_{-2}f_2f_3\Ib\big).
\ee
To investigate its status as a Laurent polynomial, we employ the same strategy. We evaluate $\Tb^{1,0}_0\Tb^{0,1}$ at $u = \xi_N$ and find
\begin{alignat}{2}
\Tb^{1,0}_0\Tb^{0,1}_2\Big|_{u = \xi_N} &= \ \
\psset{unit=1.1cm}
\begin{pspicture}[shift=-0.9](-0.3,0)(4.3,2.0)
\facegrid{(0,0)}{(4,2)}
\psarc[linewidth=0.025]{-}(0,0){0.16}{0}{90}
\psarc[linewidth=0.025]{-}(1,0){0.16}{0}{90}
\psarc[linewidth=0.025]{-}(3,0){0.16}{0}{90}
\psarc[linewidth=0.025]{-}(0,1){0.16}{0}{90}
\psarc[linewidth=0.025]{-}(1,1){0.16}{0}{90}
\psarc[linewidth=0.025]{-}(3,1){0.16}{0}{90}
\psline[linewidth=1.5pt,linecolor=blue,linestyle=dashed,dash=2pt 2pt]{-}(0,0.5)(-0.3,0.5)
\psline[linewidth=1.5pt,linecolor=blue,linestyle=dashed,dash=2pt 2pt]{-}(4,0.5)(4.3,0.5)
\psline[linewidth=1.5pt,linecolor=blue,linestyle=dashed,dash=2pt 2pt]{-}(0,1.5)(-0.3,1.5)
\psline[linewidth=1.5pt,linecolor=blue,linestyle=dashed,dash=2pt 2pt]{-}(4,1.5)(4.3,1.5)
\rput(2.5,0.5){$\ldots$}\rput(2.5,1.5){$\ldots$}
\rput(0.5,0.5){\small$\xi_{N,1}$}\rput(1.5,0.5){\small$\xi_{N,2}$}\rput(3.5,0.5){$0$}
\rput(0.5,1.5){\small$(\xi_{N,1})_3$}\rput(1.5,1.5){\small$(\xi_{N,2})_3$}\rput(3.5,1.5){$3\lambda$}
\end{pspicture}
\ \ = \ \
\psset{unit=1.1cm}
\begin{pspicture}[shift=-0.9](-0.3,0)(4.3,2.0)
\facegrid{(0,0)}{(4,2)}
\psarc[linewidth=0.025]{-}(0,0){0.16}{0}{90}
\psarc[linewidth=0.025]{-}(1,0){0.16}{0}{90}
\psarc[linewidth=0.025]{-}(0,1){0.16}{0}{90}
\psarc[linewidth=0.025]{-}(1,1){0.16}{0}{90}
\psline[linewidth=1.5pt,linecolor=blue,linestyle=dashed,dash=2pt 2pt]{-}(0,0.5)(-0.3,0.5)
\psline[linewidth=1.5pt,linecolor=blue,linestyle=dashed,dash=2pt 2pt]{-}(4,0.5)(4.3,0.5)
\psline[linewidth=1.5pt,linecolor=blue,linestyle=dashed,dash=2pt 2pt]{-}(0,1.5)(-0.3,1.5)
\psline[linewidth=1.5pt,linecolor=blue,linestyle=dashed,dash=2pt 2pt]{-}(4,1.5)(4.3,1.5)
\rput(2.5,0.5){$\ldots$}\rput(2.5,1.5){$\ldots$}
\rput(0.5,0.5){\small$\xi_{N,1}$}\rput(1.5,0.5){\small$\xi_{N,2}$}
\rput(0.5,1.5){\small$(\xi_{N,1})_3$}\rput(1.5,1.5){\small$(\xi_{N,2})_3$}
\rput(3,0){\loopid}
\rput[90](3,1){\loopej}
\end{pspicture}
\nonumber\\[0.3cm]
& = \prod_{j=1}^{N-1}s_2(\xi_{N,j})s_2(-\xi_{N,j})s_3(\xi_{N,j})s_3(-\xi_{N,j})\ \ 
\begin{pspicture}[shift=-0.9](-0.3,0)(4.3,2.0)
\facegrid{(0,0)}{(4,2)}
\psline[linewidth=1.5pt,linecolor=blue,linestyle=dashed,dash=2pt 2pt]{-}(0,0.5)(-0.3,0.5)
\psline[linewidth=1.5pt,linecolor=blue,linestyle=dashed,dash=2pt 2pt]{-}(4,0.5)(4.3,0.5)
\psline[linewidth=1.5pt,linecolor=blue,linestyle=dashed,dash=2pt 2pt]{-}(0,1.5)(-0.3,1.5)
\psline[linewidth=1.5pt,linecolor=blue,linestyle=dashed,dash=2pt 2pt]{-}(4,1.5)(4.3,1.5)
\rput(2.5,0.5){$\ldots$}\rput(2.5,1.5){$\ldots$}
\rput(0,0){\loopid}\rput(0,1){\loopej}
\rput(1,0){\loopid}\rput(1,1){\loopej}
\rput(3,0){\loopid}\rput(3,1){\loopej}
\end{pspicture}
\\[0.3cm]
& = f_{-3}f_{-2}f_2f_3 \Ib\Big|_{u = \xi_N}\nonumber
\end{alignat}
where we applied the push-through property for the dashed arc $N-1$ times. The numerator of \eqref{eq:T11} thus vanishes at $u = \xi_N$ as does $f_0$, making $\Tb^{1,1}(u)$ regular at this point. The proof is repeated for $u = \xi_j$ and shows that $\Tb^{1,1}(u)$ is indeed a Laurent polynomial in $z$. Subtracting the degrees between the numerator and denominator, we find that $\Tb^{1,1}(u)$ is a centered Laurent polynomial of degree $3N$. 

The difference of polynomial degrees between $\Tb^{2,0}(u)$ and $\Tb^{1,1}(u)$ has its origin in the fact that $3\lambda$ is simultaneously a degeneration point of the face operator and the special value appearing in the crossing symmetry. In multiplying two fundamental transfer tangles, if the difference of spectral parameters is $3\lambda$, there is a single set of values $u = \xi_j$ where simplifications due to push-through properties occur. In contrast, if the difference is $2\lambda$, there are two such families: $u = \xi_j$ and $u = \lambda + \xi_j$.

Let us write down the full set of fusion hierarchy functional equations before proceeding to establish the required polynomial properties in the next subsection.
An inconvenient consequence of having transfer tangles with different polynomial degrees is that the fusion hierarchy equations must be formulated separately depending on the polynomial degree of the tangles that are involved. 
For the $A_2^{(2)}$ loop models there are four cases:\\
\mbox{}\qquad (i) near the corner of the $s\ell(3)$ weight lattice:
\begin{subequations}
\label{eq:FHs}
\begin{alignat}{2}
\Tb^{1,0}_0 \Tb^{1,0}_2 &= f_{-1}f_0\Tb^{2,0}_0 + \sigma f_{-3}f_2 \Tb^{0,1}_0 \label{eq:FH.corner1}, \\[0.1cm]
\Tb^{1,0}_0 \Tb^{0,1}_2 &= f_0 \Tb^{1,1}_0 + f_{-3} f_{-2} f_2 f_3  \Ib, \label{eq:FH.corner2}\\[0.1cm]
\Tb^{0,1}_0 \Tb^{0,1}_2 & = f_0 f_1 \Tb^{0,2}_{0} + \sigma f_{-2} f_3 \Tb^{1,0}_2,\label{eq:FH.corner3}
\end{alignat}
\qquad (ii) on one of the two boundary segments:
\begin{alignat}{2}
\Tb^{m,0}_0 \Tb^{1,0}_{2m} &= f_{2m-3} f_{2m-2} \Tb^{m+1,0}_0 + \sigma f_{2m} \Tb^{m-1,1}_0,\qquad &&m > 1,\label{eq:FH.bdy1}\\[0.1cm]
\Tb^{0,1}_0 \Tb^{0,n}_2 &= f_0 f_1 \Tb^{0,n+1}_0 + \sigma f_{-2} \Tb^{1,n-1}_2, \qquad &&n > 1,\label{eq:FH.bdy2}
\end{alignat}
\qquad (iii) on the segments adjacent to the boundary ones:
\begin{alignat}{2}
\Tb^{m,0}_0 \Tb^{0,1}_{2m} &= f_{2m-2} \Tb^{m,1}_0 + f_{2m}f_{2m+1}\Tb^{m-1,0}_0,\qquad &&m>1,\label{eq:FH.adj.bdy1}\\[0.1cm]
\Tb^{1,0}_0 \Tb^{0,n}_2 &= f_0 \Tb^{1,n}_0 + f_{-3}f_{-2} \Tb^{0,n-1}_4,\qquad &&n>1,\label{eq:FH.adj.bdy2}
\end{alignat}
\qquad (iv) everywhere else:
\be
\Tb^{m,0}_0 \Tb^{0,n}_{2m} = f_{2m-2} \Tb^{m,n}_0 + \Tb^{m-1,0}_0 \Tb^{0,n-1}_{2m+2}, \qquad m,n>1.\label{eq:FH.bulk}
\ee
\end{subequations}

These relations have the same underlying $s\ell(3)$ structure as for the $A_2^{(1)}$ models. They also define $\Tb^{m,n}(u)$ uniquely as polynomials in the two fundamental transfer tangles, evaluated at different shifts of the spectral parameter. As in (1.33) of \cite{ZPG1995}, these polynomials can be written in terms of formal determinants:
\begingroup
\allowdisplaybreaks
\begin{subequations}
\label{eq:FH.dets}
\begin{alignat}{2}
\bigg(\prod_{k=-1}^{2m-2}f_k\bigg)\,\Tb^{m,0}_0 &= 
\begin{pmatrix}
\Tb^{1,0}_{2m-2} & \Tb^{0,1}_{2m-4} & \sigma f_{2m-8} f_{2m-3}\hspace{-0.3cm} \\[0.15cm]
\sigma f_{2m-7} f_{2m-2} & \ddots& \ddots& \ddots\\[0.15cm]
0& \ddots & \ddots& \Tb^{0,1}_{2} & \sigma f_{-2}f_3 \\[0.15cm]
0&0& \sigma f_{-1}f_4\hspace{-0.3cm}& \Tb^{1,0}_{2} & \Tb^{0,1}_{0} \\[0.15cm]
0&0&0& \sigma f_{-3}f_2& \Tb^{1,0}_{0}\end{pmatrix},
\label{eq:Tm0}
\\[0.3cm]
\bigg(\prod_{k=0}^{2n-3}f_k\bigg)\,\Tb^{0,n}_0 &= 
\begin{pmatrix}
\Tb^{0,1}_{2n-2} & \hspace{-0.4cm}\sigma f_{2n-6} f_{2n-1}\hspace{-0.6cm} &0&0&0 \\[0.15cm]
\Tb^{1,0}_{2n-2} & \ddots & \ddots &0&0 \\[0.15cm]
\sigma f_{2n-7} f_{2n-2} & \ddots & \ddots & \hspace{-0.4cm} \sigma f_{0}  f_{5}&0 \\[0.15cm]
0&\ddots& \Tb^{1,0}_{4} & \Tb^{0,1}_{2} & \sigma f_{-2} f_{3} \\[0.15cm]
0&0& \sigma f_{-1}f_{4} & \Tb^{1,0}_{2} & \Tb^{0,1}_{0} \end{pmatrix},
\label{eq:T0n}
\\[0.3cm]
\bigg(\prod_{\substack{k=-1\\[0.1cm]k \neq 2m-3,2m-1}}^{2m+2n-3}\hspace{-0.2cm}f_k\hspace{0.1cm}\bigg)\,\Tb^{m,n}_0 &= \begin{pmatrix} \ \ \ 
\begin{pspicture}(0,0)(0,0)
\pspolygon(-0.3,0.2)(2.7,0.2)(2.7,-1)(-0.3,-1)
\rput(1.2,-0.4){${n\times n}$}
\end{pspicture} &  &  & 0 & 0 & 0 \\ 
 &  &  & 0 & 0 & 0 \\
 &  &  & \sigma f_{2m-4}f_{2m+1} & 0 & 0 \\
0 & 0 & \sigma f_{2m-5}f_{2m} & 
 &  &  \\
0 & 0 & 0 &  &  &  \\
0 & 0 & 0 &  &  & 
\begin{pspicture}(0,0)(0,0)
\rput(-2.4,1){\pspolygon(-0.3,0.2)(2.7,0.2)(2.7,-1)(-0.3,-1)\rput(1.2,-0.4){${m\times m}$}}
\end{pspicture} \ \ \ \\
\end{pmatrix},
\label{eq:Tmn}
\end{alignat}
\end{subequations}
\endgroup
where the $m\times m$ box should be replaced by the matrix on the right side of \eqref{eq:Tm0}, and the $n\times n$ box should be replaced by the matrix on the right side of \eqref{eq:T0n} with $u$ shifted by $2m\lambda$. 

\subsection{Polynomial properties}

In this subsection, we show that each $\Tb^{m,n}(u)$ is a  Laurent polynomial in $z$. The proof is inductive on $m+n$, and has  been obtained previously for $m+n = 1,2$. In showing that $\Tb^{m,n}(u)$ with $m+n >2$ is a Laurent polynomial, we assume that this property has been shown for all pairs $(m',n')$ with $m'+n'<m+n$. 

Let us start with $\Tb^{m,0}(u)$ and $\Tb^{0,n}(u)$. It is clear from the determinant expressions that $\Tb^{m,0}(u) = \Tb^{0,m}(u-\lambda)$ for all $m$. With $m$ changed for $m-1$ in \eqref{eq:FH.bdy1} and \eqref{eq:FH.bdy2}, these fusion hierarchy relations give us two expressions for $\Tb^{m,0}(u)$. The first one expresses $\Tb^{m,0}(u)$ as $(f_{2m-5}f_{2m-4})^{-1}$ times a combination (sums and products) of three transfer tangles of the form $\Tb^{m',n'}(u)$ with $m'+n' < m+n$. This combination is a Laurent polynomial by the inductive assumption. The second relation writes $\Tb^{0,m}(u)$ as $(f_{-1}f_0)^{-1}$ times a different Laurent polynomial. The poles of $\Tb^{m,n}(u)$ are therefore contained in the intersection of the zeros of the functions $f_{2m-5}f_{2m-4}$ and $f_{-1}f_0$. For generic values of $\lambda$ and $\xi_j$, this is an empty set, implying that $\Tb^{m,0}(u)$ is indeed a Laurent polynomial. By continuity, this property extends to non-generic values of the parameters.

A similar idea is used for $\Tb^{m,n}(u)$ with $m,n\ge1$ and $m+n>2$. In this case, \eqref{eq:FH.adj.bdy1}, \eqref{eq:FH.adj.bdy2} and \eqref{eq:FH.bulk} tell us that the potential poles are at values of $u$ where $f_{2m-2} = 0$. Using the determinant expression \eqref{eq:Tmn}, we can derive two more recursive formulas for $\Tb^{m,n}(u)$. For instance, we have
\begin{subequations}
\label{eq:extraFHR}
\begin{alignat}{2}
f_{-1}f_0f_{1}\Tb^{m,n}_0 = f_1 \Tb^{1,0}_0\Tb^{m-1,n}_2 - \sigma f_{-3}\Tb^{0,1}_0\Tb^{m-2,n}_4 + \sigma f_{-3}f_{-2}f_{-1}\Tb^{m-3,n}_6, \qquad m>3,
\label{eq:extraFHR1}
\end{alignat}
and
\begin{alignat}{2}
f_{2m+2n-3}f_{2m+2n-4}f_{2m+2n-5}\Tb^{m,n}_0 &= f_{2m+2n-5} \Tb^{m,n-1}_0\Tb^{0,1}_{2m+2n-2} - \sigma f_{2m+2n-1} \Tb^{m,n-2}_0\Tb^{1,0}_{2m+2n-2}\nonumber\\[0.15cm] &+ \sigma f_{2m+2n-1}f_{2m+2n-2}f_{2m+2n-3}\Tb^{m,n-3}_0,\qquad n>3.
\label{eq:extraFHR2}
\end{alignat}
\end{subequations}
For $m>3$, the poles of $\Tb^{m,n}(u)$ are at values of $u$ where $f_{-1}f_0f_{1} = 0$. For generic values of the parameters $\lambda$ and $\xi_j$, there are no values of $u$ satisfying this identity as well as $f_{2m-2}=0$. The inductive hypothesis thus leads us to conclude that $\Tb^{m,n}(u)$ has no poles for $m>3$. The same argument holds for $n>3$, using the zeros of the $f$ functions on the left side of \eqref{eq:extraFHR2}.

The cases $m=3$ and $n=3$ are treated separately. Identities similar to \eqref{eq:extraFHR1} and \eqref{eq:extraFHR2} hold for $m=3$ and $n=3$, but with some differences in the $f$ functions due to the different polynomial degrees of some of the fused transfer tangles involved. These are given in \eqref{eq:T3n} and \eqref{eq:Tm3}. In each case, the right side is a Laurent polynomial because of the induction hypothesis, whereas the $f$ functions on the left side have no common zeros with $f_{m-2}$, and the proof follows.

This ends the proof of our claim that $\Tb^{m,n}(u)$ is a Laurent polynomial. The degrees of these centered Laurent polynomials can be deduced from the fusion hierarchy equations, again using an inductive argument. As announced previously, these degrees are $2N$ for the transfer tangles whose label $(m,n)$ lives on the two boundary edges of the $s\ell(3)$ weight lattice, and $3N$ in all other cases. The fused transfer tangles defined through the fusion hierarchy equations satisfy
\be
[\Tb^{m,n}(u),\Tb^{m',n'}(v)] = 0 
\ee
and 
\be
\Tb^{m,0}(u+\pi) = \Tb^{m,0}(u), \qquad \Tb^{0,n}(u+\pi) = \Tb^{0,n}(u), \qquad \Tb^{m,n}(u+\pi) = \sigma\, \Tb^{m,n}(u).
\ee
This immediately follows from the same relations satisfied by $\Tb(u)$.

\subsection{Braid limits}

The two fundamental braid operators are obtained as the $u \to \pm \ir \infty$ limits of the face operators:
\be
\begin{pspicture}[shift=-.4](1,1)
\facegrid{(0,0)}{(1,1)}
\rput(.5,.5){$\pm \infty$}
\psarc[linewidth=0.025]{-}(0,0){0.16}{0}{90}
\end{pspicture}
\ = 
\lim_{u \to \pm \ir \infty} \frac{\eE^{\mp \ir (\pi - 2 \lambda)}}{\rho_8(u)}\ \ 
\begin{pspicture}[shift=-.4](1,1)
\facegrid{(0,0)}{(1,1)}
\rput(.5,.5){$u$}
\psarc[linewidth=0.025]{-}(0,0){0.16}{0}{90}
\end{pspicture}
 \  =  \ 
\begin{pspicture}[shift=-.40](0,0)(1,1)
\facegrid{(0,0)}{(1,1)}
\rput[bl](0,0){\loopa}
\end{pspicture}
\  +  \ 
\begin{pspicture}[shift=-.40](0,0)(1,1)
\facegrid{(0,0)}{(1,1)}
\rput[bl](0,0){\loopg}
\end{pspicture}
\  +  \
\begin{pspicture}[shift=-.40](0,0)(1,1)
\facegrid{(0,0)}{(1,1)}
\rput[bl](0,0){\loopf}
\end{pspicture} 
\  - \eE^{\pm 2 \ir \lambda} \  
\begin{pspicture}[shift=-.40](0,0)(1,1)
\facegrid{(0,0)}{(1,1)}
\rput[bl](0,0){\looph}
\end{pspicture}
\ - \eE^{\mp 2 \ir \lambda} \
\begin{pspicture}[shift=-.40](0,0)(1,1)
\facegrid{(0,0)}{(1,1)}
\rput[bl](0,0){\loopi}
\end{pspicture}\ \ .
\ee
The braid operators satisfy two push-through properties, one for arcs and one for vacancies
\be
\label{eq:braid.ids}
\begin{pspicture}[shift=-.4](0,0)(2,1.5)
\facegrid{(0,0)}{(2,1)}
\rput(.5,.5){$\pm \infty$}\rput(1.5,.5){$\pm \infty$}
\psarc[linewidth=0.025]{-}(0,0){0.16}{0}{90}\psarc[linewidth=0.025]{-}(1,0){0.16}{0}{90}
\psarc[linewidth=1.5pt,linecolor=blue]{-}(1,1){0.5}{0}{180}
\end{pspicture}
\ \ = \ \ 
\begin{pspicture}[shift=-.4](0,0)(2,1.5)
\facegrid{(0,0)}{(2,1)}
\psarc[linewidth=1.5pt,linecolor=blue,linestyle=dashed, dash=2pt 2pt]{-}(1,1){0.5}{0}{180}
\psarc[linewidth=1.5pt,linecolor=blue,linestyle=dashed, dash=2pt 2pt]{-}(0,1){0.5}{-90}{0}
\psarc[linewidth=1.5pt,linecolor=blue,linestyle=dashed, dash=2pt 2pt]{-}(2,1){0.5}{180}{-90}
\psarc[linewidth=1.5pt,linecolor=blue]{-}(1,0){0.5}{0}{180}
\end{pspicture}\ \ ,
\qquad\qquad
\begin{pspicture}[shift=-.4](0,0)(1,1)
\facegrid{(0,0)}{(1,1)}
\rput(.5,.5){$\pm \infty$}
\psarc[linewidth=0.025]{-}(0,0){0.16}{0}{90}
\pscircle[fillstyle=solid,fillcolor=black](0.5,1){0.06}
\end{pspicture}
\ \ = \ \ 
\begin{pspicture}[shift=-.4](0,0)(1,1)
\facegrid{(0,0)}{(1,1)}
\pscircle[fillstyle=solid,fillcolor=black](0.5,1){0.06}
\pscircle[fillstyle=solid,fillcolor=black](0.5,0){0.06}
\psline[linewidth=1.5pt,linecolor=blue,linestyle=dashed, dash=2pt 2pt]{-}(0,0.5)(1,0.5)
\end{pspicture}\ \ .
\ee
The fundamental braid transfer tangle is defined as
\be
\psset{unit=1.1cm}
\Tb_{\!\pm\infty} = 
\lim_{u \to \pm \ir \infty} \frac{\eE^{\mp\ir(\pi -2\lambda)N}}{f_{-2}f_{-3}}\, \Tb(u)
= \ \
\begin{pspicture}[shift=-0.4](-0.3,0)(4.3,1.0)
\facegrid{(0,0)}{(4,1)}
\psarc[linewidth=0.025]{-}(0,0){0.16}{0}{90}
\psarc[linewidth=0.025]{-}(1,0){0.16}{0}{90}
\psarc[linewidth=0.025]{-}(3,0){0.16}{0}{90}
\psline[linewidth=1.5pt,linecolor=blue,linestyle=dashed,dash=2pt 2pt]{-}(0,0.5)(-0.3,0.5)
\psline[linewidth=1.5pt,linecolor=blue,linestyle=dashed,dash=2pt 2pt]{-}(4,0.5)(4.3,0.5)
\rput(2.5,0.5){$\ldots$}
\rput(0.5,.5){$\pm \infty$}
\rput(1.5,.5){$\pm \infty$}
\rput(3.5,.5){$\pm \infty$}
\end{pspicture}
\ \ .
\ee
It follows from the push-through properties \eqref{eq:braid.ids} that $\Tb_{\infty}$ and $\Tb_{-\infty}$ are in the center of $\pdTL_N(\alpha,\beta)$. 

In the standard modules, the braid transfer tangles are proportional to the identity matrix. Indeed, in computing the action of $\Tb_{\!\pm \infty}$ on a given link state, all the arcs and vacancies push through. The result is then independent of the organisation of the arcs and vacancies, and depends only on the number of defects. The unique eigenvalue is given by
\be
\label{eq:stan.braid}
\Tb_{\!\pm \infty}\Big|_{\stan_{N,d}} = 
\left\{\begin{array}{cl}
\alpha +1, & d=0,\\[0.2cm]
\omega\, \eE^{\mp \ir(\pi - 2 \lambda)d} + 1 + \omega^{-1} \eE^{\pm \ir(\pi - 2 \lambda)d},& d\ge 1.
\end{array}\right.
\ee
In principle, $\alpha$ and $\omega$ are independent parameters. However, working sector-by-sector, only one parameter enters each sector. So it is convenient to parameterise $\alpha$ as
\be
\label{eq:alpha.omega}
\alpha = \omega + \omega^{-1},
\ee
as with this convention, the second line of the right side of \eqref{eq:stan.braid} holds for $d=0$ as well.

The fused braid transfer tangles are defined as the $u\to \pm \ir \infty$ limit of the fused transfer tangles:
\be
\label{eq:braid.T.def}
\Tb^{m,0}_{\!\pm \infty} = \Tb^{0,m}_{\!\pm \infty} = \lim_{u \to \pm \ir \infty} \frac{\eE^{\mp \ir (\pi-2\lambda)Nm}}{f_{-3}f_{-2}}\, \Tb^{m,0}(u), \qquad
\Tb^{m,n}_{\!\pm \infty} = \lim_{u \to \pm \ir \infty} \frac{\eE^{\mp \ir (\pi-2\lambda)N(m+n)}}{f_{-3}f_{-2}f_{2m-1}}\, \Tb^{m,n}(u).
\ee
The fusion hierarchy relations that they satisfy are obtained from the $u\to\pm \ir \infty$ limit of \eqref{eq:FHs}. There are no issues here with polynomial degrees that vary according to the fusion label, so these relations are expressed in a unified manner:
\begin{subequations}
\begin{alignat}{2}
\Tb^{m,0}_{\!\pm \infty}\Tb^{1,0}_{\!\pm \infty} &=\Tb^{m-1,1}_{\!\pm \infty} + \Tb^{m+1,0}_{\!\pm \infty},\\[0.1cm]
\Tb^{0,1}_{\!\pm \infty}\Tb^{0,n}_{\!\pm \infty} &= \Tb^{1,n-1}_{\!\pm \infty} + \Tb^{0,n+1}_{\!\pm \infty},\\[0.1cm]
\Tb^{m,0}_{\!\pm \infty}\Tb^{0,n}_{\!\pm \infty} &= \Tb^{m,n}_{\!\pm \infty} + \Tb^{m-1,0}_{\!\pm \infty}\Tb^{0,n-1}_{\!\pm \infty}.\label{eq:FHbraid}
\end{alignat}
\end{subequations}
The fused braid transfer tangles are thus polynomials in $\Tb_{\!\pm\infty}$, which can be expressed in terms of determinants similarly to \eqref{eq:FH.dets}. On the standard modules, $\Tb^{m,n}_{\!\pm \infty}$ is thus represented by a diagonal matrix with a single eigenvalue. In particular, we have
\be
\label{eq:fused.braid.eig}
\Tb^{m,0}_{\!\pm\infty}\Big|_{\stan_{N,d}} = \Tb^{0,m}_{\!\pm\infty}\Big|_{\stan_{N,d}} = U_m(\omega\, \eE^{\mp \ir (\pi - 2 \lambda)d}, 1)
\ee
where the $s\ell(3)$ Chebyshev polynomials~\cite{DFZ90} are
\be
U_m(y_1,y_2) = \frac{y_1^{m+2}(y_2-y_3)+y_2^{m+2}(y_3-y_1)+y_3^{m+2}(y_1-y_2)}{(y_1-y_2)(y_1-y_3)(y_2-y_3)} , \qquad y_1 y_2 y_3 = 1.
\ee
The values for $\Tb^{m,n}_{\!\pm \infty}\big|_{\stan_{N,d}}$ with $m,n\neq0$ are obtained from  \eqref{eq:FHbraid} and  \eqref{eq:fused.braid.eig}.

\subsection[$T$- and $Y$-systems]{$\boldsymbol T$- and $\boldsymbol Y$-systems}

The $T$-system is a certain set of bilinear functional equations satisfied by the fused transfer tangles. For the $A_2^{(2)}$ loop models, the $T$-system only involves fused transfer tangles where the second fusion label is zero and takes the form
\be
\label{eq:Trelations}
\Tb^{m,0}_0 \Tb^{m,0}_2 = \sigma^m f_{-3}f_{2m}\Tb^{m,0}_1 + \Tb^{m+1,0}_0 \Tb^{m-1,0}_2, \qquad m \ge 0.
\ee
Here we use the conventions
\be
\Tb^{0,0}_k = f_{k-3}f_{k-2}\Ib, \qquad \Tb^{-1,0}_k = 0.\ee 
For $m=0$, \eqref{eq:Trelations} holds trivially. For $m=1$, it is equivalent to the fusion hierarchy relation \eqref{eq:FH.corner1}.
For $m > 1$, the $T$-system equations are a special case of a two-parameter family of bilinear functional equations proved in \cref{sec:T.system.proof}.

The $Y$-system is a set of functional equations in the quantities $\tb^m(u)$ defined as the ratio of the two terms on the right side of the $T$-system equation:
\be
\tb^m(u) = \tb^m_0 = \frac{\Tb^{m+1,0}_0 \Tb^{m-1,0}_2}{\sigma^m f_{-3}f_{2m}\Tb^{m,0}_1}, \qquad m\ge 0.
\ee
The lower index again serves to indicate a shift in the spectral parameter: $\tb^m_k = \tb^m(u\!+\!k\lambda)$. As defined here, these objects are not tangles because of the inverses in the denominators. Similarly, the $Y$-system equations given below are not actually equalities between tangles. These equations are expressed in the manner which is most useful for working with transfer matrix eigenvalues in particular representations of $\pdTL_N(\alpha,\beta)$. Any such given equation can be turned into an equality between tangles by substituting in the definition for $\tb^m(u)$ and multiplying the left and right sides by any transfer tangles that appear in the denominators.

The $T$-system equations thus translate to
\be
\Tb^{m,0}_0 \Tb^{m,0}_2 =  \sigma^m f_{-3}f_{2m}\Tb^{m,0}_1(\Ib + \tb^m_0), \qquad m \ge 0.
\ee
Explicitly, the $Y$-system of the generic $A_2^{(2)}$ loop models takes the form
\be
\label{eq:Ysys}
\tb^m_0\tb^m_2 = \frac{(\Ib+\tb^{m-1}_2)(\Ib+\tb^{m+1}_0)}{\Ib+(\tb^{m}_1)^{-1}}, \qquad m \ge 1.
\ee
The proof of these equations readily follows from the $T$-system equations:
\begin{alignat}{2}
\tb^m_0 \tb^m_2 &= \frac{(\Tb^{m-1,0}_2\Tb^{m-1,0}_4)(\Tb^{m+1,0}_0\Tb^{m+1,0}_2)}{f_{-3}f_{-1}f_{2m}f_{2m+2}\Tb^{m,0}_1\Tb^{m,0}_3} \nonumber\\&= 
\frac{(\sigma^{m-1}f_{-1}f_{2m} \Tb^{m-1,0}_3+\Tb^{m,0}_2\Tb^{m-2,0}_4)(\sigma^{m+1}f_{-3}f_{2m+2}\Tb^{m+1,0}_1+\Tb^{m+2,0}_0\Tb^{m,0}_2)}{f_{-3}f_{-1}f_{2m}f_{2m+2}(\sigma^mf_{-2}f_{2m+1}\Tb^{m,0}_2+\Tb^{m+1,0}_1\Tb^{m-1,0}_3)} \nonumber\\&
= \frac{\displaystyle\bigg(\Ib+\frac{\Tb^{m,0}_2\Tb^{m-2,0}_4}{\sigma^{m-1}f_{-1}f_{2m}\Tb^{m-1,0}_3}\bigg)\bigg(\Ib+\frac{\Tb^{m+2,0}_0\Tb^{m,0}_2}{\sigma^{m+1}f_{-3}f_{2m+2}\Tb^{m+1,0}_1}\bigg)}{\displaystyle\bigg(\Ib+\frac{\sigma^mf_{-2}f_{2m+1}\Tb^{m,0}_2}{\Tb^{m+1,0}_1\Tb^{m-1,0}_3}\bigg)}
\nonumber
= \frac{(\Ib+\tb^{m-1}_2)(\Ib+\tb^{m+1}_0)}{\Ib+(\tb^{m}_1)^{-1}}.
\end{alignat}
The $T$- and $Y$-systems \eqref{eq:Trelations} and \eqref{eq:Ysys} are valid for all values of $\lambda \in \mathbb R$ and are  infinite systems of equations.

\section{Closure relations at roots of unity}\label{sec:closures}

In this section, we specialise the spectral parameter $\lambda$ to a root of unity value which, for convenience, we now parameterise in terms of two coprime integers $a$ and $b$ as
\be
\label{eq:lambda.ab}
\lambda = \frac{(b-a)\pi}{2b}, \qquad \textrm{gcd}(a,b)=1.
\ee
This parameterisation is such that 
\be
\label{eq:syms}
f_{k+2b} = \sigma^{b-a} f_k, \qquad \Tb^{m,0}_{k+2b} = \Tb^{m,0}_{k}, \qquad \Tb^{0,n}_{k+2b} = \Tb^{0,n}_{k},\qquad \Tb^{m,n}_{k+2b} = \sigma^{b-a}\Tb^{m,n}_{k}, \qquad m,n\ge 1.
\ee 
Comparing with the parameterisation \eqref{eq:lambda.pp'} for $\lambda$, we see that $a$ and $b$ are simply related to the coprime integers $p$ and $p'$ of the dilute logarithmic minimal model $\mathcal {DLM}(p,p')$ by
\be
\begin{array}{lll}
p\textrm{ odd}: \quad& a = p, & b = 2p',\\[0.1cm]
p\textrm{ even}: \quad & a = p/2, & b = p'.\\
\end{array}
\ee
In this section, we use strictly the weights with the normalization $\sin2\lambda\sin3\lambda$ as in \eqref{eq:weights}. We consider $b\ge 2$ and avoid the special cases $\lambda = \frac\pi 3,\frac \pi 2, \frac{2\pi}3$.

\subsection{Closure of the fusion hierarchy}\label{sec:FHclosure}

The fusion hierarchy equations \eqref{eq:FHs} allow one to construct recursively the fused transfer tangles $\Tb^{m,n}(u)$, for all values of $\lambda$. At roots of unity, there exist extra relations, the {\it closure relations}, which take the form of linear relations between the various fused transfer tangles. Closure relations have been previously derived for the $A_1^{(1)}$ loop models~\cite{MDPR2014} and $A_2^{(1)}$ loop 
models~\cite{MDPR2018}. The $A_2^{(2)}$ case is similar to the $A_2^{(1)}$ case in its underlying $s\ell(3)$ structure, but with the added complication that the fused transfer tangles do not all have the same polynomial degrees in $z$. As a result, the closure relations take slightly different forms depending on the polynomial degrees of the tangles involved.

With $\lambda$ as in \eqref{eq:lambda.ab}, the first closure relations pertain to $\Tb^{b,0}(u)$ and $\Tb^{0,b}(u)$. For $b\ge 4$, these relations take the form
\begin{subequations}
\label{eq:closure}
\be
\hspace{0.8cm}b \ge 4:\qquad\left\{
\begin{array}{l}
f_{-1}\Tb^{b,0}_0 = \sigma^{b-a-1}\, \Tb^{b-2,1}_2 - \sigma f_{-3} \Tb^{b-3,0}_4 + f_{-3}f_{-2}f_{-1}\Jb,\label{eq:closure.a}\\[0.1cm]
f_{-3}\Tb^{0,b}_0 = \sigma\, \Tb^{1,b-2}_0 -\sigma f_{-1} \Tb^{0,b-3}_2+ f_{-3}f_{-2}f_{-1}\Jb.
\end{array}\right.
\ee
The analagous closure relations for $b=2$ and $b=3$ are
\begin{alignat}{2}
&b=2:\qquad
\left\{
\begin{array}{l}
\Tb^{2,0}_0 = \Tb^{0,1}_2 + f_{-3}f_{-2}\Jb,\\[0.1cm]
\Tb^{0,2}_0 = \Tb^{1,0}_0 + f_{-2}f_{-1}\Jb,
\end{array}
\right.
\\[0.2cm]
&b=3:\qquad
\left\{
\begin{array}{l}
f_{-1}\Tb^{3,0}_0 = \sigma^a \Tb^{1,1}_2 - \sigma f_{-5}f_{-4}f_{-3}\Ib + f_{-3}f_{-2}f_{-1}\Jb,\\[0.1cm]
f_{-3}\Tb^{0,3}_0 = \sigma\, \Tb^{1,1}_0 -\sigma f_{-1} f_0 f_{1} \Ib+ f_{-3}f_{-2}f_{-1}\Jb.
\end{array}
\right.
\end{alignat}
\end{subequations}
In these relations, the tangle $\Jb$ is an element of $\pdTL_N(\alpha,\beta)$ which is independent of $u$. It is expressed in terms of the braid fused transfer tangles as
\be
\label{eq:J}
\sigma^a\Jb = \Tb^{b,0}_{\!\pm\infty} - \Tb^{b-2,1}_{\!\pm\infty} + \Tb^{b-3,0}_{\!\pm\infty},
\ee 
and is therefore in the center of $\pdTL_N(\alpha,\beta)$. It is diagonal on the standard modules, with the unique eigenvalue
\be
\label{eq:J.eigs}
\sigma^a\Jb\Big|_{\stan_{N,d}} = (-1)^{ad}(\omega^{b}+\omega^{-b})+1.
\ee
This results holds for $d=0$ with the convention \eqref{eq:alpha.omega}. The proof of the closure relations is given in \cref{sec:FHclosure.proof}.

\begin{figure}
\begin{center}
$
\psset{unit=1}
\begin{pspicture}[shift=-0.9](0,-1.0)(12,6)
\pspolygon[fillstyle=solid,fillcolor=lightlightblue](0,0)(3.0,4.242)(9.0,4.242)(6.0,0)
\psline[linecolor=red,linewidth=0.1cm]{-}(2.0,4.242)(11,4.242)
\psline[linecolor=red,linewidth=0.1cm]{-}(2.0,4.242)(5.5,-0.707)
\psline[linecolor=red,linewidth=0.1cm]{-}(-1.5,-0.707)(7.5,-0.707)
\psline[linecolor=red,linewidth=0.1cm]{-}(-1.5,-0.707)(3,5.656)
\psline[linecolor=red,linewidth=0.1cm]{-}(5.5,-0.707)(10,5.656)
\multiput(-1.5,-0.707)(0.5,0.707){10}{\multiput(0,0)(1,0){10}{\psdot(0,0)}}
\psset{linewidth=0.1pt}
\multiput(0,0)(0.5,0.707){8}{\multiput(0,0)(1,0){8}{\psline{-}(0,0)(0.5,0.707)\psline{-}(0.5,0.707)(1,0)\psline{-}(0,0)(1,0)\psline{-}(1,0)(1.5,0.707)\psline{-}(0.5,0.707)(1.5,0.707)}}
\pscircle[linewidth=1.5pt,linecolor=black](3.5,4.949){.2}
\pscircle[linewidth=1.5pt,linecolor=black](3.5,3.535){.2}
\pscircle[linewidth=1.5pt,linecolor=black](2.0,2.828){.2}
\pscircle[linewidth=1.5pt,linecolor=black](0,0){.2}
\rput(0,0.25){\scriptsize$(0,0)$}
\rput(1,0.25){\scriptsize$(0,1)$}
\rput(2,0.25){\scriptsize$(0,2)$}
\rput(0.5,0.957){\scriptsize$(1,0)$}
\rput(1.5,0.957){\scriptsize$(1,1)$}
\rput(1,1.664){\scriptsize$(2,0)$}
\rput(2.0,3.288){\scriptsize$(b\!-\!3,0)$}
\rput(3.5,5.35){\scriptsize$(b\!,0)$}
\rput(3.5,3.94){\scriptsize$(b\!-\!2,1)$}
\end{pspicture}
$
\caption{
The restricted set of $s\ell(3)$ weights for $\lambda = \frac{(b-a)\pi}{2b}$ consists of the nodes in the shaded region. The terms entering the closure relation \eqref{eq:closure.a} are encircled and are related to one another via reflections along the red critical lines and translations into the elementary domain. The term at the origin is the one corresponding to $\Jb$.
}
\label{fig:folding}
\end{center}
\end{figure}
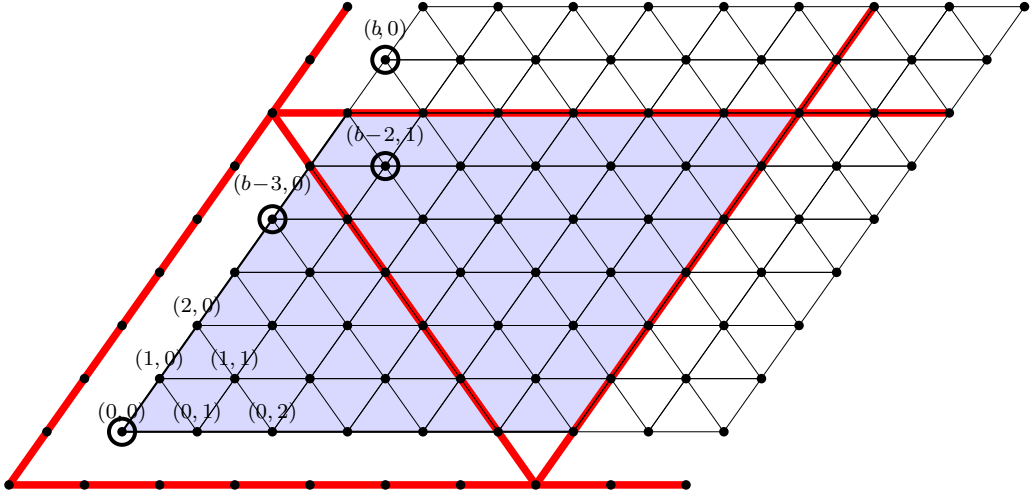

In fact, there exist many more closure relations satisfied the fused transfer tangles. The full picture is the following. For a given choice of $a$ and $b$ with $\lambda$ fixed as in \eqref{eq:lambda.ab}, let us define the {\it restricted set} of $s\ell(3)$ weights, namely those labeled by pairs $(m,n)$ with $0\le m,n\le b-1$. All the other transfer tangles $\Tb^{m,n}(u)$, with $m$ or $n$ larger than $b-1$, can be expressed as a linear combination of tangles in the restricted set. In the $s\ell(3)$ weight lattice, the nodes corresponding to the restricted set form a rhombus. The tangles $\Tb^{b,0}(u)$ and $\Tb^{0,b}(u)$, appearing in the closure relations \eqref{eq:closure}, are then the tangles that lie just outside of this rhombus. There is an elegant geometrical interpretation of the different terms appearing in this relation, which is described in \cref{fig:folding}. The situation is therefore similar to what was found for the $A_2^{(1)}$ model. One difference is that the $A_2^{(2)}$ model exhibits an extra ${\Bbb Z}_2$ folding. The nodes on the bottom boundary edge of the rhombus are identified to those of the diagonal edge via the relation $\Tb^{0,m}_0 = \Tb^{m,0}_1$.

For later use, it is useful to know the form of the extra closure relations for all the nodes that lie just outside of the rhombus. These take slightly different forms depending on the polynomial degrees of the tangles involved.
For $1\le k \le b-4$, the extra closure relations are of the form
\begin{subequations}
\be
1\le k \le b-4:\qquad
\left\{\begin{array}{l}
\Tb^{b,k}_0 = \sigma\, \Tb^{b-2,k+1}_2 - \sigma^{b-a-1} f_{-3} \Tb^{b-3-k,0}_{2k+4} +\sigma^{b-a} \Jb f_{-3} \Tb^{0,k}_0,\\[0.15cm] 
\Tb^{k,b}_0 = \sigma\, \Tb^{k+1,b-2}_0 - \sigma f_{2k-1} \Tb^{0,b-3-k}_{2k+2} + \Jb f_{2k-1} \Tb^{k,0}_0.
\end{array}\right.
\ee
Likewise, for $k=b-3,b-2,b-1$, we find
\begin{alignat}{2}
\label{eq:more.clo3}
k=b-3:\qquad&
\left\{\begin{array}{l}
\Tb^{b,b-3}_0 = \sigma\, \Tb^{b-2,b-2}_2 - \sigma^{b-a-1} f_{-5}f_{-4}f_{-3}  \Ib +\sigma^{b-a} \Jb f_{-3} \Tb^{0,b-3}_0,\\[0.15cm]
\Tb^{b-3,b}_0 = \sigma\, \Tb^{b-2,b-2}_0 - \sigma f_{-7} f_{-6} f_{-5} \Ib + \Jb f_{-7} \Tb^{b-3,0}_0,
\end{array}\right.
\\[0.3cm]
\label{eq:more.clo2}
k=b-2:\qquad&
\left\{\begin{array}{l}
\Tb^{b,b-2}_0 = \sigma\, \Tb^{b-2,b-1}_2 +\sigma^{b-a} \Jb f_{-3} \Tb^{0,b-2}_0,\\[0.15cm] 
\Tb^{b-2,b}_0 = \sigma\, \Tb^{b-1,b-2}_0 + \sigma^{b-a}\Jb f_{-5} \Tb^{b-2,0}_0,
\end{array}\right.
\\[0.3cm]
\label{eq:more.clo1}
k=b-1:\qquad&
\left\{\begin{array}{l}
\Tb^{b,b-1}_0 = \sigma\, \Tb^{b-2,b}_2 +\sigma^{b-a} \Jb f_{-3} \Tb^{0,b-1}_0,\\[0.15cm]
\Tb^{b-1,b}_0 = \sigma\, \Tb^{b,b-2}_0 + \sigma^{b-a}\Jb f_{-3} \Tb^{b-1,0}_0.
\end{array}\right.
\end{alignat}
\end{subequations}
These relations are proved in \cref{sec:FHclosure.proof}. They are afterwards used in \cref{sec:Y.closure} to prove the closure relations of the $Y$-system of the next section.

\subsection[Closure of the $Y$-system]{Closure of the $\boldsymbol Y$-system}

At roots of unity, the infinite $Y$-system reduces to a finite system of equations. The extra relation takes the form
\be\label{eq:Y.closure}
\Ib+\tb^{b-1}_0 = \frac{\displaystyle \Ib + \sigma^{a-1} \Jb \frac{\Tb^{b-2,0}_2}{\Tb^{b-1,0}_1} +\sigma^a  \Jb \bigg(\frac{\Tb^{b-2,0}_2}{\Tb^{b-1,0}_1}\bigg)^{\!2} + \sigma \bigg(\frac{\Tb^{b-2,0}_2}{\Tb^{b-1,0}_1}\bigg)^{\!3}}{\bigg(\displaystyle\Ib - \frac{\Tb^{b-2,0}_1\Tb^{b-2,0}_2}{\Tb^{b-1,0}_0\Tb^{b-1,0}_1}\bigg)\bigg(\displaystyle\Ib - \frac{\Tb^{b-2,0}_2\Tb^{b-2,0}_3}{\Tb^{b-1,0}_1\Tb^{b-1,0}_2}\bigg)}.
\ee
This relation follows from the closure relations of the fusion hierarchy relation. It is in fact equivalent to a quartic relation satisfied by the fused transfer tangles, and the proof is given in \cref{sec:Y.closure}.

\begin{figure} 
\centering
$
\psset{unit=2.2}
\begin{pspicture}[shift=-0.9](0,-1.5)(6,1.5)
\psline[linecolor=blue]{-}(0,0)(4,0)
\multiput(0,0)(1,0){5}{\pscurve[linecolor=red](0,0)(-0.1,0.4)(0,0.8)(0.1,0.4)(0,0)}
\rput(5.25,0.3){\pscurve[linecolor=red](0,0)(-0.05,0.2)(0,0.4)(0.05,0.2)(0,0)}
\rput(4.9,0.8){\pscurve[linecolor=red](0,0)(-0.05,0.2)(0,0.4)(0.05,0.2)(0,0)}
\rput(5.6,0.8){\pscurve[linecolor=red](0,0)(-0.05,0.2)(0,0.4)(0.05,0.2)(0,0)}
\rput(5.25,-0.5){
\pscurve[linecolor=blue,linewidth=0.185cm](0,0)(-0.1,-0.4)(0,-0.8)(0.1,-0.4)(0,0)
\pscurve[linecolor=white,linewidth=0.135cm](0,0)(-0.1,-0.4)(0,-0.8)(0.1,-0.4)(0,0)
\pscurve[linecolor=blue,linewidth=0.085cm](0,0)(-0.1,-0.4)(0,-0.8)(0.1,-0.4)(0,0)
\pscurve[linecolor=white,linewidth=0.03cm](0,0)(-0.1,-0.4)(0,-0.8)(0.1,-0.4)(0,0)
}
\psline[linecolor=red]{-}(5.25,0.3)(4.9,0.8)(5.6,0.8)(5.25,0.3)
\psline[linecolor=blue]{-}(4,0)(5.25,0.3)
\psline[linecolor=blue]{-}(4,0)(4.9,0.8)
\psline[linecolor=blue]{-}(4,0)(5.6,0.8)
\psline[linecolor=black,linewidth=0.085cm]{-}(4,0)(5.25,-0.5)\psline[linecolor=white,linewidth=0.035cm]{-}(4,0)(5.25,-0.5)
\psline[linecolor=black,linewidth=0.085cm]{-}(5.25,0.3)(5.25,-0.5)\psline[linecolor=white,linewidth=0.035cm]{-}(5.25,0.3)(5.25,-0.5)
\psline[linecolor=black,linewidth=0.085cm]{-}(4.9,0.8)(5.25,-0.5)\psline[linecolor=white,linewidth=0.035cm]{-}(4.9,0.8)(5.25,-0.5)
\psline[linecolor=black,linewidth=0.085cm]{-}(5.6,0.8)(5.25,-0.5)\psline[linecolor=white,linewidth=0.035cm]{-}(5.6,0.8)(5.25,-0.5)
\multiput(0,0)(1,0){5}{\pscircle[linewidth=1.5pt,linecolor=black,fillstyle=solid,fillcolor=white](0,0){.07}}
\pscircle[linewidth=1.5pt,linecolor=black,fillstyle=solid,fillcolor=white](5.25,-0.5){.07}
\pscircle[linewidth=1.5pt,linecolor=black,fillstyle=solid,fillcolor=white](5.25,0.3){.07}
\pscircle[linewidth=1.5pt,linecolor=black,fillstyle=solid,fillcolor=white](4.9,0.8){.07}
\pscircle[linewidth=1.5pt,linecolor=black,fillstyle=solid,fillcolor=white](5.6,0.8){.07}
\rput(0,-0.175){\scriptsize$\tb^{1}$}\rput(1,-0.175){\scriptsize$\tb^{2}$}\rput(2,-0.175){\scriptsize$...$}\rput(2.95,-0.175){\scriptsize$\tb^{b-3}$}\rput(3.95,-0.175){\scriptsize$\tb^{b-2}$}
\rput(5.73,0.88){\scriptsize$\xb$}
\rput(5.03,0.88){\scriptsize$\xb$}
\rput(5.38,0.3){\scriptsize$\xb$}
\rput(5.38,-0.42){\scriptsize$\yb$}
\end{pspicture}
$
\caption{The TBA diagram of the $Y$-system for the dilute $A^{(2)}_2$ model at the root of unity $\lambda = \frac{\pi(b-a)}{2b}$. The blue and red lines respectively indicate contributions to the $Y$-system in the numerator and denominator. Black lines indicate contributions that appear in the denominator for one relation but in the numerator in the other. The black edges are all doubled, and the edge from $\yb$ to itself is quadrupled. }
\label{fig:DynkinY}
\end{figure}
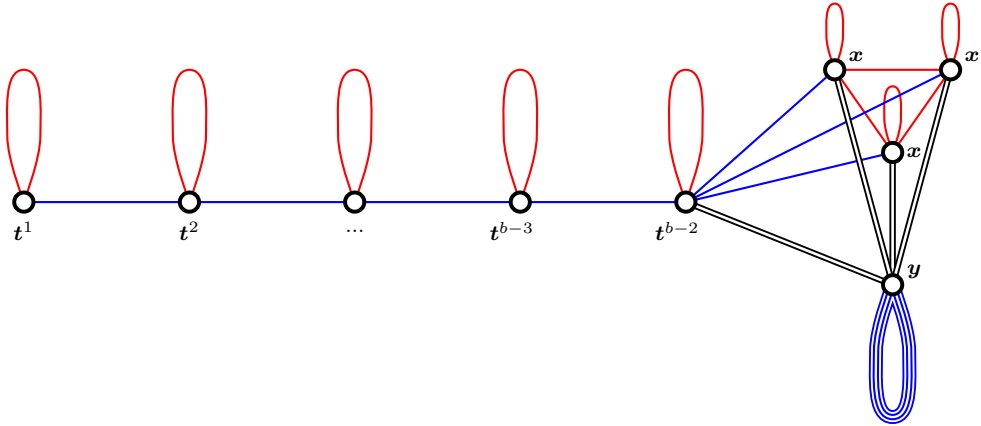

We rewrite this system of equation using the notations
\be
\Jb = \sigma^a(\eE^{\ir \Lambdab} + \Ib + \eE^{-\ir \Lambdab}), \qquad \xb_0 = \sigma \frac{\Tb^{b-2,0}_2}{\Tb^{b-1,0}_1}, 
\qquad \yb_0 = -\xb_{-1}\xb_0.
\ee
The eigenvalues of $\eE^{\ir \Lambdab}$ on the standard modules are easily read off from \eqref{eq:J.eigs}. The closure relation \eqref{eq:Y.closure} is then expressed in a product form:
\begin{subequations}
\label{rootY}
\be
\Ib+\tb^{b-1}_0 = \frac{(\Ib + \eE^{\ir \Lambdab} \xb_0)(\Ib + \xb_0)(\Ib + \eE^{-\ir \Lambdab} \xb_0)}{(\Ib + \yb_0)(\Ib + \yb_1)}.
\ee
Moreover, we find
\be
\xb_0 \xb_2 = \frac{(\Ib + \tb^{b-2}_2)(\Ib+\yb_1)(\Ib+\yb_2)}{(\Ib + \eE^{\ir \Lambdab} \xb_1)(\Ib + \xb_1^{-1})(\Ib + \eE^{-\ir \Lambdab} \xb_1)}
\ee
and
\be
\yb_0 \yb_2 = \frac{(\Ib + \tb^{b-2}_2)(\Ib + \tb^{b-2}_3)(\Ib+\yb_1)(\Ib+\yb_2)^2(\Ib+\yb_3)}{(\Ib + \eE^{\ir \Lambdab} \xb_1)(\Ib + \xb_1^{-1})(\Ib + \eE^{-\ir \Lambdab} \xb_1)(\Ib + \eE^{\ir \Lambdab} \xb_2)(\Ib + \xb_2^{-1})(\Ib + \eE^{-\ir \Lambdab} \xb_2)}.
\ee
\end{subequations}
These are obtained directly from \eqref{eq:Trelations} and \eqref{eq:Y.closure}. The last three equations along with \eqref{eq:Ysys} for $m = 1, \dots, b-2$ constitute the closed $Y$-system of the $A_2^{(2)}$ loop models at the root of unity $\lambda = \frac{(b-a)\pi}{2b}$. The TBA diagram of this $Y$-system is illustrated in \cref{fig:DynkinY}. It is apparent that this diagram is the $\mathbb Z_2$-folding of the similar diagram found for the $A_2^{(1)}$ models.

\goodbreak
\section{Conclusion}

In this paper, we have derived the fusion hierarchy, $T$- and $Y$-system of functional equations for the dilute $A_2^{(2)}$ loop models. In the generic case with $\frac{\lambda}{\pi}\notin{\Bbb Q}$, these sets of equations do not close finitely. For the roots of unity case $\frac{\lambda}{\pi}\in{\Bbb Q}$, we established finite closure and derived closed finite $T$- and $Y$-systems.
As in the $A_2^{(1)}$ case it is expected that, at roots of unity, the same $T$- and $Y$-systems apply to the $A_2^{(2)}$ vertex models and also to the $A_2^{(2)}$ RSOS models but with finite truncation. This thus completes the derivation of $T$- and $Y$-systems for the trinity of $A_1^{(1)}$, $A_2^{(1)}$ and $A_2^{(2)}$ models. We stress that our $A_2^{(2)}$ $T$- and $Y$-systems are algebraic and hold for all $\lambda$ and therefore for all critical branches, including the non-compact regime~\cite{VJS2014} with $\lambda\in[\frac{\pi}{2},\frac{2\pi}{3}]$.

It is noteworthy that our derivation of the fusion hierarchy relations for the $A_2^{(2)}$ model avoids the construction of Wenzl-Jones projectors. In a sense, our derivation is elegant in that it only uses the polynomial properties of the transfer tangles, their commutativity via the Yang-Baxter equation, and the push-through properties satisfied by the triangle operators. It would nevertheless be interesting to understand the complete construction of the Wenzl-Jones projectors, beyond the partial construction given in \cref{sec:partial.WJ}. It is clear from the local relations given there that the underlying algebraic structure is similar but not identical to the $s\ell(3)$ spider algebra \cite{Kuperberg96} that came up in our previous investigation of the dilute $A_2^{(1)}$ models \cite{MDPR2018}. This question is relevant, as we expect that this algebraic structure also plays a role in the conformal description of these lattice models in the scaling limit.

In principle, through the associated TBA integral equations, the $Y$-systems allow for the complete solution of the spectra and conformal data of the $A_2^{(2)}$ loop models. In the context of the $A_1^{(1)}$ model, this was achieved~\cite{MDKP2017} for critical bond percolation on the square lattice. For the $A_2^{(2)}$ case, this is of particular interest for the model ${\cal DLM}(2,3)$, which corresponds to critical site percolation on the triangular lattice. The prevailing folklore dictates that the lattice structure is irrelevant and that bond and site percolation should lie in the same universality class. So the exact solution of site percolation on the triangular lattice holds out the prospect to test the validity of universality in the framework of Logarithmic CFT (LCFT). This is nontrivial since, in LCFT, two models can share the same conformal dimensions and critical exponents but differ in the details of the Jordan cells appearing in their reducible yet indecomposable representations. 

It is also relevant to point out here that, in the continuum scaling limit, the loop segments of the dilute $A_2^{(2)}$ loop models become~\cite{SAPR2012} fractal curves. Indeed, the conformal loop ensemble of these curves is described by $\mbox{SLE}_{\kappa}$~\cite{Schramm2000} with
\begin{eqnarray}
\kappa=\frac{4p'}{p}
\end{eqnarray}
For site percolation on the triangular lattice, $\kappa=6$~\cite{Smirnov2001,SLE} corresponding to planar brownian motion.

It is known~\cite{Zam1989,FZ1990,WNS1992,WPSN1994,BNW94,Suzuki1998} that the critical $A_2^{(2)}$ RSOS model with $\lambda=\frac{5\pi}{16}$ is described in the continuum scaling limit by an $E_8$ coset CFT and that, off-criticality, it lies in the universality class of the Ising model in a magnetic field.
So it is an interesting question as to whether there are any remnants of the $E_8$ structure to be found in the dilute logarithmic Ising model ${\cal DLM}(3,4)$ with $c=\frac12$, $\lambda=\frac{5\pi}{16}$ and $\beta=\sqrt{2}$.

\goodbreak
\section*{Acknowledgments} 

AMD is an FNRS Postdoctoral Researcher under the project CR28075116. He acknowledges support from the EOS-contract O013018F.
AMD acknowledges the hospitality and support of the University of Melbourne where part of this work was done. PAP thanks Philippe Ruelle for hospitality during a visit to the Universit\'e Catholique de Louvain. The authors thank Yvan Saint-Aubin and J\o rgen Rasmussen for useful discussions.

\goodbreak
\appendix

\section{Wenzl-Jones projectors}\label{sec:partial.WJ}

\subsection{Local triangle relations}\label{sec:more.triangles}

In this appendix, we give a construction of the Wenzl-Jones projectors for the $s\ell(3)$ weights $(m,0)$, $(m,1)$, $(0,n)$ and $(1,n)$ for $m,n \ge 1$. We start by defining two families of triangle operators:
\begin{subequations}
\label{eq:triangles.m}
\begin{alignat}{2}
\psset{unit=0.8}
\begin{pspicture}[shift=-.40](0,0)(2,1)
\pspolygon[fillstyle=solid,fillcolor=lightlightblue](0,1)(1,0)(2,1)
\rput(1,0.6){$_m$}
\psarc[linewidth=0.025]{-}(1,0){0.16}{45}{135}
\end{pspicture} 
\ &= 
\kappa_1(m)\
\psset{unit=0.8}
\begin{pspicture}[shift=-.40](0,0)(2,1)
\pspolygon[fillstyle=solid,fillcolor=lightlightblue](0,1)(1,0)(2,1)
\psarc[linewidth=1.5pt,linecolor=blue](1,0){.707}{45}{135}
\end{pspicture} 
\ + \kappa_2(m) \
\begin{pspicture}[shift=-.40](0,0)(2,1)
\pspolygon[fillstyle=solid,fillcolor=lightlightblue](0,1)(1,0)(2,1)
\psline[linewidth=1.5pt,linecolor=blue](0.5,0.5)(1,1)
\end{pspicture} 
\ + \kappa_3(m)\
\begin{pspicture}[shift=-.40](0,0)(2,1)
\pspolygon[fillstyle=solid,fillcolor=lightlightblue](0,1)(1,0)(2,1)
\psline[linewidth=1.5pt,linecolor=blue](1.5,0.5)(1,1)
\end{pspicture}
\ + \kappa_4(m)\
\begin{pspicture}[shift=-.40](0,0)(2,1)
\pspolygon[fillstyle=solid,fillcolor=lightlightblue](0,1)(1,0)(2,1)
\end{pspicture}
\ ,
\\[0.3cm]
\psset{unit=0.8}
\begin{pspicture}[shift=-.40](0,0)(2,1)
\pspolygon[fillstyle=solid,fillcolor=lightlightblue](0,1)(1,0)(2,1)
\rput(1,0.4){$_m$}
\rput(1,0){\wobblyarc{0.71}{45}{135}}
\end{pspicture} 
\ &= \epsilon_1(m)\
\psset{unit=0.8}
\begin{pspicture}[shift=-.40](0,0)(2,1)
\pspolygon[fillstyle=solid,fillcolor=lightlightblue](0,1)(1,0)(2,1)
\psarc[linewidth=1.5pt,linecolor=blue](1,0){.707}{45}{135}
\end{pspicture} 
\ + \epsilon_2(m) \
\begin{pspicture}[shift=-.40](0,0)(2,1)
\pspolygon[fillstyle=solid,fillcolor=lightlightblue](0,1)(1,0)(2,1)
\end{pspicture}
\ .
\end{alignat}
\end{subequations}
The prefactors are given by
\begin{subequations}
\label{eq:prefactors}
\begin{alignat}{3}
\kappa_1(m) &= - \frac{[2m][2m+1]}{[2m+2][2m+3]}, \qquad &\kappa_2(m) &= - \frac{[2m]}{[2m+2]}, 
\\[0.2cm]
\kappa_3(m) &= \frac{[2m][2m+1]}{[2m-1][2m+2]},\qquad &\kappa_4(m) &=  \frac{[1][2][2m]}{[2m-1][2m+2][2m+3]},
\\[0.2cm]
\epsilon_1(m) &= \frac{[2m][2m+1]}{[2m+3][2m+4]}, \qquad &\epsilon_2(m) &= \frac{[2m][2m+1][2m+5]}{[2m-1][2m+3][2m+4]},
\end{alignat}
\end{subequations}
where
\be
[m] = x^m - x^{-m}, \qquad x = \eE^{\ir \lambda}.
\ee
These triangle operators satisfy the local relations
\begin{subequations}
\label{eq:spider.ids}
\begin{alignat}{2}
&
\psset{unit=0.8}
\begin{pspicture}[shift=-.90](0,-1)(2,1)
\pspolygon[fillstyle=solid,fillcolor=lightlightblue](0,1)(1,0)(2,1)
\rput(1,0.6){$_1$}
\psarc[linewidth=0.025]{-}(1,0){0.16}{45}{135}
\pspolygon[fillstyle=solid,fillcolor=lightlightblue](0,-1)(1,0)(2,-1)
\rput(1,-0.6){\specialcircle{0.075}}
\psarc[linewidth=1.5pt,linecolor=blue,linestyle=dashed,dash=2pt 2pt](1,0){.707}{-45}{45}
\psarc[linewidth=1.5pt,linecolor=blue,linestyle=dashed,dash=2pt 2pt](1,0){.707}{135}{-135}
\end{pspicture} \ = \ \
\begin{pspicture}[shift=-0.65](0,0)(0.75,1.5)
\pspolygon[fillstyle=solid,fillcolor=lightlightblue](0,0)(0.75,0)(0.75,1.5)(0,1.5)
\psline[linewidth=1.5pt,linecolor=blue,linestyle=dashed,dash=2pt 2pt](0.375,0)(0.375,1.5)
\end{pspicture}\ \ ,
\qquad 
&&
\psset{unit=0.8}
\begin{pspicture}[shift=-.90](-1,-1)(2,1)
\pspolygon[fillstyle=solid,fillcolor=lightlightblue](0,1)(1,0)(2,1)
\pspolygon[fillstyle=solid,fillcolor=lightlightblue](0,1)(1,0)(-1,0)
\rput(0,0.4){\specialcircle{0.075}}
\pspolygon[fillstyle=solid,fillcolor=lightlightblue](0,-1)(1,0)(-1,0)
\rput(1,0.6){$_{m+1}$}
\psarc[linewidth=0.025]{-}(1,0){0.16}{45}{135}
\pspolygon[fillstyle=solid,fillcolor=lightlightblue](0,-1)(1,0)(2,-1)
\rput(1,-0.6){\specialcircle{0.075}}
\psarc[linewidth=0.025]{-}(0,-1){0.16}{45}{135}
\rput(0,-0.4){$_m$}
\psarc[linewidth=1.5pt,linecolor=blue,linestyle=dashed,dash=2pt 2pt](1,0){.707}{-45}{45}
\end{pspicture} \ = - \ \ 
\begin{pspicture}[shift=-.90](0,-1)(2,1)
\pspolygon[fillstyle=solid,fillcolor=lightlightblue](0,0)(1,1)(2,0)(1,-1)
\psarc[linewidth=1.5pt,linecolor=blue,linestyle=dashed,dash=2pt 2pt](0,0){.707}{-45}{45}
\psarc[linewidth=1.5pt,linecolor=blue,linestyle=dashed,dash=2pt 2pt](2,0){.707}{135}{-135}
\end{pspicture} 
\ \ +  \ \ 
\begin{pspicture}[shift=-.90](-1,-1)(2,1)
\pspolygon[fillstyle=solid,fillcolor=lightlightblue](0,1)(1,0)(2,1)
\rput(1,0.6){$_{m+1}$}
\psarc[linewidth=0.025]{-}(1,0){0.16}{45}{135}
\pspolygon[fillstyle=solid,fillcolor=lightlightblue](0,-1)(1,0)(2,-1)
\pspolygon[fillstyle=solid,fillcolor=lightlightblue](1,0)(0,1)(-1,0)(0,-1)
\rput(1,-0.6){\specialcircle{0.075}}
\psarc[linewidth=1.5pt,linecolor=blue,linestyle=dashed,dash=2pt 2pt](-1,0){.707}{-45}{45}
\psarc[linewidth=1.5pt,linecolor=blue,linestyle=dashed,dash=2pt 2pt](1,0){.707}{-45}{45}
\psarc[linewidth=1.5pt,linecolor=blue,linestyle=dashed,dash=2pt 2pt](1,0){.707}{135}{-135}
\end{pspicture} 
\ \ + \ \
\begin{pspicture}[shift=-.90](0,-1)(2,1)
\pspolygon[fillstyle=solid,fillcolor=lightlightblue](0,0)(1,1)(2,0)(1,-1)
\psarc[linewidth=1.5pt,linecolor=blue,linestyle=dashed,dash=2pt 2pt](1,1){.707}{-135}{-45}
\rput(1,-1){\wobblyarc{.707}{45}{135}}
\rput(1,-0.6){$_{m}$}
\end{pspicture}\ \ ,
\\
&
\psset{unit=0.8}
\begin{pspicture}[shift=-.90](0,-1)(2,1)
\pspolygon[fillstyle=solid,fillcolor=lightlightblue](0,1)(1,0)(2,1)
\rput(1,0.4){$_1$}
\rput(1,0){\wobblyarc{0.71}{45}{135}}
\pspolygon[fillstyle=solid,fillcolor=lightlightblue](0,-1)(1,0)(2,-1)
\psarc[linewidth=1.5pt,linecolor=blue,linestyle=dashed,dash=2pt 2pt](1,0){.707}{135}{45}
\end{pspicture} \ = 1,
\qquad 
&&
\psset{unit=0.8}
\begin{pspicture}[shift=-.90](-1,-1)(2,1)
\pspolygon[fillstyle=solid,fillcolor=lightlightblue](0,1)(1,0)(2,1)
\pspolygon[fillstyle=solid,fillcolor=lightlightblue](0,1)(1,0)(-1,0)
\rput(0,0.4){\specialcircle{0.075}}
\pspolygon[fillstyle=solid,fillcolor=lightlightblue](0,-1)(1,0)(-1,0)
\rput(1,0.4){\scriptsize$_{m+1}$}
\rput(1,0){\wobblyarc{0.71}{45}{135}}
\pspolygon[fillstyle=solid,fillcolor=lightlightblue](0,-1)(1,0)(2,-1)
\rput(1,-0.6){\specialcircle{0.075}}
\psarc[linewidth=0.025]{-}(0,-1){0.16}{45}{135}
\rput(0,-0.4){$_m$}
\psarc[linewidth=1.5pt,linecolor=blue,linestyle=dashed,dash=2pt 2pt](1,0){.707}{-45}{45}
\end{pspicture} \ = \ \ 
\begin{pspicture}[shift=-0.65](0,0)(0.75,1.5)
\pspolygon[fillstyle=solid,fillcolor=lightlightblue](0,0)(0.75,0)(0.75,1.5)(0,1.5)
\psline[linewidth=1.5pt,linecolor=blue,linestyle=dashed,dash=2pt 2pt](0.375,0)(0.375,1.5)
\end{pspicture}
\ \ \times \Bigg( \ \ 
\begin{pspicture}[shift=-.90](0,-1)(2,1)
\pspolygon[fillstyle=solid,fillcolor=lightlightblue](0,1)(1,0)(2,1)
\rput(1,0.4){\scriptsize$_{m+1}$}
\rput(1,0){\wobblyarc{0.71}{45}{135}}
\pspolygon[fillstyle=solid,fillcolor=lightlightblue](0,-1)(1,0)(2,-1)
\psarc[linewidth=1.5pt,linecolor=blue,linestyle=dashed,dash=2pt 2pt](1,0){.707}{135}{45}
\end{pspicture} -1
\Bigg).
\end{alignat}
\end{subequations}

We note that there exists a second solution for the prefactors in \eqref{eq:triangles.m} that satisfies the local relations \eqref{eq:spider.ids}. It is given by
\begin{subequations}
\label{eq:prefactors.prime}
\begin{alignat}{3}
\bar\kappa_1(m) &= - \frac{[2m]\{2m+1\}}{[2m+2]\{2m+3\}}, \qquad &\bar\kappa_2(m) &=  \frac{[2m]}{[2m+2]}, 
\\[0.2cm]
\bar\kappa_3(m) &= \frac{[2m]\{2m+1\}}{\{2m-1\}[2m+2]},\qquad &\bar\kappa_4(m) &=  \frac{[1][2][2m]}{\{2m-1\}[2m+2]\{2m+3\}},
\\[0.2cm]
\bar\epsilon_1(m) &= - \frac{[2m]\{2m+1\}}{\{2m+3\}[2m+4]}, \qquad &\bar\epsilon_2(m) &= \frac{[2m]\{2m+1\}\{2m+5\}}{\{2m-1\}\{2m+3\}[2m+4]},
\end{alignat}
\end{subequations}
where $\{m\} = x^m + x^{-m}$. The two solutions are similar but nevertheless appear to be inequivalent.

\subsection{Definition and properties of the projectors}

We denote by $P^{m,n}$ the Wenzl-Jones projector with label $(m,n)$ and draw it as a pink rectangle in the diagrammatic calculus:
$P^{m,n} = \
\begin{pspicture}[shift=-0.05](0,0)(1.5,0.3)
\pspolygon[fillstyle=solid,fillcolor=pink](0,0)(1.5,0)(1.5,0.3)(0,0.3)(0,0)\rput(0.75,0.15){$_{m,n}$}\psline{-}(0.1,0)(0.1,0.1)(0,0.1)
\end{pspicture}\ $.
The marker in the lower corner indicates the orientation of the diagrams and serves as a reminder that these tangles are not invariant under reflections or rotations.
The definition of the Wenzl-Jones projectors $P^{m,0}$ and $P^{m,1}$ is recursive and uses the triangle operators introduced in \cref{sec:more.triangles}.
The initial condition is
\be
\begin{pspicture}[shift=-0.05](0,0)(1.5,0.3)
 \pspolygon[fillstyle=solid,fillcolor=pink](0,0)(1.5,0)(1.5,0.3)(0,0.3)(0,0)\rput(0.75,0.15){$_{1,0}$}\psline{-}(0.1,0)(0.1,0.1)(0,0.1)
 \end{pspicture} \ \ = \ \
\psset{unit=0.8} 
 \begin{pspicture}[shift=-0.65](0,0)(0.75,1.5)
\pspolygon[fillstyle=solid,fillcolor=lightlightblue](0,0)(0.75,0)(0.75,1.5)(0,1.5)
\psline[linewidth=1.5pt,linecolor=blue,linestyle=dashed,dash=2pt 2pt](0.375,0)(0.375,1.5)
\end{pspicture}\ \ ,
\ee
and the general recursive definition is
\begin{subequations}
\label{eq:P.def}
\begin{alignat}{2}
&\label{eq:Pm0rec}
\begin{pspicture}[shift=-0.05](0,0)(2.1,0.3)
\pspolygon[fillstyle=solid,fillcolor=pink](0,0)(2.1,0)(2.1,0.3)(0,0.3)(0,0)\rput(1.05,0.15){$_{m,0}$}\psline{-}(0.1,0)(0.1,0.1)(0,0.1)
 \end{pspicture} \ \ = \ \ 
\begin{pspicture}[shift=-0.05](0,0)(2.1,0.3)
\pspolygon[fillstyle=solid,fillcolor=pink](0,0)(1.8,0)(1.8,0.3)(0,0.3)(0,0)\rput(0.9,0.15){$_{m-1,0}$}\psline{-}(0.1,0)(0.1,0.1)(0,0.1)
\psline[linewidth=1.5pt,linecolor=blue,linestyle=dashed,dash=2pt 2pt](2.0,0)(2.0,0.3)
\end{pspicture} \ \ -  \ \ 
\begin{pspicture}[shift=-0.6](0,0)(2.1,1.4)
\pspolygon[fillstyle=solid,fillcolor=pink](0,0)(1.8,0)(1.8,0.3)(0,0.3)(0,0)\rput(0.9,0.15){$_{m-1,0}$}\psline{-}(0.1,0)(0.1,0.1)(0,0.1)
\pspolygon[fillstyle=solid,fillcolor=lightlightblue](1.8,0.3)(2.2,0.7)(1.4,0.7)\psarc[linewidth=0.025]{-}(1.8,0.3){0.10}{45}{135}
\rput(1.8,0.60){\tiny$_{m\!-\!1}$}
\pspolygon[fillstyle=solid,fillcolor=lightlightblue](1.8,1.1)(2.2,0.7)(1.4,0.7)
\rput(1.8,0.85){\specialcircle{0.045}}
\rput(0,1.1){\pspolygon[fillstyle=solid,fillcolor=pink](0,0)(1.8,0)(1.8,0.3)(0,0.3)(0,0)\rput(0.9,0.15){$_{m-1,0}$}\psline{-}(0.1,0)(0.1,0.1)(0,0.1)}
\psline[linewidth=1.5pt,linecolor=blue,linestyle=dashed,dash=2pt 2pt](0.2,0.3)(0.2,1.1)
\psline[linewidth=1.5pt,linecolor=blue,linestyle=dashed,dash=2pt 2pt](0.6,0.3)(0.6,1.1)
\psline[linewidth=1.5pt,linecolor=blue,linestyle=dashed,dash=2pt 2pt](1.2,0.3)(1.2,1.1)
\psline[linewidth=1.5pt,linecolor=blue,linestyle=dashed,dash=2pt 2pt](1.6,0.3)(1.6,0.5)
\psline[linewidth=1.5pt,linecolor=blue,linestyle=dashed,dash=2pt 2pt](1.6,0.9)(1.6,1.1)
\psline[linewidth=1.5pt,linecolor=blue,linestyle=dashed,dash=2pt 2pt](2.0,0)(2.0,0.5)
\psline[linewidth=1.5pt,linecolor=blue,linestyle=dashed,dash=2pt 2pt](2.0,0.9)(2.0,1.4)
\rput(0.9,0.7){$...$}
\end{pspicture}\ \ ,\qquad m\ge2,
\\[0.2cm]
&\label{eq:Pm1rec}
\begin{pspicture}[shift=-0.05](0,0)(2.1,0.3)
\pspolygon[fillstyle=solid,fillcolor=pink](0,0)(2.1,0)(2.1,0.3)(0,0.3)(0,0)\rput(1.05,0.15){$_{m,1}$}\psline{-}(0.1,0)(0.1,0.1)(0,0.1)
 \end{pspicture} \ \ = \ \ 
\begin{pspicture}[shift=-0.05](0,0)(2.1,0.3)
\pspolygon[fillstyle=solid,fillcolor=pink](0,0)(1.8,0)(1.8,0.3)(0,0.3)(0,0)\rput(0.9,0.15){$_{m,0}$}\psline{-}(0.1,0)(0.1,0.1)(0,0.1)
\psline[linewidth=1.5pt,linecolor=blue,linestyle=dashed,dash=2pt 2pt](2.0,0)(2.0,0.3)
\end{pspicture} \ \ -  \ \ 
\begin{pspicture}[shift=-0.6](0,0)(2.1,1.4)
\pspolygon[fillstyle=solid,fillcolor=pink](0,0)(1.8,0)(1.8,0.3)(0,0.3)(0,0)\rput(0.9,0.15){$_{m,0}$}\psline{-}(0.1,0)(0.1,0.1)(0,0.1)
\pspolygon[fillstyle=solid,fillcolor=lightlightblue](1.8,0.3)(2.2,0.7)(1.4,0.7)
\rput(1.8,0.47){\tiny$_{m}$}
\rput(1.8,0.3){\wobblyarcthree{0.3}{45}{135}}
\pspolygon[fillstyle=solid,fillcolor=lightlightblue](1.8,1.1)(2.2,0.7)(1.4,0.7)
\psarc[linewidth=1.5pt,linecolor=blue,linestyle=dashed,dash=2pt 2pt](1.8,1){.2}{-135}{-45}
\rput(0,1.1){\pspolygon[fillstyle=solid,fillcolor=pink](0,0)(1.8,0)(1.8,0.3)(0,0.3)(0,0)\rput(0.9,0.15){$_{m,0}$}\psline{-}(0.1,0)(0.1,0.1)(0,0.1)}
\psline[linewidth=1.5pt,linecolor=blue,linestyle=dashed,dash=2pt 2pt](0.2,0.3)(0.2,1.1)
\psline[linewidth=1.5pt,linecolor=blue,linestyle=dashed,dash=2pt 2pt](0.6,0.3)(0.6,1.1)
\psline[linewidth=1.5pt,linecolor=blue,linestyle=dashed,dash=2pt 2pt](1.2,0.3)(1.2,1.1)
\psline[linewidth=1.5pt,linecolor=blue,linestyle=dashed,dash=2pt 2pt](1.6,0.3)(1.6,0.5)
\psline[linewidth=1.5pt,linecolor=blue,linestyle=dashed,dash=2pt 2pt](1.6,0.9)(1.6,1.1)
\psline[linewidth=1.5pt,linecolor=blue,linestyle=dashed,dash=2pt 2pt](2.0,0)(2.0,0.5)
\psline[linewidth=1.5pt,linecolor=blue,linestyle=dashed,dash=2pt 2pt](2.0,0.9)(2.0,1.4)
\rput(0.9,0.7){$...$}
\end{pspicture}\ \ ,\qquad m\ge1.
\end{alignat}
\end{subequations}

It follows from these definitions and the local relations \eqref{eq:spider.ids} that these tangles are indeed projectors. For $P^{m,0}$, the proof uses the same arguments as those given in Proposition A.1 of \cite{MDPR2018}. It is recursive on $m$ and shows that the  following identities hold:
\be
\label{eq:props.Pm0}
\begin{pspicture}[shift=-0.75](0,-0.7)(1.5,0.7)
\pspolygon[fillstyle=solid,fillcolor=pink](0,0)(1.5,0)(1.5,0.3)(0,0.3)(0,0)\rput(0.75,0.15){$_{m,0}$}\psline{-}(0.1,0)(0.1,0.1)(0,0.1)
\rput(1.0,0){\pspolygon[fillstyle=solid,fillcolor=lightlightblue](0,0)(0.4,-0.4)(-0.4,-0.4)
\rput(0,-0.25){\specialcircle{0.045}}
}
\psline[linewidth=1.5pt,linecolor=blue,linestyle=dashed,dash=2pt 2pt](0.8,0)(0.8,-0.2)
\psline[linewidth=1.5pt,linecolor=blue,linestyle=dashed,dash=2pt 2pt](1.2,0)(1.2,-0.2)
 \end{pspicture} \ = 0, \qquad \quad
\begin{pspicture}[shift=-0.2](0,0)(1.5,0.6)
\pspolygon[fillstyle=solid,fillcolor=pink](0,0)(1.5,0)(1.5,0.3)(0,0.3)(0,0)\rput(0.75,0.15){$_{m,0}$}\psline{-}(0.1,0)(0.1,0.1)(0,0.1)
\rput(0,0.3){\pspolygon[fillstyle=solid,fillcolor=pink](0,0)(1.2,0)(1.2,0.3)(0,0.3)(0,0)\rput(0.6,0.15){$_{n,0}$}\psline{-}(0.1,0)(0.1,0.1)(0,0.1)}
\end{pspicture} 
\ \ = \ \
\begin{pspicture}[shift=-0.2](0,0)(1.5,0.6)
\pspolygon[fillstyle=solid,fillcolor=pink](0,0)(1.2,0)(1.2,0.3)(0,0.3)(0,0)\rput(0.6,0.15){$_{n,0}$}\psline{-}(0.1,0)(0.1,0.1)(0,0.1)
\rput(0,0.3){\pspolygon[fillstyle=solid,fillcolor=pink](0,0)(1.5,0)(1.5,0.3)(0,0.3)(0,0)\rput(0.75,0.15){$_{m,0}$}\psline{-}(0.1,0)(0.1,0.1)(0,0.1)}
\end{pspicture} 
\ \ = \ \
\begin{pspicture}[shift=-0.05](0,0)(1.5,0.3)
\pspolygon[fillstyle=solid,fillcolor=pink](0,0)(1.5,0)(1.5,0.3)(0,0.3)(0,0)\rput(0.75,0.15){$_{m,0}$}\psline{-}(0.1,0)(0.1,0.1)(0,0.1)
\end{pspicture}\ , \qquad \quad m \ge n \ge 1.
\ee
Likewise, for $P^{m,1}$, the local relations \eqref{eq:spider.ids} imply that
\be
\label{eq:props.Pm1}
\begin{pspicture}[shift=-0.75](0,-0.7)(1.5,0.3)
\pspolygon[fillstyle=solid,fillcolor=pink](0,0)(1.5,0)(1.5,0.3)(0,0.3)(0,0)\rput(0.75,0.15){$_{m,1}$}\psline{-}(0.1,0)(0.1,0.1)(0,0.1)
\rput(1.2,0){\pspolygon[fillstyle=solid,fillcolor=lightlightblue](0,0)(0.4,-0.4)(-0.4,-0.4)
\psarc[linewidth=1.5pt,linecolor=blue,linestyle=dashed,dash=2pt 2pt](0,0){0.2}{180}{0}
}
\end{pspicture} \ = 0,
\qquad\quad
\begin{pspicture}[shift=-0.2](0,0)(1.5,0.6)
\pspolygon[fillstyle=solid,fillcolor=pink](0,0)(1.5,0)(1.5,0.3)(0,0.3)(0,0)\rput(0.75,0.15){$_{m,1}$}\psline{-}(0.1,0)(0.1,0.1)(0,0.1)
\rput(0,0.3){\pspolygon[fillstyle=solid,fillcolor=pink](0,0)(1.5,0)(1.5,0.3)(0,0.3)(0,0)\rput(0.75,0.15){$_{m,1}$}\psline{-}(0.1,0)(0.1,0.1)(0,0.1)}
\end{pspicture} \ \ = \ \
\begin{pspicture}[shift=-0.05](0,0)(1.5,0.3)
\pspolygon[fillstyle=solid,fillcolor=pink](0,0)(1.5,0)(1.5,0.3)(0,0.3)(0,0)\rput(0.75,0.15){$_{m,1}$}\psline{-}(0.1,0)(0.1,0.1)(0,0.1)
\end{pspicture}\ ,
\ee
where the dashed loop segment in the first diagram connects the last two nodes of the projector.

Theses projectors are used to build fused face operators. For general values of $(m,n)$, these are defined as
\be
\psset{unit=1.2}
\begin{pspicture}[shift=-.40](0,0)(1,1)
\facegrid{(0,0)}{(1,1)}
\psarc[linewidth=0.025]{-}(0,0){0.16}{0}{90}
\rput(.5,.7){\tiny $(m,n)$}
\rput(.5,.40){$u$}
\end{pspicture} \ \ = \bigg(\prod_{\substack{k=-1\\[0.1cm] k \neq 2m-3,2m-1}}^{2m+2n-3} \displaystyle\frac{1}{s_k(u)} \bigg)\ \ 
\begin{pspicture}[shift=-3.9](-0.3,0)(1.3,8)
\facegrid{(0,0)}{(1,8)}
\pspolygon[fillstyle=solid,fillcolor=pink](0,0.1)(-0.3,0.1)(-0.3,7.9)(0,7.9)
\pspolygon[fillstyle=solid,fillcolor=pink](1,0.1)(1.3,0.1)(1.3,7.9)(1,7.9)\rput(-0.12,4){\rput{90}(0,0){$_{m,n}$}}\rput(1.18,4){\rput{90}(0,0){$_{m,n}$}}
\rput(0,0.1){\psline{-}(-0.1,0)(-0.1,0.1)(0,0.1)}\rput(1.3,0.1){\psline{-}(-0.1,0)(-0.1,0.1)(0,0.1)}
\psarc[linewidth=0.025]{-}(0,0){0.16}{0}{90}\rput(.5,.5){$u_0$}
\rput(0,1){\psarc[linewidth=0.025]{-}(0,0){0.16}{0}{90}\rput(.5,.5){$u_2$}}
\rput(0.5,2.6){$\vdots$}
\rput(0,3){\psarc[linewidth=0.025]{-}(0,0){0.16}{0}{90}\rput(.5,.5){$u_{2m-2}$}}
\rput(0,4){\psarc[linewidth=0.025]{-}(0,0){0.16}{0}{90}\rput(.5,.5){$u_{2m+1}$}}
\rput(0,5){\psarc[linewidth=0.025]{-}(0,0){0.16}{0}{90}\rput(.5,.5){$u_{2m+3}$}}
\rput(0.5,6.6){$\vdots$}
\rput(0,7){\psarc[linewidth=0.025]{-}(0,0){0.16}{0}{90}\rput(.5,.5){\tiny$u_{2m\!+\!2n\!-\!1}$}}
\end{pspicture}\ \ .
\ee
For $(m,0)$ and $(m,1)$ it follows, from the properties of the projectors and the decompositions of the elementary face operators at $u = 0, 2\lambda, 3\lambda$, that the corresponding fused face operators are centered Laurent polynomials of respective degrees $2$ and $3$. 

The fused transfer tangles $\Tb^{m,0}(u)$ and $\Tb^{m,1}(u)$ are then defined from the fused face operators as in \eqref{eq:fusedTmn}. The fusion hierarchy relations \eqref{eq:FH.corner1}, \eqref{eq:FH.corner2}, \eqref{eq:FH.bdy1} and \eqref{eq:FH.adj.bdy1} are derived using the recursive definition of the projectors and the properties \eqref{eq:props.Pm0} and \eqref{eq:props.Pm1}. The proof is diagrammatic and closely follows the proof for the $A_2^{(1)}$ loop model in Appendix C.1 of \cite{MDPR2018}.

The projectors with labels $P^{0,n}$ and $P^{1,n}$ are obtained as the reflection of the projectors $P^{n,0}$ and $P^{n,1}$ about a vertical axis using the symmetry
\be
\label{eq:P0n}
\begin{pspicture}[shift=-0.05](0,0)(1.5,0.3)
\pspolygon[fillstyle=solid,fillcolor=pink](0,0)(1.5,0)(1.5,0.3)(0,0.3)(0,0)\rput(0.75,0.15){$_{0,n}$}\psline{-}(0.1,0)(0.1,0.1)(0,0.1)
\end{pspicture}
\ = \
\begin{pspicture}[shift=-0.05](0,0)(1.5,0.3)
\pspolygon[fillstyle=solid,fillcolor=pink](0,0)(1.5,0)(1.5,0.3)(0,0.3)(0,0)\rput(0.75,0.15){$_{n,0}$}\psline{-}(1.4,0)(1.4,0.1)(1.5,0.1).
\end{pspicture}\, .
\ee 
With this definition, the projectors $P^{0,n}$ and $P^{1,n}$ satisfy recursive relations given by the reflections of \eqref{eq:P.def} via a vertical axis. These allow us to derive the fusion hierarchy relations \eqref{eq:FH.corner3}, \eqref{eq:FH.bdy2} and \eqref{eq:FH.adj.bdy2}. The proof again follows the same proof for the $A_2^{(1)}$ loop model given in Appendix C.1 of \cite{MDPR2018}.

For the construction of Wenzl-Jones projectors to be complete, one needs to define the projectors $P^{m,n}$ with $m,n \ge 2$. Inspired from the same construction for the $A_2^{(1)}$ model, one expects that the first projector in this set, $P^{2,2}$, is of the form
\be
\label{eq:P22}
\begin{pspicture}[shift=-0.05](0,0)(1.5,0.3)
\pspolygon[fillstyle=solid,fillcolor=pink](0,0)(1.5,0)(1.5,0.3)(0,0.3)(0,0)\rput(0.75,0.15){$_{2,2}$}\psline{-}(0.1,0)(0.1,0.1)(0,0.1)
\end{pspicture}
\ = \
\begin{pspicture}[shift=-0.05](0,0)(1.7,0.3)
\pspolygon[fillstyle=solid,fillcolor=pink](0,0)(0.8,0)(0.8,0.3)(0,0.3)(0,0)\rput(0.4,0.15){$_{2,0}$}\psline{-}(0.1,0)(0.1,0.1)(0,0.1)
\rput(0.9,0){\pspolygon[fillstyle=solid,fillcolor=pink](0,0)(0.8,0)(0.8,0.3)(0,0.3)(0,0)\rput(0.4,0.15){$_{2,0}$}\psline{-}(0.1,0)(0.1,0.1)(0,0.1)}
\end{pspicture}
\ - \
\begin{pspicture}[shift=-1.05](0,0)(1.7,2.3)
\pspolygon[fillstyle=solid,fillcolor=pink](0,0)(0.8,0)(0.8,0.3)(0,0.3)(0,0)\rput(0.4,0.15){$_{2,0}$}\psline{-}(0.1,0)(0.1,0.1)(0,0.1)
\rput(0.9,0){\pspolygon[fillstyle=solid,fillcolor=pink](0,0)(0.8,0)(0.8,0.3)(0,0.3)(0,0)\rput(0.4,0.15){$_{2,0}$}\psline{-}(0.1,0)(0.1,0.1)(0,0.1)}
\rput(0,2.0){\pspolygon[fillstyle=solid,fillcolor=pink](0,0)(0.8,0)(0.8,0.3)(0,0.3)(0,0)\rput(0.4,0.15){$_{2,0}$}\psline{-}(0.1,0)(0.1,0.1)(0,0.1)
\rput(0.9,0){\pspolygon[fillstyle=solid,fillcolor=pink](0,0)(0.8,0)(0.8,0.3)(0,0.3)(0,0)\rput(0.4,0.15){$_{2,0}$}\psline{-}(0.1,0)(0.1,0.1)(0,0.1)}
\psarc[linewidth=1.5pt,linecolor=blue,linestyle=dashed,dash=2pt 2pt](0.85,0){0.25}{180}{0}
}
\psline[linewidth=1.5pt,linecolor=blue,linestyle=dashed,dash=2pt 2pt]{-}(0.2,0.3)(0.2,2.0)
\psline[linewidth=1.5pt,linecolor=blue,linestyle=dashed,dash=2pt 2pt]{-}(1.4,0.3)(1.4,2.0)
\rput(0.85,0.3){\wobblyarc{0.25}{0}{180}}
\rput(0.85,0.7){\scriptsize $_?$}
\end{pspicture}
\ + \
\begin{pspicture}[shift=-1.05](0,0)(1.7,2.3)
\pspolygon[fillstyle=solid,fillcolor=pink](0,0)(0.8,0)(0.8,0.3)(0,0.3)(0,0)\rput(0.4,0.15){$_{2,0}$}\psline{-}(0.1,0)(0.1,0.1)(0,0.1)
\rput(0.9,0){\pspolygon[fillstyle=solid,fillcolor=pink](0,0)(0.8,0)(0.8,0.3)(0,0.3)(0,0)\rput(0.4,0.15){$_{2,0}$}\psline{-}(0.1,0)(0.1,0.1)(0,0.1)}
\rput(0,2.0){\pspolygon[fillstyle=solid,fillcolor=pink](0,0)(0.8,0)(0.8,0.3)(0,0.3)(0,0)\rput(0.4,0.15){$_{2,0}$}\psline{-}(0.1,0)(0.1,0.1)(0,0.1)
\rput(0.9,0){\pspolygon[fillstyle=solid,fillcolor=pink](0,0)(0.8,0)(0.8,0.3)(0,0.3)(0,0)\rput(0.4,0.15){$_{2,0}$}\psline{-}(0.1,0)(0.1,0.1)(0,0.1)}
\psarc[linewidth=1.5pt,linecolor=blue,linestyle=dashed,dash=2pt 2pt](0.85,0){0.25}{180}{0}
\psarc[linewidth=1.5pt,linecolor=blue,linestyle=dashed,dash=2pt 2pt](0.85,0){0.65}{180}{0}
}
\rput(0.85,0.3){\wobblyarc{0.25}{0}{180}}
\rput(0.85,0.3){\wobblyarc{0.65}{0}{180}}
\rput(0.85,0.7){\scriptsize $_?$}
\rput(0.85,1.1){\scriptsize $_?$}
\end{pspicture}
\ ,
\ee
where each wavy loop segment with a question mark is an unknown linear combination of the diagrams 
$\psset{unit=0.3}
\begin{pspicture}[shift=-.20](0,0)(2,1)
\pspolygon[fillstyle=solid,fillcolor=lightlightblue](0,1)(1,0)(2,1)
\psarc[linewidth=1.5pt,linecolor=blue](1,0){.707}{45}{135}
\end{pspicture}$
and 
$\psset{unit=0.3}
\begin{pspicture}[shift=-.20](0,0)(2,1)
\pspolygon[fillstyle=solid,fillcolor=lightlightblue](0,1)(1,0)(2,1)
\end{pspicture}$. 
This tangle should satisfy the relations
\be
\label{eq:P22.rels}
\begin{pspicture}[shift=-0.75](0,-0.7)(1.5,0.3)
\pspolygon[fillstyle=solid,fillcolor=pink](0,0)(1.5,0)(1.5,0.3)(0,0.3)(0,0)\rput(0.75,0.15){$_{2,2}$}\psline{-}(0.1,0)(0.1,0.1)(0,0.1)
\rput(0.75,0){\pspolygon[fillstyle=solid,fillcolor=lightlightblue](0,0)(0.4,-0.4)(-0.4,-0.4)
\psarc[linewidth=1.5pt,linecolor=blue,linestyle=dashed,dash=2pt 2pt](0,0){0.2}{180}{0}
}
\end{pspicture} \ = 0,
\qquad\quad
\begin{pspicture}[shift=-0.2](0,0)(1.5,0.6)
\pspolygon[fillstyle=solid,fillcolor=pink](0,0)(1.5,0)(1.5,0.3)(0,0.3)(0,0)\rput(0.75,0.15){$_{2,2}$}\psline{-}(0.1,0)(0.1,0.1)(0,0.1)
\rput(0,0.3){\pspolygon[fillstyle=solid,fillcolor=pink](0,0)(1.5,0)(1.5,0.3)(0,0.3)(0,0)\rput(0.75,0.15){$_{2,2}$}\psline{-}(0.1,0)(0.1,0.1)(0,0.1)}
\end{pspicture} \ \ = \ \
\begin{pspicture}[shift=-0.05](0,0)(1.5,0.3)
\pspolygon[fillstyle=solid,fillcolor=pink](0,0)(1.5,0)(1.5,0.3)(0,0.3)(0,0)\rput(0.75,0.15){$_{2,2}$}\psline{-}(0.1,0)(0.1,0.1)(0,0.1)
\end{pspicture}\ ,
\ee
where the dashed arc in the first diagram connects the second and third nodes of $P^{2,2}$. Its recursive definition should ultimately allow us to derive the fusion hierarchy relation \eqref{eq:FH.bulk} for $\Tb^{2,2}(u)$.

Allowing for all three unknown wavy arcs in \eqref{eq:P22} to be distinct linear combinations, we find that there are no solutions for these linear combinations that simultaneously satisfy \eqref{eq:P22.rels} and are consistent with our construction of the projectors $P^{2,0}$ and $P^{0,2}$ in \eqref{eq:Pm0rec} and \eqref{eq:P0n}. In searching for a resolution to this problem, we have also considered the possibility of selecting different solutions, namely \eqref{eq:prefactors} or \eqref{eq:prefactors.prime}, for the projectors $P^{2,0}$ and $P^{0,2}$ appearing in \eqref{eq:P0n}. But this failed as well. It therefore remains an open problem to construct a complete set of Wenzl-Jones projectors for the $A_2^{(2)}$ loop models.

\section{Proof of the functional equations}

\subsection[$T$-system equations]{$\boldsymbol T$-system equations}\label{sec:T.system.proof}

In this section, we prove the $T$-system equations \eqref{eq:Trelations} valid for all $\lambda$. Actually, these relations belong to a larger two parameter $T$-system of bilinear equations as in the following proposition.
\begin{Proposition}
For $m\ge 1$, we have
\begin{subequations}
\label{eq:moreTT}
\begin{alignat}{2}
\Tb^{m,0}_0 \Tb^{m,0}_2 &= \sigma^m f_{-3}f_{2m}\Tb^{0,m}_0 + \Tb^{m+1,0}_0 \Tb^{m-1,0}_2,\label{eq:sameTT}
\\[0.1cm]
\Tb^{m,0}_0 \Tb^{m-k,0}_{2k+2} &= \sigma^{m-k}f_{2m} \Tb^{k,m-k}_0 + \Tb^{m+1,0}_0 \Tb^{m-1-k,0}_{2k+2}, \qquad  1\le k < m.\label{eq:moreTTk}
\end{alignat}
\end{subequations}
\end{Proposition}
\proof
For $m=1$, the only relation to prove is \eqref{eq:sameTT} and it is identical to the fusion hierarchy relation \eqref{eq:FH.corner1}. For $m=2$ and $m=3$, there are respectively two and three identities to establish. These are proved straightforwardly by expanding each fused transfer tangle in terms of the fundamental transfer tangle via \eqref{eq:FH.dets} and verifying that the left and right sides of the equations coincide. For $m>3$, the proof is inductive on $m$ and assumes that it holds for $m\!-\!1$, $m\!-\!2$ and $m\!-\!3$. Applying \eqref{eq:extraFHR1}, we find
\begin{alignat}{2}
&f_{-1}f_0f_1\Tb^{m,0}_0 \Tb^{m-k,0}_{2k+2} = (f_1\Tb^{1,0}_0 \Tb^{m-1,0}_2 - \sigma f_{-3}\Tb^{0,1}_0 \Tb^{m-2,0}_4 + \sigma f_{-3}f_{-2}f_{-1} \Tb^{m-3,0}_6)\Tb^{m-k,0}_{2k+2}
\nonumber\\[0.1cm]
&\hspace{0.2cm}= f_1\Tb^{1,0}_0 (\Tb^{m-1,0}_0\Tb^{m-1-(k-1),0}_{2(k-1)+2})_2 - \sigma f_{-3} \Tb^{0,1}_0(\Tb^{m-2,0}_0 \Tb^{m-2-(k-2),0}_{2(k-2)+2})_4\nonumber \\ &\hspace{0.2cm}
+ \sigma f_{-3}f_{-2}f_{-1} (\Tb^{m-3,0}_0\Tb^{m-3-(k-3),0}_{2(k-3)+2})_6
\label{eq:fffTT}
\end{alignat}
where a parenthesis with a subscript indicates that the $u$-arguments of its entire content are shifted accordingly by the specified multiples of $\lambda$. Each of the parentheses in the last two lines is of the form $\Tb^{m',0}_0 \Tb^{m'-k',0}_{2k'+2}$ with $1\le m'<m$ and $k<m'$. However, for $k=0,1,2,3$, some values of $k'$ are negative, forcing us to treat separately these cases and the case $k>3$. Let us start with the latter. For all three parentheses, we apply \eqref{eq:moreTTk} via the induction hypothesis for $m\!-\!1$, $m\!-\!2$ and $m\!-\!3$ and find
\begingroup
\allowdisplaybreaks
\begin{alignat}{2}
f_{-1}f_0f_1\Tb^{m,0}_0 \Tb^{m-k,0}_{2k+2} &= f_1\Tb^{1,0}_0 (\sigma^{m-k}f_{2m} \Tb^{k-1,m-k}_2 + \Tb^{m,0}_2 \Tb^{m-1-k,0}_{2k+2}) 
\nonumber\\[0.1cm]&\hspace{0.5cm}
- \sigma f_{-3} \Tb^{0,1}_0(\sigma^{m-k}f_{2m} \Tb^{k-2,m-k}_4 + \Tb^{m-1,0}_4 \Tb^{m-1-k,0}_{2k+2}) 
\nonumber\\[0.1cm]&\hspace{0.5cm}
+ \sigma f_{-3}f_{-2}f_{-1} (\sigma^{m-k}f_{2m} \Tb^{k-3,m-k}_6 + \Tb^{m-2,0}_6 \Tb^{m-1-k,0}_{2k+2}) 
\nonumber\\[0.1cm]
& = \sigma^{m-k} f_{2m} (f_1\Tb^{1,0}_0\Tb^{k-1,m-k}_2 - \sigma f_{-3}\Tb^{0,1}_0 \Tb^{k-2,m-k}_4 +  \sigma f_{-3}f_{-2}f_{-1} \Tb^{k-3,m-k}_3) \nonumber\\[0.1cm]&\hspace{0.5cm}
+ (f_1\Tb^{1,0}_0\Tb^{m,0}_2 - \sigma f_{-3} \Tb^{0,1}_0\Tb^{m-1,0}_4+\sigma f_{-3}f_{-2}f_{-1}\Tb^{m-2,0}_6)\Tb^{m-1-k,0}_{2k+2}
\nonumber\\[0.1cm]& = f_{-1}f_0f_1(\sigma^{m-k}f_{2m} \Tb^{k,m-k}_0 + \Tb^{m+1,0}_0\Tb^{m-1-k,0}_{2k+2})
\end{alignat}
\endgroup
where we used \eqref{eq:extraFHR1} at the last step. This completes the proof of the induction hypothesis for $m>3$, $k > 3$.

The same proofs for the special cases $k=0,1,2,3$ require the following identities:
\begin{subequations}
\begin{alignat}{3}
f_{-1}f_0 \Tb^{2,n}_0 &= \Tb^{1,0}_0 \Tb^{1,n}_2 - \sigma f_{-3} \Tb^{0,1}_0 \Tb^{0,n}_4, \qquad &n \ge 1,
\label{eq:T2n}\\[0.2cm]
f_{-1}f_0f_1\Tb^{3,n}_0 &= f_1 \Tb^{1,0}_0 \Tb^{2,n}_2 - \sigma f_{-3}\Tb^{0,1}_0\Tb^{1,n}_4 + \sigma f_{-3}f_{-2}f_{-1}f_3 \Tb^{0,n}_6, \qquad &n \ge 1.
\label{eq:T3n}
\end{alignat}
These are obtained from the determinant formula \eqref{eq:Tmn} and are extensions of \eqref{eq:extraFHR1} to $m=2$ and $m=3$. For completeness, we also write down the similar equations extending \eqref{eq:extraFHR2} to $n=2$ and $n=3$:
\begin{alignat}{3}
f_{2m}f_{2m+1}\Tb^{m,2}_0 &= \Tb^{m,1}_0 \Tb^{0,1}_{2m+2} -\sigma f_{2m+3}\Tb^{m,0}_0\Tb^{1,0}_{2m+2}\qquad\qquad &m \ge 1,\label{eq:Tm2}
\\[0.2cm]
f_{2m+1}f_{2m+2}f_{2m+3}\Tb^{m,3}_0 &= f_{2m+1}\Tb^{m,2}_0 \Tb^{0,1}_{2m+4} -\sigma f_{2m+5}\Tb^{m,1}_0\Tb^{1,0}_{2m+4} \label{eq:Tm3}
\nonumber\\[0.1cm]
&+ \sigma f_{2m-1} f_{2m+3} f_{2m+4} f_{2m+5} \Tb^{m,0}_0\qquad\qquad &m \ge 1.
\end{alignat}
\end{subequations}

For $k=3$, the last parenthesis in \eqref{eq:fffTT} is of the form $\Tb^{m',0}_0 \Tb^{m'-k',0}_{2k'+2}$ with $m' = m-3$ and $k'=0$, so in applying the induction hypothesis, one uses \eqref{eq:sameTT} instead of \eqref{eq:moreTTk}. This yields
\begin{alignat}{2}
f_{-1}f_0f_1\Tb^{m,0}_0 \Tb^{m-3,0}_{8} &= f_1\Tb^{1,0}_0 (\sigma^{m-3}f_{2m} \Tb^{2,m-3}_2 + \Tb^{m,0}_2 \Tb^{m-4,0}_{8}) 
\nonumber\\[0.1cm]&\hspace{0.5cm}
- \sigma f_{-3} \Tb^{0,1}_0(\sigma^{m-3}f_{2m} \Tb^{1,m-3}_4 + \Tb^{m-1,0}_4 \Tb^{m-4,0}_{8}) 
\nonumber\\[0.1cm]&\hspace{0.5cm}
+ \sigma f_{-3}f_{-2}f_{-1} (\sigma^{m-3}f_3f_{2m} \Tb^{0,m-3}_6 + \Tb^{m-2,0}_6 \Tb^{m-4,0}_{8}) 
\nonumber\\[0.1cm]
& = \sigma^{m-3} f_{2m} (f_1\Tb^{1,0}_0\Tb^{2,m-3}_2 - \sigma f_{-3}\Tb^{0,1}_0 \Tb^{1,m-3}_4 +  \sigma f_{-3}f_{-2}f_{-1}f_3 \Tb^{0,m-3}_6) \nonumber\\[0.1cm]&\hspace{0.5cm}
+ (f_1\Tb^{1,0}_0\Tb^{m,0}_2 - \sigma f_{-3} \Tb^{0,1}_0\Tb^{m-1,0}_4+\sigma f_{-3}f_{-2}f_{-1}\Tb^{m-2,0}_6)\Tb^{m-4,0}_{8}
\nonumber\\[0.1cm]&= f_{-1}f_0f_1(\sigma^{m-3}f_{2m} \Tb^{3,m-3}_0 + \Tb^{m+1,0}_0\Tb^{m-4,0}_{8}),
\end{alignat}
where we used \eqref{eq:extraFHR1} and, at the last step, \eqref{eq:T3n}.

For $k=2$, the second parenthesis is modified using \eqref{eq:sameTT}, which yields
\begin{alignat}{2}
f_{-1}f_0f_1\Tb^{m,0}_0 \Tb^{m-2,0}_{6} &= f_1\Tb^{1,0}_0 (\sigma^{m-2}f_{2m} \Tb^{1,m-2}_2 + \Tb^{m,0}_2 \Tb^{m-3,0}_{6}) 
\nonumber\\[0.1cm]&\hspace{0.5cm}
- \sigma f_{-3} \Tb^{0,1}_0(\sigma^{m-2}f_{2m}f_1 \Tb^{0,m-2}_4 + \Tb^{m-1,0}_4 \Tb^{m-3,0}_{4}) 
+ \sigma f_{-3}f_{-2}f_{-1} \Tb^{m-3,0}_6\Tb^{m-2,0}_6
\nonumber\\[0.1cm]
& = \sigma^{m-2} f_{2m} (f_1\Tb^{1,0}_0\Tb^{1,m-2}_2 - \sigma f_{-3}f_1\Tb^{0,1}_0 \Tb^{0,m-2}_4) \nonumber\\[0.1cm]&\hspace{0.5cm}
+ (f_1\Tb^{1,0}_0\Tb^{m,0}_2 - \sigma f_{-3} \Tb^{0,1}_0\Tb^{m-1,0}_4+\sigma f_{-3}f_{-2}f_{-1}\Tb^{m-2,0}_6)\Tb^{m-3,0}_{6}
\nonumber\\[0.1cm]&=f_{-1}f_0f_1(\sigma^{m-2}f_{2m} \Tb^{2,m-2}_0 + \Tb^{m+1,0}_0\Tb^{m-3,0}_{6}),
\end{alignat}
where we used \eqref{eq:extraFHR1} and, at the last step, \eqref{eq:T2n}.

For $k=1$, we again start from \eqref{eq:fffTT} and find
\begin{alignat}{2}
f_{-1}f_0f_1\Tb^{m,0}_0 \Tb^{m-1,0}_4 &= f_1\Tb^{1,0}_0 (\sigma^{m-1}f_{2m} \Tb^{0,m-1}_2 + \Tb^{m,0}_2 \Tb^{m-2,0}_4) 
- \sigma f_{-3} \Tb^{0,1}_0\Tb^{m-1,0}_4\Tb^{m-2,0}_4\nonumber\\[0.1cm]&
+ \sigma f_{-3}f_{-2}f_{-1} (\Tb^{m-2,0}_4\Tb^{m-2,0}_6-\sigma^{m-2}f_1 f_{2m}\Tb^{0,m-2}_4)
\end{alignat}
where the last parenthesis was obtained by applying \eqref{eq:sameTT} via the induction hypothesis, with $k=0$ and $m$ replaced by $m-2$. This yields
\begin{alignat}{2}
f_{-1}f_0f_1\Tb^{m,0}_0 \Tb^{m-1,0}_4 &= \sigma^{m-1}f_{-1}f_1f_{2m} (\Tb^{1,0}_0 \Tb^{0,m-1}_2 - f_{-3}f_{-2}\Tb^{0,m-2}_4)
\nonumber\\[0.1cm]
&\hspace{0.5cm}+ (f_1\Tb^{1,0}_0\Tb^{m,0}_2 - \sigma f_{-3} \Tb^{0,1}_0\Tb^{m-1,0}_4+\sigma f_{-3} f_{-2} f_{-1} \Tb^{m-2,0}_6)\Tb^{m-2,0}_4
\nonumber\\[0.1cm]
& = f_{-1}f_0f_1 (\sigma^{m-1}f_{2m}\Tb^{1,m-1}_0 + \Tb^{m+1,0}_0 \Tb^{m-2,0}_4)
\end{alignat}
where we used \eqref{eq:FH.adj.bdy2} and, at the last step, \eqref{eq:extraFHR1}.

The final step is to check that \eqref{eq:sameTT} is also satisfied via the induction hypothesis. This corresponds to the case $k=0$ in \eqref{eq:fffTT}. We find
\begin{alignat}{2}
f_{-1}f_0f_1\Tb^{m,0}_0 \Tb^{m,0}_2 &= f_1\Tb^{1,0}_0 \Tb^{m,0}_2\Tb^{m-1,0}_2 - \sigma f_{-3} \Tb^{0,1}_0(\Tb^{m-1,0}_2\Tb^{m-1,0}_4-\sigma^{m-1}f_{-1}f_{2m}\Tb^{0,m-1}_2)
\nonumber\\[0.1cm]&
+ \sigma f_{-3}f_{-2}f_{-1} (\Tb^{m-1,0}_2\Tb^{m-2,0}_6-\sigma^{m-2} f_{2m}\Tb^{1,m-2}_0).
\end{alignat}
The last parenthesis on the first line was obtained by applying \eqref{eq:sameTT} via the induction hypothesis, with $m$ replaced by $m-1$. Likewise, the parenthesis on the second line was obtained by applying \eqref{eq:moreTTk} with $k=1$ and $m$ replaced by $m-1$. This yields
\begin{alignat}{2}
f_{-1}f_0f_1\Tb^{m,0}_0 \Tb^{m,0}_2 &= \sigma^m f_{-3}f_{-1}f_{2m}(\Tb^{0,1}_0\Tb^{0,m-1}_2 - \sigma f_{-2}\Tb^{1,m-2}_0)
\nonumber\\[0.1cm]
&+ (f_1 \Tb^{1,0}_0\Tb^{m,0}_2-\sigma f_{-3}\Tb^{0,1}_0 \Tb^{m-1,0}_4 + \sigma f_{-3}f_{-2}f_{-1}\Tb^{m-2,0}_6)\Tb^{m-4,0}_2
\nonumber\\[0.1cm]
& = f_{-1}f_0f_1(\sigma^m f_{-3}f_{2m}\Tb^{0,m}_0 + \Tb^{m+1,0}_0\Tb^{m-1,0}_2)
\end{alignat}
where we used \eqref{eq:FH.bdy2} and, at the last step, \eqref{eq:extraFHR1}. The induction hypothesis has been verified in all cases, so this completes the proof of the proposition.
\eproof

\subsection{Closure of the fusion hierarchy}\label{sec:FHclosure.proof}

In this section, we prove the closure relations of the fusion hierarchy at roots of unity specifed by \eqref{eq:lambda.ab}.
\begin{Proposition}
\label{prop:closure}
For $b \ge 4$, we have
\begin{subequations}
\begin{alignat}{2}
&f_{-1}\Tb^{b,0}_0 = \sigma^{b-a+1}\, \Tb^{b-2,1}_2 - \sigma f_{-3} \Tb^{b-3,0}_4 + f_{-1}f_{-2}f_{-3}\Jb,\label{eq:FHCa}\\[0.1cm]
&f_{-3}\Tb^{0,b}_0 = \sigma\, \Tb^{1,b-2}_0 -\sigma f_{-1} \Tb^{0,b-3}_2+ f_{-1}f_{-2}f_{-3}\Jb,\label{eq:FHCb}
\end{alignat}
\end{subequations}
where $\Jb$ is given by \eqref{eq:J}.
\end{Proposition}
\proof
We prove only \eqref{eq:FHCa}, as the proof of \eqref{eq:FHCb} uses the same ideas. Let us first note that the left and right sides are centered Laurent polynomials in $z$ of degree $3N$. It therefore suffices to prove that the equality holds at $6N+1$ distinct values. We consider the points $u=\xi_j-2\lambda$ and $u=\xi_j-2\lambda+\pi$, for $j=1, \dots, N$, where the function $f_{-2}$ vanishes. For these $2N$ values, we find
\begin{alignat}{2}
f_{-1}f_0f_1 \Tb^{b,0}_0 &\overset{\textrm{\tiny\eqref{eq:extraFHR1}}}{=} f_1 \Tb^{1,0}_0\Tb^{b-1,0}_2 - \sigma f_{-3} \Tb^{0,1}_0 \Tb^{b-2,0}_4
\overset{\textrm{\tiny\eqref{eq:syms}}}{=} f_1 \Tb^{b-1,0}_2\Tb^{1,0}_{2b} - \sigma f_{-3}  \Tb^{b-2,0}_4 \Tb^{0,1}_{2b}
\nonumber\\
&\overset{\textrm{\tiny\eqref{eq:FHs}}}{=} f_1(f_{2b-3}f_{2b-2}\Tb^{b,0}_2 + \sigma f_{2b}\Tb^{b-2,1}_2) - \sigma f_{-3} (f_{2b-2}\Tb^{b-2,1}_4 + f_{2b}f_{2b+1}\Tb^{b-3,0}_4)
\nonumber\\
&\overset{\textrm{\tiny\eqref{eq:syms}}}{=}f_0f_1(\sigma^{b-a-1}\Tb^{b-2,1}_2 - \sigma f_{-3}\Tb^{b-3,0}_4).
\label{eq:f2=0}
\end{alignat}
Dividing out by $f_0f_1$, we obtain precisely \eqref{eq:FHCa} specialised to $f_{-2}=0$. 

We consider $2N$ more values, namely $u=\xi_j-3\lambda$ and $u=\xi_j-3\lambda+\pi$, where the function $f_{-3}$ vanishes. At these points, the same derivation as in \eqref{eq:f2=0} yields
\be
f_{-1}f_0f_1\Tb^{b,0}_0 = f_0f_1(\sigma^{b-a-1}\, \Tb^{b-2,1}_2).
\ee
This reproduces \eqref{eq:FHCa} specialised to $f_{-3}=0$.

We next specialise to the values $u=\xi_j-\lambda$ and $u=\xi_j-\lambda+\pi$, for which $f_{-1}$ vanishes. At these $2N$ values, we have 
\begin{alignat}{2}
\sigma^{b-a} f_{-4} \Tb^{b-2,1}_2 &\overset{\textrm{\tiny\eqref{eq:syms}}}{=} f_{2b-4}\Tb^{b-2,1}_2 \overset{\textrm{\tiny\eqref{eq:FH.adj.bdy1}}}{=} \Tb^{b-2,0}_2 \Tb^{0,1}_{2b-2} - f_{2b-2}f_{2b-1}\Tb^{b-3,0}_2
\nonumber\\
&\overset{\textrm{\tiny\eqref{eq:syms}}}{=} \Tb^{0,1}_{-2}\Tb^{b-2,0}_2 \overset{\textrm{\tiny\eqref{eq:extraFHR1}}}{=} f_{-4}f_{-3} \Tb^{b-3,0}_4.
\end{alignat}
Dividing out by $f_{-4}$, we see that this equality precisely reproduces \eqref{eq:FHCa} specialised to $f_{-1} = 0$.

We have thus checked that the identity holds at $6N$ distinct points. The last value of $u$ can be chosen arbitrarily, with $\Jb$ fixed accordingly as a constant tangle. Clearly, the value of $\Jb$ does not depend on the specific value of $u$ that we choose to fix it. A convenient choice is to fix $\Jb$ such that the equality holds in the braid limit. This is achieved by multiplying both sides of \eqref{eq:FHCa} by $\eE^{-\ir(\pi-2\lambda)Nb}/(f_{-3}f_{-2}f_{-1})$, taking the limit $u \to \ir \infty$ and applying \eqref{eq:braid.T.def}. This calculation is straightforward and yields \eqref{eq:J} as announced.
\eproof

The analogous closure relations for the cases $b=2$ and $b=3$ are proven in the same way. These relations are equalities between centered Laurent polynomials in $z$ of degrees $2N$ for $b=2$ and $3N$ for $b=3$. In the former case, one checks that the equality holds at $4N+1$ distinct values.

We now proceed with the proof of the closure relations for $\Tb^{b,k}(u)$ and $\Tb^{k,b}(u)$ with $k \ge 1$.
\begin{Proposition}
\label{prop:closurek}
For $1 \le k \le b-4$, we have
\begin{subequations}
\begin{alignat}{2}
\Tb^{b,k}_0 &= \sigma\, \Tb^{b-2,k+1}_2 - \sigma^{b-a-1} f_{-3} \Tb^{b-3-k,0}_{2k+4} +\sigma^{b-a} f_{-3} \Jb\,  \Tb^{0,k}_0,\label{eq:extra.closure.k.a}\\[0.1cm] 
\Tb^{k,b}_0 &= \sigma\, \Tb^{k+1,b-2}_0 - \sigma f_{2k-1} \Tb^{0,b-3-k}_{2k+2} + f_{2k-1}\Jb\,  \Tb^{k,0}_0.\label{eq:extra.closure.k.b}
\end{alignat}
\end{subequations}
\end{Proposition}
\proof
We prove only \eqref{eq:extra.closure.k.a}, as \eqref{eq:extra.closure.k.b} follows by the same arguments. The proof of \eqref{eq:extra.closure.k.a} is inductive on $k$. The inductive assumption is that it holds for $k-1$, $k-2$ and $k-3$. We thus show that the relation holds for $k=1,2,3$, and then that the inductive assumption is satisfied for $k \ge 4$. We start with the latter. For $k\ge 4$, we have
\begin{alignat}{2}
f_{2b+2k-5}f_{2b+2k-4}f_{2b+2k-3} \Tb^{b,k}_0 &\overset{\textrm{\tiny\eqref{eq:extraFHR2}}}{=} f_{2b+2k-5} \Tb^{b,k-1}_0 \Tb^{0,1}_{2b+2k-2} - \sigma f_{2b+2k-1}\Tb^{b,k-2}_0 \Tb^{1,0}_{2b+2k-2} 
\nonumber\\[0.1cm]
&\hspace{0.25cm}+ \sigma f_{2b+2k-3}f_{2b+2k-2}f_{2b+2k-1}\Tb^{b,k-3}_0
\nonumber\\[0.1cm]
&\hspace{-3.5cm}\overset{\textrm{\tiny\eqref{eq:extra.closure.k.a}}}{=} f_{2b+2k-5}(\sigma\, \Tb^{b-2,k}_2 - \sigma^{b-a-1} f_{-3}\Tb^{b-2-k,0}_{2k+2}+\sigma^{b-a} f_{-3} \Jb\, \Tb^{0,k-1}_0)\Tb^{0,1}_{2k-2} 
\nonumber\\[0.1cm]
&\hspace{-3.25cm}- \sigma f_{2b+2k-1} (\sigma\, \Tb^{b-2,k-1}_2 - \sigma^{b-a-1}f_{-3}\Tb^{b-1-k,0}_{2k} + \sigma^{b-a}f_{-3} \Jb\, \Tb^{0,k-2}_0)\Tb^{1,0}_{2k-2}
\nonumber\\[0.1cm]
&\hspace{-3.25cm} 
+ \sigma f_{2b+2k-3}f_{2b+2k-2}f_{2b+2k-1}(\sigma\, \Tb^{b-2,k-2}_2 - \sigma^{b-a-1}f_{-3}\Tb^{b-k,0}_{2k-2} + \sigma^{b-a}f_{-3}\Jb\, \Tb^{0,k-3}_0)
\nonumber\\[0.1cm]
&\hspace{-3.4cm} \overset{\textrm{\tiny\eqref{eq:extraFHR}}}{=} f_{2b+2k-5}f_{2b+2k-4}f_{2b+2k-3} (\sigma\, \Tb^{b-2,k+1}_2 - \sigma^{b-a-1}f_{-3}\Tb^{b-3-k,0}_{2k+4}+\sigma^{b-a}f_{-3}\Jb\,\Tb^{0,k}_0),\label{eq:fffTbk0}
\end{alignat}
which indeed confirms the inductive hypothesis.

For $k=3$, the proof is similar, but requires that we use \eqref{eq:Tm3} and \eqref{eq:FHCa}:
\begin{alignat}{2}
f_{2b+1}f_{2b+2}f_{2b+3} \Tb^{b,3}_0 &\overset{\textrm{\tiny\eqref{eq:Tm3}}}{=} f_{2b+1} \Tb^{b,2}_0 \Tb^{0,1}_{2b+4} - \sigma f_{2b+5}\Tb^{b,1}_0 \Tb^{1,0}_{2b+4} + \sigma f_{2b+3}f_{2b+4}f_{2b+5}f_{2b-1}\Tb^{b,0}_0
\nonumber\\[-0.0cm]
&\hspace{-0.07cm}\overset{\substack{\textrm{\tiny\eqref{eq:FHCa}}\\\textrm{\tiny\eqref{eq:extra.closure.k.a}}\\}}{=} f_{2b+1}(\sigma\, \Tb^{b-2,3}_2 - \sigma^{b-a-1} f_{-3}\Tb^{b-4,0}_{6}+\sigma_{b-a} f_{-3} \Jb\, \Tb^{0,1}_0)\Tb^{0,1}_{4} 
\nonumber\\[0.1cm]
&\hspace{0.20cm}- \sigma f_{2b+5} (\sigma\, \Tb^{b-2,2}_2 - \sigma^{b-a-1}f_{-3}\Tb^{b-5,0}_{8} + \sigma^{b-a}f_{-3} \Jb\, \Tb^{0,2}_0)\Tb^{1,0}_{4} 
\nonumber\\[0.1cm]
&\hspace{0.20cm}+ \sigma^{b-a-1} f_{2b+3}f_{2b+4}f_{2b+5}(\sigma\, \Tb^{b-2,1}_2 - \sigma^{b-a-1}f_{-3}\Tb^{b-3,0}_{4} + \sigma^{b-a}f_{-3}\Jb\, \Tb^{0,0}_0)
\nonumber\\[0.1cm]
&\hspace{0.03cm}\overset{\textrm{\tiny\eqref{eq:extraFHR}}}{=} f_{2b+1}f_{2b+2}f_{2b+3} (\sigma\, \Tb^{b-2,4}_2 - \sigma^{b-a-1}f_{-3}\Tb^{b-6,0}_{10}+\sigma^{b-a}f_{-3}\Jb\,\Tb^{0,3}_0),
\end{alignat}
which confirms the result for this case.

For $k=2$, we find
\begin{alignat}{2}
f_{2b-1}f_{2b}f_{2b+1} \Tb^{b,2}_0 &\overset{\textrm{\tiny\eqref{eq:Tm2}}}{=} f_{2b-1} (\Tb^{b,1}_0 \Tb^{0,1}_{2b+2}-\sigma f_{2b+3}\Tb^{b,0}_0 \Tb^{1,0}_{2b+2})
\nonumber\\[0.1cm]
&\hspace{-0.07cm}\overset{\substack{\textrm{\tiny\eqref{eq:FHCa}}\\\textrm{\tiny\eqref{eq:extra.closure.k.a}}\\}}{=} f_{2b-1} (\sigma\, \Tb^{b-2,2}_2 - \sigma^{b-a-1}f_{-3}\Tb^{b-4,0}_6+\sigma^{b-a}f_{-3}\Jb\,\Tb^{0,1}_0)\Tb^{0,1}_2
\nonumber\\[0.1cm]
&\hspace{0.20cm} - \sigma f_3 (\sigma^{b-a-1} \Tb^{b-2,1}_2 - \sigma f_{-3}\Tb^{b-3,0}_4 + f_{-3}f_{-2}f_{-1}\Jb)\Tb^{1,0}_2 
\nonumber\\[0.1cm]
&\hspace{0.20cm} = f_{2b-1}f_{2b}f_{2b+1}(\sigma\, \Tb^{b-2,3}_2 - \sigma^{b-a-1} f_{-3} \Tb^{b-5,0}_8 + f_{-3}\sigma^{b-a}\Jb\, \Tb^{0,2}_0)
\end{alignat}
where we used \eqref{eq:FH.corner3}, \eqref{eq:extraFHR1} and \eqref{eq:Tm3} at the last equality. This proves the result for this case.

Finally, for $k=1$, we find
\begin{alignat}{2}
f_{2b-2}f_{2b-1} \Tb^{b,1}_0 &\overset{\textrm{\tiny\eqref{eq:FH.adj.bdy1}}}{=} f_{2b-1} (\Tb^{b,0}_0 \Tb^{0,1}_{2b}- f_{2b}f_{2b+1}\Tb^{b-1,0}_0)
\nonumber\\[0.1cm]
&\hspace{-1.1cm}\overset{\textrm{\tiny\eqref{eq:FHCa}}}{=} \sigma^{b-a}(\sigma^{b-a-1}\Tb^{b-2,1}_2 - \sigma f_{-3}\Tb^{b-3,0}_4 + f_{-3}f_{-2}f_{-1}\Jb) \Tb^{0,1}_0 - \sigma^{b-a}f_{-1}f_0f_1 \Tb^{b-1,0}_0
\nonumber\\[0.1cm]
&\hspace{-1.03cm}\overset{\textrm{\tiny\eqref{eq:Tm2}}}{=} \sigma(f_{2b-2} f_{2b-1}\Tb^{b-2,2}_2 + \sigma f_{2b+1} \Tb^{b-2,0}_2 \Tb^{1,0}_{2b}) - \sigma^{b-a-1} f_{-3} \Tb^{b-3,0}_4 \Tb^{0,1}_0 
\nonumber\\[0.1cm]
&\hspace{-0.875cm} + \sigma^{b-a} f_{-3}f_{-2}f_{-1}\Jb\, \Tb^{0,1}_0 - \sigma^{b-a}f_{-1}f_0f_1 \Tb^{b-1,0}_0 
\nonumber\\[0.1cm]
&\hspace{-1.1cm}\overset{\textrm{\tiny\eqref{eq:extraFHR1}}}{=} f_{2b-2}f_{2b-1}(\sigma\, \Tb^{b-2,2}_2 - \sigma^{b-a-1}f_{-3}\Tb^{b-4,0}_6 + \sigma^{b-a}f_{-3}\Jb\,\Tb^{0,1}_0).
\end{alignat}
This ends the proof of the last case.
\eproof

\noindent The proofs of the closure relation for the special cases \eqref{eq:more.clo3}, \eqref{eq:more.clo2} and \eqref{eq:more.clo1} follow the same steps as in \eqref{eq:fffTbk0}.

\subsection[Closure of the $Y$-system]{Closure of the $\boldsymbol Y$-system}\label{sec:Y.closure}
In this section, we prove the closure relations for the $Y$-system at roots of unity. This is achieved via the following proposition. The steps of the proof follow those of Proposition C.4 of \cite{MDPR2018}.

\begin{Proposition}
For $\lambda = \frac{(b-a)\pi}{2b}$, we have
\begin{alignat}{2}
\Big(\Tb^{b-1,0}_0&\Tb^{0,b-1}_0-\Tb^{b-2,0}_2\Tb^{0,b-2}_0\Big)\Big(\Tb^{b-1,0}_2\Tb^{0,b-1}_0-\Tb^{b-2,0}_2\Tb^{0,b-2}_2\Big) = \nonumber\\ &f_{-3}f_{-2} \Big( \sigma^{a-1}(\Tb^{0,b-1}_0)^3 +  \Jb (\Tb^{0,b-1}_0)^2 \Tb^{b-2,0}_2 + \sigma \Jb\, \Tb^{0,b-1}_0 (\Tb^{b-2,0}_2)^2 + \sigma^a(\Tb^{b-2,0}_2)^3 \Big).
\label{eq:monster}
\end{alignat}
\end{Proposition}
\proof
We first consider the factor $\Tb^{b-1,0}_0\Tb^{b-1,0}_2(\Tb^{0,b-1}_0)^2$ appearing on the left side: 
\begin{alignat}{2}
\Tb^{b-1,0}_0\Tb^{b-1,0}_2(\Tb^{0,b-1}_0)^2 &\overset{\textrm{\tiny\eqref{eq:Trelations}}}{=} (\sigma^{b-1} f_{-3}f_{2b-2}\Tb^{b-1,0}_1 + \Tb^{b,0}_0\Tb^{b-2,0}_2)(\Tb^{0,b-1}_0)^2
\nonumber\\[0.1cm]
&\hspace{-0.1cm}\overset{\textrm{\tiny\eqref{eq:FH.bulk}}}{=} \sigma^{a-1}f_{-3}f_{-2}(\Tb^{0,b-1}_0)^3 + \Tb^{b-2,0}_2 \Tb^{0,b-1}_0(f_{2b-2}\Tb^{b,b-1}_0 + \Tb^{b-1,0}_0 \Tb^{0,b-2}_{2b+2}).
\end{alignat}
The last term is $\Tb^{b-1,0}_0 \Tb^{0,b-1}_0\Tb^{b-2,0}_2\Tb^{0,b-2}_2$ and appears on the left side of \eqref{eq:monster}. We then use the closure relations to find
\begin{alignat}{2}
\Tb^{b-1,0}_0&\Tb^{b-1,0}_2(\Tb^{0,b-1}_0)^2 - \Tb^{b-1,0}_0 \Tb^{0,b-1}_0\Tb^{b-2,0}_2\Tb^{0,b-2}_2 &
\nonumber\\[0.1cm]
&\overset{\textrm{\tiny\eqref{eq:more.clo1}}}{=} \sigma^{a-1} f_{-3}f_{-2}(\Tb^{0,b-1}_0)^3 + \sigma^{b-a} f_{-2}\Tb^{b-2,0}_2 \Tb^{0,b-1}_0 (\sigma\, \Tb^{b-2,b}_2 + \sigma^{b-a} f_{-3} \Jb\, \Tb^{0,b-1}_0)&
\nonumber\\[0.1cm]
&\overset{\textrm{\tiny\eqref{eq:more.clo2}}}{=} \sigma^{a-1} f_{-3}f_{-2}(\Tb^{0,b-1}_0)^3 + \sigma^{b-a-1} f_{-2}\Tb^{b-2,0}_2 \Tb^{0,b-1}_0 (\sigma \, \Tb^{b-1,b-2}_2 + \sigma^{b-a} f_{-3} \Jb \, \Tb^{b-2,0}_2) 
\nonumber\\[0.1cm]
&\hspace{0.75cm} + f_{-3}f_{-2} \Jb\, \Tb^{b-2,0}_2 (\Tb^{0,b-1}_0)^2
\nonumber\\[0.1cm]
&\hspace{0.15cm} = f_{-3}f_{-2} \big(\sigma^{a-1} (\Tb^{0,b-1}_0)^3 + \Jb (\Tb^{0,b-1}_0)^2 \Tb^{b-2,0}_2 + \sigma \Jb \, \Tb^{0,b-1}_0 (\Tb^{b-2,0}_2)^2 \big) 
\nonumber\\[0.1cm]
&\hspace{0.75cm} + \sigma^{b-a}f_{-2} \Tb^{b-2,0}_2\Tb^{0,b-1}_0 \Tb^{b-1,b-2}_2.
\end{alignat}
The first three terms appear explicitly on the right side of \eqref{eq:monster}. For the last term, we find
\begin{alignat}{2}
\sigma^{b-a}f_{-2} \Tb^{b-2,0}_2\Tb^{0,b-1}_0 \Tb^{b-1,b-2}_2 &\overset{\textrm{\tiny\eqref{eq:FH.bulk}}}{=} \Tb^{b-2,0}_2\Tb^{0,b-1}_0 (\Tb^{b-1,0}_2 \Tb^{0,b-2}_0 - \Tb^{b-2,0}_2 \Tb^{0,b-3}_2)
\nonumber\\[0.1cm]
&\hspace{-3cm}\overset{\textrm{\tiny\eqref{eq:Trelations}}}{=} \Tb^{b-2,0}_2 \Tb^{0,b-2}_0 \Tb^{b-1,0}_2\Tb^{0,b-1}_0 - \Tb^{b-2,0}_2 \Tb^{b-2,0}_2(\Tb^{b-2,0}_1 \Tb^{b-2,0}_3 - \sigma^a f_{-3}f_{-2} \Tb^{b-2,0}_2).
\end{alignat}
These are precisely the missing terms, with the first two terms belonging to the left side of \eqref{eq:monster} and the third belonging to the right side.
\eproof

The closure relation \eqref{eq:Y.closure} for the $Y$-system is then easily obtained from \eqref{eq:monster}. One factorises $\Tb^{b-1,0}_0\Tb^{b-1,0}_2(\Tb^{0,b-1}_0)^2$ from the left side of \eqref{eq:monster}, divides throughout by $\sigma^{a-1}(\Tb^{0,b-1}_0)^3$, recalls the folding property $\Tb^{0,m}_k = \Tb^{m,0}_{k+1}$ and applies
\be
\frac{\Tb^{b-1,0}_0\Tb^{b-1,0}_2}{\sigma^{b-1} f_{-3}f_{2b-2}\Tb^{b-1,0}_1} = \Ib+\tb^{b-1}_0.
\ee
The required result follows readily.


\end{document}